\documentclass[prb,aps,nofootinbib,amssymb,twocolumn,superscriptaddress,10pt,]{revtex4-2}

\usepackage[T1]{fontenc}
\usepackage[dvipsnames]{xcolor}
\usepackage{amsfonts}
\usepackage{amsmath,amssymb,bbold,bm}
\usepackage{array}
\usepackage{color}
\usepackage{comment}
\usepackage{chemformula}
\usepackage{dcolumn}
\usepackage{float,latexsym}
\usepackage{graphicx}
\usepackage{gensymb}
\usepackage{hyperref}
\usepackage{cleveref}
\usepackage{ifthen}
\usepackage{makecell}
\usepackage{multirow}
\usepackage{mathtools}
\usepackage{nicematrix}
\usepackage{physics}
\usepackage{pifont}
\usepackage{placeins}
\usepackage{tabularx}
\usepackage{siunitx}
\usepackage{tikz}
\usepackage[compat=1.1.0]{tikz-feynman}
\usepackage[normalem]{ulem}
\usepackage{xstring}
\usepackage{xr}
\usepackage{mathrsfs}

\usetikzlibrary{decorations.markings}
\usetikzlibrary{positioning}
\usetikzlibrary{arrows.meta}
\usetikzlibrary{calc}
\tikzfeynmanset{warn luatex=false}

\makeatletter
\def\maketag@@@#1{\hbox{\m@th\normalfont\normalsize#1}}
\makeatother

\crefname{appendix}{Appendix}{Appendices}
\crefname{equation}{Eq.}{Eqs.}
\crefname{figure}{Fig.}{Figs.}
\crefname{table}{Table}{Tables}
\crefname{section}{Section}{Sections}
\crefname{enumi}{Point}{Points}
\creflabelformat{appendix}{[#2#1#3]}
\crefmultiformat{figure}{Figs.~#2#1\xdef\mycreffirstarg{#1}#3}%
 { and~#2#1#3}{, #2#1#3}{ and~#2#1#3}
\makeatletter     
\renewcommand\onecolumngrid{% <<<<<<
\do@columngrid{one}{\@ne}%
\def\set@footnotewidth{\onecolumngrid}% <<<<<<<<<<<<<<<<
\def\footnoterule{\kern-6pt\hrule width 1.5in\kern6pt}%
}

\makeatletter
\AtBeginDocument{\let\LS@rot\@undefined}
\makeatother

\renewcommand{\vec}{\mathbf}
\newcommand{\cre}[2]{{#1}^\dagger_{#2}}
\newcommand{\des}[2]{{#1}_{#2}}

\newcommand{\normord}[1]{:\mathrel{#1}:}

\newcommand{\bb}{\mathbf{b}}

\newcommand{\vG}{\mathbf{G}}
\newcommand{\vQ}{\mathbf{Q}}
\newcommand{\vR}{\mathbf{R}}
\newcommand{\vrr}{\mathbf{r}}
\newcommand{\vK}{\mathbf{K}}

\definecolor{colorhhy}{rgb}{0.9, 0.17, 0.31}

\newcommand{\titlePaper}{Pairing from repulsive interactions in twisted WSe2 from functional renormalization}

\newcommand{\TUM}{\affiliation{School of Natural Sciences, Technische Universit\"at München, D-85748 Garching, Germany}}
\newcommand{\MPI}{\affiliation{Max Planck Institute for Solid State Research, D-70569 Stuttgart, Germany}}
\newcommand{\RUB}{\affiliation{Theoretical Physics III, Ruhr-University Bochum, D-44801 Bochum, Germany}}
\newcommand{\Donostia}{\affiliation{Donostia International Physics Center (DIPC), Paseo Manuel de Lardizábal. 20018, San Sebastián, Spain}}
\newcommand{\Princeton}{\affiliation{Department of Physics, Princeton University, Princeton, New Jersey 08544, USA}}
\newcommand{\IKERBASQUE}{\affiliation{IKERBASQUE, Basque Foundation for Science, 48013 Bilbao, Spain}}
\newcommand{\Beijing}{\affiliation{Beijing National Laboratory for Condensed Matter Physics and Institute of Physics, Chinese Academy of Sciences, Beijing 100190, China}}
\newcommand{\ChineseAcademy}{\affiliation{University of Chinese Academy of Sciences, Beijing 100049, China}}
\newcommand{\PeierlsCentre}{\affiliation{Rudolf Peierls Centre for Theoretical Physics, University of Oxford, Oxford OX1 3PU, United Kingdom}}
\newcommand{\CornellSolid}{\affiliation{Laboratory of Atomic and Solid State Physics, Cornell University, Ithaca, NY, USA}}
\newcommand{\CornellNano}{\affiliation{Kavli Institute at Cornell for Nanoscale Science, Ithaca, NY, USA}}
\newcommand{\CornellApplied}{\affiliation{School of Applied and Engineering Physics, Cornell University, Ithaca, NY, USA}}
\newcommand{\MPIHH}{\affiliation{Max Planck Institute for the Structure and Dynamics of Matter, D-22761 Hamburg, Germany}}

\newcommand{\paperAuthors}{% alphabetical order for now
    \author{Hannes Braun} \MPI\TUM
    \author{Dumitru C\u{a}lug\u{a}ru} \PeierlsCentre
    \author{Pietro M. Bonetti} \MPI
    \author{Haoyu Hu} \Princeton \Donostia
    \author{Yan Zhang} \Beijing \ChineseAcademy
    \author{Yi Jiang} \Donostia
    \author{Quansheng~Wu} \Beijing \ChineseAcademy
    \author{Kin Fai Mak} \CornellSolid \CornellNano \CornellApplied \MPIHH
    \author{Jie Shan} \CornellSolid \CornellNano \CornellApplied \MPIHH
    \author{B. Andrei Bernevig} \Donostia \Princeton \IKERBASQUE
    \author{Michael M. Scherer} \RUB   
    \author{Laura Classen} \MPI\TUM
}

\newcommand{\vk}{\mathbf{k}}
\newcommand{\vq}{\mathbf{q}}
\def\abs#1{\left|{#1}\right|}          			% absolute value, \abs{x} gives |x|
\newcolumntype{P}[1]{>{\raggedleft\arraybackslash}p{#1}}
\newcolumntype{R}[1]{>{\centering\arraybackslash}p{#1}}

\let\oldAA\AA
\renewcommand{\AA}{\text{\normalfont\oldAA}}

\newcommand{\tTot}{\text{tot}}
\newcommand{\tMBZ}{\text{MBZ}}
\newcommand{\tInt}{\text{int}}

\newcommand{\expec}[1]{\left\langle #1 \right\rangle}

\crefname{appendix}{Appendix}{Appendices}
\crefname{equation}{Eq.}{Eqs.}
\crefname{figure}{Fig.}{Figs.}
\crefname{table}{Table}{Tables}
\crefname{section}{Section}{Sections}
\creflabelformat{appendix}{[#2#1#3]}

\makeatletter
\AtBeginDocument{\let\LS@rot\@undefined}
\makeatother

\makeatletter
\renewcommand\onecolumngrid{% <<<<<<
\do@columngrid{one}{\@ne}%
\def\set@footnotewidth{\onecolumngrid}% <<<<<<<<<<<<<<<<
\def\footnoterule{\kern-6pt\hrule width 1.5in\kern6pt}%
}

\newcommand{\siSection}{appendix}

\allowdisplaybreaks
\begin{document}
\title{\titlePaper}
\paperAuthors
\let\oldaddcontentsline\addcontentsline
% !TEX root = ./main.tex

\begin{abstract}
Motivated by the observation of superconductivity in twisted bilayer WSe$_2$ over an extended range of displacement fields, reaching from the correlated insulator down to zero field, we analyze the correlated electron phases of this system at hole filling $\nu = -1$ and twist angle $3.65^\circ$ using Hartree--Fock, random phase approximation, and functional renormalization group calculations. 
We derive a continuum model with gate-screened Coulomb interaction from first principles and compare its instabilities to those of a realistic three-orbital Wannier model with a Hubbard interaction and an artificial single-orbital Hubbard model. 
In both cases, we find a superconducting instability of mixed $d$-wave singlet and $p$-wave triplet character, arising adjacent to inter-valley-coherent spin-density-wave order in the phase diagram spanned by displacement field and interaction strength. 
Our additional analysis of the pairing mechanism points to inter-valley coherent spin fluctuations as the pairing glue.
The extent of the pairing regime and the associated critical temperatures depend on the model input, and we trace differences to the interaction range entering the projected interaction.
Within the continuum model, the gate-screened Coulomb interaction yields pairing down to zero displacement field, bringing theory into agreement with experiment.
\end{abstract}

\maketitle

\textit{Introduction.}~~
Moir\'e heterostructures have emerged as a tunable platform for engineering strongly
correlated quantum phases, including superconductors that appear in close proximity to correlated
insulators~\cite{balentsSuperconductivityStrongCorrelations2020, behuraMoirePhysicsTwisted2021, kennesMoireHeterostructuresCondensedmatter2021, kwanMeanfieldModelingMoire2025, liTwistronicsMoireExcitonic2024, liuDevelopingGrapheneBasedMoire2022, melladoMagneticMoireSystems2025, nuckollsMicroscopicPerspectiveMoire2024, renReviewMoireofmoireSuperlattice2025, zhaiTwistronicsMoireSuperlattice2025}. 
Superconductivity was observed early in graphene-based systems~\cite{Cao.2018,Yankowitz.2019,Stepanov.2020,Lu.2019,Hao.2021,ParkSCtrilayer2021,Park.2022,Zhang.2022}
and has more recently been reported in twisted transition-metal dichalcogenide (TMD)
homobilayers; a robust superconducting region appears adjacent to correlated
insulating or metallic states in the phase diagrams of twisted bilayer WSe$_2$
(tWSe$_2$)~\cite{Wang.2020,xiaSuperconductivityTwistedBilayer2025,guoSuperconductivity50degTwisted2025,xiaBandwidthtunedMottTransition2026,Guo.2026,FOU24,GHI21} and twisted bilayer MoTe$_2$~\cite{Sun.2026,xu2026signaturesunconventionalsuperconductivitynear,chang2025evidencecompetinggroundstates,PhysRevX.13.031037,PhysRevX.13.031037,Cai2023,Zeng2023,Park2023MoTe2,li2025universal} as a function of electron density, twist angle, and displacement field. 
Superconductivity in tWSe$_2$ evolves smoothly with twist angle between $3.6^\circ$--$4.7^\circ$~\cite{xiaBandwidthtunedMottTransition2026} and $3.8^\circ$--$5.0^\circ$~\cite{Guo.2026}. 
At the lower end of this window---$3.6^\circ$ in Ref.~\cite{xiaBandwidthtunedMottTransition2026} and $3.8^\circ$ in Ref.~\cite{Guo.2026}---it emerges at hole filling $\nu=-1$ at finite displacement field. 
However, the extent of the superconducting region differs between experiments; it goes down from the insulator to zero field at $3.6^\circ$ in \cite{xiaBandwidthtunedMottTransition2026}, while it is confined to a small region near the insulating state at $3.8^\circ$ in \cite{Guo.2026}.

Motivated by these experimental findings, we study electronically driven superconductivity in tWSe$_2$ and show how sensitively it depends on theoretical modeling. 
Since the pairing mechanism in moir\'e systems remains debated, such an analysis is of broader relevance. 
The band structure, wave functions, and electronic interactions of tWSe$_2$ are faithfully captured using a first-principles continuum model and a gate-screened Coulomb interaction, when the parameters of the moir\'e Hamiltonian are accurately included~\cite{PhysRevB.109.045147,PhysRevLett.132.036501,ZhangPolarization2024,devakulMagicTwistedTransition2021,PhysRevB.108.085117,Kormanyos_2015}.
For many-body calculations, a tight-binding model with extended Hubbard interactions obtained from the continuum model is often more useful. 
As a result, different models have been used in many insightful studies of correlated states in tWSe$_2$~\cite{christos2025approximate,zhu2025superconductivity,guerciTopologicalSuperconductivityRepulsive2024,kleblCompetitionDensityWaves2023,kimTheoryCorrelatedInsulators2025,tuoTheoryTopologicalSuperconductivity2025,fischerTheoryIntervalleycoherentAFM2025,zangHartreeFockStudyMoire2021,WAN24d,munoz-segoviaTwistangleEvolutionIntervalleycoherent2025,PEN26,JMU26,BRI26,pppy-tmlw,ZAN21}.  
Most recently, an extensive analysis of competing ordering tendencies within an \textit{ab-initio}-derived tight-binding model~\cite{fischerTheoryIntervalleycoherentAFM2025} led to a phase diagram consistent with one of the experiments~\cite{Guo.2026}, including the superconducting region in the observed range of finite displacement fields, charge carrier densities, and twist angles.
Interestingly, the superconducting region extending down to zero displacement field observed in another experiment~\cite{xiaBandwidthtunedMottTransition2026} remains to be understood.

While the agreement between continuum and tight-binding models can be made excellent, common approximations such as retaining only a few bands and short-range interactions lead to small differences that can affect the complex interplay of fluctuations driving a pairing instability.
For example, extended Hubbard interactions can enlarge the pairing regime in a simplified triangular-lattice model for moir\'e TMDs~\cite{schererChiralSuperconductivityEnhanced2022,PhysRevB.106.125141}, but reduce it in a realistic tight-binding model for tWSe$_2$~\cite{fischerTheoryIntervalleycoherentAFM2025}.
Apart from the interaction range, differences in quantum geometry~\cite{yu2024quantum,Verma2025} of continuum and tight-binding models can also affect pairing properties~\cite{hu2025geometric,peotta2015superfluidity,hu2019geometric,xie2020topology,hanQuantumGeometricNesting2024, herzog-arbeitmanManybodySuperconductivityTopological2026, herzog-arbeitmanSuperfluidWeightBounds2022, hofmannSuperconductivityPseudogapPhase2020, huhtinenRevisitingFlatBand2022, lauUniversalSuppressionSuperfluid2022, mitscherlingBoundResistivityFlatband2022, rossiQuantumMetricCorrelated2021, tormaSuperconductivitySuperfluidityQuantum2022, vermaOpticalSpectralWeight2021, yuQuantumGeometryQuantum2025}. 
For example, it was pointed out that the pairing potential from a screened Coulomb interaction is directly related to the topological properties of the topmost band in tWSe$_2$~\cite{guerciTopologicalSuperconductivityRepulsive2024}.
More generally, the pairing interaction from an electronic mechanism must depend on the quantum-geometric properties of the particle-hole channel~\cite{QuGeomNest_Kivelson,QuGeomNest_Savary}, because it is obtained by projecting the particle-hole channel onto the pairing channel.
Therefore, a treatment of superconductivity within a first-principles continuum model for tWSe$_2$ with a realistic screened Coulomb interaction, using a method that treats the pairing and particle-hole channels on an equal footing, is needed as a reference for comparison with descendant tight-binding models.

Here we provide an analysis of competing phases, including superconductivity, in tWSe$_2$ within a continuum model extracted from density functional theory~(DFT) using Hartree--Fock~(HF), 
random phase approximation~(RPA), and functional renormalization group~(FRG) calculations. 
We focus on a twist angle of $3.65^\circ$ and map out the phase diagram as a function of displacement field and interaction strength. 
We compare the instabilities obtained from a projected screened Coulomb interaction, a projected Hubbard interaction, and a constant interaction in the topmost valence band.

We find that the screened long-range Coulomb interaction produces a pairing instability down to zero displacement field, in contrast to the Hubbard and constant interactions, which lead to a superconducting dome following the density-of-states peak.
We argue that this reconciles the experimental observation~\cite{xiaBandwidthtunedMottTransition2026} with theoretical modeling.

\begin{figure}
    \centering    
    \includegraphics[width=\columnwidth]{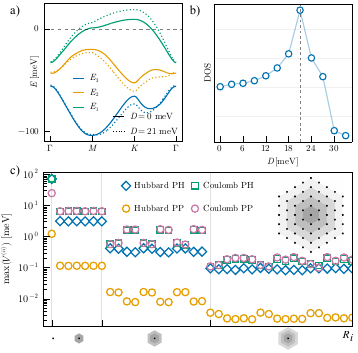}
    \caption{
        (a)~Band dispersion of the three topmost moir\'e bands in valley~$K$ at two values of the displacement field $D$. 
        Bands in valley $K'$ are related by time reversal.
        (b)~Density of states (DOS) of topmost band $E^\sigma_1$ at filling $\nu=-1$ as a function of~$D$.
        (c)~Range of projected Coulomb and Hubbard interactions: maximal magnitude of the Fourier-transformed bare interaction $V^{X,ii}_{\sigma\sigma'}$ in particle-hole (X=PH) and particle-particle (X=PP) channels at distance $\textbf{R}_i$ (\cref{eq:app:Vii} in \cite{SM}) for $D=0$~meV and $10/\epsilon=1.6$ (Coulomb) or $\alpha=1.67$ (Hubbard). 
        The vectors $\mathbf{R}_i$ are defined on the triangular moir\'e lattice numbered along hexagonal shells with increasing distance as depicted in the inset, and accordingly 
        on the x-axis.
        }
    \label{fig:model}
\end{figure}

\textit{Model.}~~ 
We construct the continuum model for tWSe$_2$ at twist angle $\theta = 3.65^\circ$ as a
function of the displacement field $D$ using the average of the parameters obtained from DFT
calculations at two adjacent commensurate angles ($3.48^\circ$ and $3.89^\circ$), see Sec.~\ref{sec:dft_continuum} in \cite{SM}.

Due to strong Ising spin-orbit coupling, spin and valley are locked, i.e., states in valley $K$ ($K'$) are fully spin-up (-down), which we label by $\sigma = (K,\uparrow), (K',\downarrow)$. 
We show the three highest bands in valley $K$ in \cref{fig:model}~(a).
We focus on the topmost bands $E_1^{\sigma}(\vk)$ at filling $\nu=-1$, where superconductivity was observed. 
The displacement field changes the band energies and wave functions. 
In particular, it tunes a Van Hove singularity (VHS) through the Fermi level at $D_\text{VH}\approx 21$~meV, see Fig.~\ref{fig:model}~(b). 
Furthermore, the highest valence bands possess a nontrivial topology with a Chern number sequence that depends on the displacement field  (\cref{fig:Gap_chern_number_ele_field_DFT} in \cite{SM}), obstructing a single-orbital description of the topmost band.
Instead, the topology and symmetry properties of the continuum model allow us to formulate different types of Wannierized models that correctly capture the energies and wave functions of the topmost valence band in each valley. 
As we describe in detail in~\cite{SM}, we build a three-orbital model on a triangular lattice (3Tri; Sec.~\ref{app:DFT3Tri} in \cite{SM}), a model with one orbital on a triangular lattice and two orbitals on a honeycomb lattice (1Tri2Hon; Sec.~\ref{app:DFT1tri2hon} in \cite{SM}), and a two-orbital model on a honeycomb lattice (2Hon; Sec.~\ref{app:DFT2hon} in \cite{SM}). In the following calculations, we focus on the 3Tri model. 
We note that the overlap between the eigenstates of the highest valence band in the continuum and 3Tri model exceeds 99\%, indicating that the 3Tri model provides a faithful description of the relevant low-energy degrees of freedom.

We equip the single-particle Hamiltonian with a gate-screened Coulomb interaction $V(\vq) = e^2/(2\epsilon_0\epsilon|\vq|\Omega_0)\tanh(|\vq|d/2)$ with dielectric constant $\epsilon$, gate distance $d$, and unit cell area $\Omega_0$. 
Projecting onto the highest valence band per valley ($b=1$) yields a general momentum-dependent interaction of the form
\begin{align}
    \mathcal V = 
    \frac{1}{2N_q}
    \sum_{\vq,\vk,\vk'}\sum_{\sigma,\sigma'}
    &
    V_{\sigma\sigma'}(\vq;\vk,\vk') \notag\\
    &\times 
    \tilde{\gamma}^\dagger_{\vk, \sigma}
    \tilde{\gamma}^{\phantom{\dagger}}_{\vk+\vq, \sigma}
    \tilde{\gamma}^\dagger_{\vk'+\vq, \sigma'}
    \tilde{\gamma}^{\phantom{\dagger}}_{\vk', \sigma'}
    \,,\label{eq:generalia}
\end{align}
where $N_q$ is the number of momentum points, and $\tilde{\gamma}_{\vk,\sigma}$ annihilates an electron in the first band $b=1$ with vector $\vk$ and spin/valley $\sigma$.
%%%%%%%%%%%%%%%%%%%%%%%%%%%
The projected interaction is genuinely long-range in real space, see Fig.~\ref{fig:model}~(c), and is given by
$V_{\sigma\sigma'}(\vq;\vk,\vk')
=\sum_{\mathbf{G}}
V(\vq+\mathbf{G}) 
\Lambda^\sigma(\vk+\vq,-\vq-\mathbf{G})
\Lambda^{\sigma'}(\vk',\vq+\mathbf{G})
$
with reciprocal lattice vectors $\mathbf{G}$. The form factor $
\Lambda^\sigma(\vk,\vq+\mathbf{G})
=
\sum_{\mathbf{Q}}\mathbf{U}^\sigma_{\mathbf{Q},1}(\vk)
(\mathbf{U}^\sigma_{\mathbf{Q-G},1}(\vk+\vq))^\ast 
$ is built from eigenstates $\mathbf{U}^\sigma_{\mathbf{Q},1}(\vk)$ with $\mathbf{Q}\in \mathcal Q_l^\sigma=\{\mathbf{G}+\sigma(-1)^l(\mathbf{K}_b-\mathbf{K}_t)\}$ and $\mathbf{K}$ points of top ($t$) and bottom ($b$) layers. 
It measures the overlap of Bloch states at different momenta and thereby also encodes the quantum geometry of the band. 
Here, we project the interaction by first transforming to the Wannier basis via $\mathbf{L}_{n,\mathbf{Q}}^\sigma$ and then  to band space via $\mathbf{Y}^\sigma_{n,1}$ using $\mathbf{U}^\sigma_{\mathbf{Q},1}(\vk)=\sum_n{\mathbf{Y}^\sigma_{n,1}\mathbf{L}^\sigma_{n,\mathbf{Q}}}$ with Wannier orbitals $n$. 

Alternatively, a common approximation keeps only an on-site Hubbard interaction $U$ in the Wannier basis corresponding to a stronlgy screened limit with purely local interaction. 
This gives $V_{\sigma\sigma'}(\vq;\vk,\vk')
=U\sum_{n,m,n',m'}\tilde\Lambda_{nm}^\sigma(\vk+\vq,-\vq)
\tilde\Lambda_{m'n'}^{\sigma'}(\vk',\vq)$ with form factors built from Wannier eigenstates $\tilde\Lambda_{nm}^{\sigma}(\vk,\vq)=\mathbf{Y}^\sigma_{n,1}(\vk)
(\mathbf{Y}^\sigma_{m,1}(\vk+\vq))^\ast$. 
We note that the Hubbard approximation also neglects the momentum and orbital dependence of the transformation $\mathbf{L}_{n,\mathbf{Q}}^\sigma$, which is controlled by the spatial extent of the Wannier orbitals and becomes small for strongly localized orbitals. 
We compare the resulting instabilities to the ones from the screened Coulomb interaction below.

Finally, as an artificial reference we also compare to a momentum-independent interaction 
$V_{\sigma\sigma'}(\vq,\vk,\vk')=g$, 
which contains no quantum-geometric contributions. 
It mimics a Hubbard interaction in a single-orbital description of the highest valence band if such a description were not topologically obstructed.

We parameterize the strength of the Hubbard interaction by $\alpha$, defined through $U= \alpha W$ (and likewise for $g$), where $W$ is the bandwidth. 
There is no unique way to set the overall interaction strength when comparing the momentum-dependent interactions. 
For concreteness, we choose to set their maxima equal. 
This means that we determine $10/\epsilon$ and $\alpha$ so that the gauge-invariant part $\max_{\sigma\sigma'}\sum_{\vk,\vk'}V_{\sigma,\sigma'}(\vq=0,\vk,\vk')$ is the same for the different models. 
The resulting $U$ is of the same order as the DFT estimates (Sec.~\ref{app:DFT3Tri} in \cite{SM}). 
Furthermore, we checked that the functional form of the displacement-field dependence of the superconducting critical temperature is insensitive to the relation between the interaction strengths.

\textit{Methods.}~~
We employ unrestricted HF (Sec.~\ref{app:sec:HF}), RPA (Sec.~\ref{sec:app:RPA}), and FRG (Sec.~\ref{sec:app:fRG_results}) calculations \cite{SM} to map out the phase diagram of ordered states as a function of displacement field~$D$ and interaction strength~$10/\epsilon$ (Coulomb interaction) or $\alpha$ (Hubbard and constant interactions).
We investigate particle-hole orders involving spin-valley or charge degrees of freedom either via RPA or via HF at zero temperature. 
We use FRG to analyze not only particle-hole orders but also particle-particle, i.e., superconducting, instabilities. 
FRG treats all interaction channels on an equal footing and can thus capture fluctuation-induced pairing~\cite{salmhoferFermionicRenormalizationGroup2001,kopietzIntroductionFunctionalRenormalization2010,metznerFunctionalRenormalizationGroup2012,platt_functional_2013,dupuis_nonperturbative_2021,honerkampRecentDevelopmentsFunctional2022}.
For FRG, we use a static, temperature-flow truncated-unity scheme in a level-two truncation for the two-particle vertex~\cite{honerkampTemperatureflowRenormalizationGroup2001,lichtensteinHighperformanceFunctionalRenormalization2017,schoberTruncatedUnityFunctionalRenormalization2018,oConsistentCombinationTruncatedunity2024}. 
Self-energy corrections are neglected, as they affect the flow of the vertex only at third order in the interaction~\cite{honerkampInteractionFlowMethod2004,metznerFunctionalRenormalizationGroup2012,husemannFrequencyDependentVertex2012,PhysRevB.96.235110,vilardiDynamicalFunctionalRenormalization2020,SciPostPhys.6.1.009,hilleQuantitativeFunctionalRenormalization2020}.

In our RPA and FRG calculations, we precisely resolve the momentum dependence of the vertex in terms of a hybrid real- and momentum-space representation~\cite{eckhardtTruncatedUnityParquet2020,husemannEfficientParametrizationVertex2009,oConsistentCombinationTruncatedunity2024,delapenaCompetingElectronicInstabilities2017,lichtensteinHighperformanceFunctionalRenormalization2017,profeTU$$^2$$FRGScalableApproach2022,schoberTruncatedUnityFunctionalRenormalization2018} so that incommensurate wave vectors can also be detected, see Sec.~\ref{sec:app:FRG} in \cite{SM} for details.

In HF, we employ a $36\times36$ discretization of the moir\'e Brillouin zone and allow for moir\'e translation-symmetry breaking in supercells of up to $6\times 6$. 
For RPA and FRG, we set a lower temperature cutoff of $T_\text{cutoff}=5\,\mathrm{mK}$, applied if the flow has not already been terminated by a divergent two-particle vertex.

\begin{figure}
    \centering
    \includegraphics[width=1.0\columnwidth]{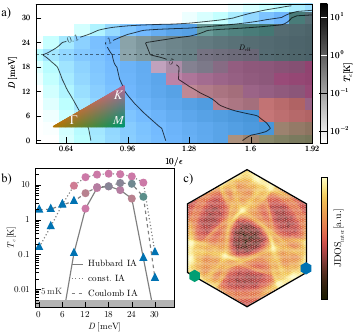}
    \caption{ (a) FRG phase diagram of interaction strength~$10/\epsilon$ and displacement field~$D$. 
    IVC-SDW instabilities are represented by the colors shown in the moir\'e BZ inset, which represent the IVC-SDW wave vector.
    Superconducting instabilities are represented by the blue tones. 
    The opacity of the coloring represents the temperature of the emerging instability $T_c$.
    (b) Line-cut through the phase diagram at $10/\epsilon=1.36$ and comparable $\alpha=0.5$ for the three different initial interactions, i.e., Coulomb/Hubbard/constant as described in the main text.
    Circles represent IVC-SDW instabilities, triangles superconducting instabilities. 
    Notably, for the projected Coulomb interaction, the superconducting $T_c$ is systematically the highest for small $D$.
    (c) Inter-valley JDOS for $D=18$~meV with wave vectors of the FRG IVC-SDW (blue marker) and the RPA IVC-SDW (green marker).
    }
    \label{fig:fRG_phasediagram}
\end{figure}

\textit{Phase diagram.}~~We present the FRG phase diagram for the continuum model and projected double-gate-screened Coulomb interaction at hole filling $\nu=-1$ in Fig.~\ref{fig:fRG_phasediagram}~(a).
We find an extended region of inter-valley-coherent spin-density-wave (IVC-SDW) order with a wave vector that depends on the displacement field. 
This order breaks translation symmetry on both the moir\'e and WSe$_{2}$ scales, while spin-valley locking makes the inter-valley coherence simultaneously an in-plane spin order. 
For sufficiently large interaction strength, $10/\epsilon\gtrsim1.8$, the IVC-SDW region spans the entire displacement field range down to $D=0$~meV.
It survives to lower values of $10/\epsilon\gtrsim 0.96$, where it becomes centered around the Van Hove displacement field $D_\text{VH}$ due to the enhancement of the critical scales by the peaked density of states (DOS).
When IVC-SDW order is suppressed by the displacement field or the interaction strength, we find superconducting instabilities.
The pairing region at $D < D_\text{VH}$ extends all the way to $D=0$ with a remarkably slowly decreasing critical temperature $T_c$ and is significantly broader in $D$ than in the Hubbard 3Tri and constant single-orbital cases, see Fig.~\ref{fig:fRG_phasediagram}~(b) and Sec.~\ref{sec:app:fRG_results_constant_interaction} and \ref{sec:app:fRG_results_hubbard_interaction} in \cite{SM}. 
Lines of constant $T_c$ run approximately parallel to the IVC-SDW phase transition, so that a higher low-temperature cutoff (modeling, for example, the experimental resolution) would lead to the pairing region wrapping around the IVC-SDW phase.
For $D>D_\text{VH}$, pairing is present only in a small sliver and is much more fragile; the critical temperature $T_c$ drops rapidly, so that superconductivity disappears for $D\gtrsim30$~meV.  
The $T_c$ behavior here resembles that found for a constant or Hubbard interaction (see Fig.~\ref{fig:fRG_phasediagram}~(b)) and follows the decreasing DOS. 
The superconducting critical temperatures, $5\,\mathrm{mK} \lesssim T_c^{\mathrm{SC}} \lesssim 2\,\mathrm{K}$, are considerably smaller than the IVC-SDW critical scales, $2\,\mathrm{K}\lesssim T_{\mathrm{IVC}}\lesssim 20\,\mathrm{K}$, 
especially for the IVC-SDW near $D_\text{VH}$.

\textit{Inter-valley coherence.}~~The wave vector $\vq$ of the IVC-SDW state depends on the displacement field, while it changes only slightly with the interaction strength (see the color code in Fig.~\ref{fig:fRG_phasediagram}~(a)).
We generally find an incommensurate wave vector, which lies in the region around~$K$. 
It is very close to $K$ for $D\approx 18$~meV, lies between $K$ and $\Gamma$ for small~$D$, and moves towards $M$ for larger $D$. At $K$, the IVC-SDW corresponds to a 120$^\circ$ antiferromagnetic order. 
We observe the shift in wave vector for all types of initial interactions (Coulomb, Hubbard, and constant), indicating the importance of Fermi surface properties in determining the wave vector. 
We can demonstrate the effect of the Fermi surface more directly by comparing to the joint DOS (JDOS) at the Fermi level 
$\mathrm{JDOS}_{\mathrm{inter/intra}}(\vq)= \int_\vk\delta(E_1^\sigma(\vk))\delta(E_1^{\bar\sigma/\sigma}(\vk+\vq))/V_{\rm{BZ}}$.
We find that the IVC-SDW wave vector coincides with a peak of the inter-valley JDOS (see Fig.~\ref{fig:fRG_phasediagram}~(c) and Fig.~\ref{fig:app:single_particle_prop} in \cite{SM}), which explains the evolution. Namely, the wave vector connects points with large overlap in the $\vq$-shifted Fermi surface at general $D$ or Van Hove points at $D_{\textrm{VH}}$.

\textit{Spin-valley order from RPA and HF.}~~
We also find IVC-SDW order with HF (Fig.~\ref{app:fig:HF_DFT} in \cite{SM}) and RPA (Fig.~\ref{fig:app:rpa_susceptibilities} in \cite{SM}) up to intermediate interaction strengths ($10/\epsilon\lesssim 1.6$ in RPA and $10/\epsilon< 1.4$ in HF). 
In RPA, the incommensurate IVC-SDW wave vector follows the FRG behavior, i.e., it lies around $K$ moving slightly towards $\Gamma$ or $M$ depending on $D$. 
In HF, for the supercell sizes we have explored, we obtain an IVC-SDW correlated insulator with a finite band gap and zero Chern number and find that its IVC-SDW wave vector is pinned to $K$ for $10/\epsilon< 1.2$ or $M$ for $10/\epsilon=1.2$. 
For large $D\gtrsim 33$~meV, the HF calculations show signatures of layer polarization.

Overall, RPA, HF, and FRG qualitatively agree on the IVC-SDW phase in the non-superconducting region of the phase diagram, which was also found in previous works~\cite{devakulMagicTwistedTransition2021,FOU24,ZAN21,WAN24d,BRI26,fischerTheoryIntervalleycoherentAFM2025,Guo.2026,kleblCompetitionDensityWaves2023,tuoTheoryTopologicalSuperconductivity2025,zangHartreeFockStudyMoire2021,MUN25,PEN26}. 
The difference in commensurability of the wave vector can be due to energetic advantages from pinning to high-symmetry points or due to the smaller resolution of tested wave vectors in HF. 
For larger interactions $10/\epsilon\gtrsim1.9$ and small displacement field $D\lesssim 6$~meV, both RPA and HF predict a spin-valley-polarized state, while FRG still yields an IVC-SDW instability, indicating that inter-channel fluctuations qualitatively affect phase boundaries.

\begin{figure}
    \centering
    \includegraphics[width=\columnwidth]{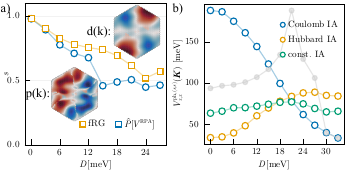}
    \caption{
    (a) Weight $s$ of the pairing eigenfunction $\Delta(\vk)=s\cdot d(\vk)t_0 +(1-s)\cdot p(\vk) t_z$ with Pauli matrices in spin/valley space $t_i=i\sigma_i\sigma_y$ for $10/\epsilon=0.8$ extracted from FRG and from an IVC-spin-fluctuation approach which projects the RPA vertex into the pairing channel $\hat P [V^{\text{RPA}}]$. 
    Inset: $d(\vk)$ and $p(\vk)$ in the Brillouin zone.
    (b) The RPA-dressed interactions in the IVC-SDW channel for Coulomb/Hubbard/constant interaction as a function of $D$. 
    The DOS is shown in gray for reference.
    }
    \label{fig:pairing}
\end{figure}

\textit{Pairing.}~~To extract the pairing symmetry, we diagonalize the singular pairing vertex close to the critical temperature, see insets of Fig.~\ref{fig:pairing}(a). 
We find a mixed pairing state with $d$-wave singlet and $p$-wave triplet symmetry, both belong to the irreducible representation~$E$ of the point group~$C_{3v}$. 
The mixing between singlet and triplet pairing is enforced by the broken SU(2) symmetry, which, in turn, is related to the absence of inversion symmetry in tWSe$_2$. 
Inversion symmetry emerges at $D=0$~meV within the first-harmonic approximation and is satisfied only approximately in the more precise DFT-derived continuum model (Sec.~\ref{sec:band_structure_and_topology} in \cite{SM}). 
Accordingly, we see in Fig.~\ref{fig:pairing}(a) that the triplet admixture increases with displacement field from being negligible at~$D=0$~meV to approximately equal weight around~$D_{\rm{VH}}$. 
In contrast, there is only a weak dependence on the interaction strength~$10/\epsilon$.

To identify the main pairing mechanism, we compare the FRG results to an IVC-spin-fluctuation approach, i.e., we run an RPA calculation for the crossed particle-hole vertex in the pairing regions of the phase diagram, project the RPA-dressed particle-hole vertex $V^{\text{RPA}}$ onto the Cooper channel, and diagonalize the resulting vertex, see App.~\ref{sec:app:pairing_mechanism} in \cite{SM}. 
We find the leading attractive eigenmode of this projected vertex reproduces the FRG pairing eigenfunction~$\Delta(\vk)$. Similarly, the singlet-triplet mixing ratio as a function of~$D$ predicted by this IVC-spin-fluctuation approximation matches the FRG result across the phase diagram (Fig.~\ref{fig:pairing}(a), Fig.~\ref{fig:app:pairing_projection_additional} in \cite{SM}). 
Thus, we conclude that the pairing mechanism is consistent with IVC-spin fluctuations between the two valleys in agreement with \cite{fischerTheoryIntervalleycoherentAFM2025}. 
We emphasize in this context that our FRG calculation goes beyond such a spin-fluctuation approach by taking all inter-channel fluctuations into account, which is necessary to validate the IVC-spin-fluctuation mechanism.

\textit{Model dependence.} 
To isolate the effect of the interaction parameterization, we compare the screened Coulomb interaction, the Hubbard interaction in the 3Tri lattice model, and the artificial single-orbital case (constant-in-momentum interaction without form factors). 

The Coulomb interaction produces a markedly more extended and robust pairing dome  than all other cases (Fig.~\ref{fig:fRG_phasediagram}(b)). 
We trace this directly to the availability of fluctuations for the pairing glue: Fig.~\ref{fig:pairing} b) shows that the RPA dressed interaction in the IVC-SDW channel, which mediates the Cooper pairing, grows towards small displacement fields in the Coulomb case, while it decreases in the Hubbard and single-orbital cases.
This increase simultaneously compensates for both the drop of the DOS away from $D_{\mathrm{VH}}$ and the nominally larger pairing repulsion of the bare Coulomb interaction (Fig.~\ref{fig:model}(c)). 
Since both models share the same band structure, the different behavior must originate from the bare interactions, namely from their range and form factors. 
The Hubbard interaction corresponds to the limit of an extremely screened, contact-like interaction and infinitely localized Wannier orbitals. 
After projection onto the topmost band, the Hubbard interaction still becomes momentum dependent and acquires non-local components. 
However, it remains more localized in real space than the projected Coulomb interaction, with substantially weaker inter-site components, see Fig.~\ref{fig:model}(c). 
The Hubbard and  constant, single-orbital interaction have the same range so that the different pairing behavior between them reveals the role of the form factors encoding the transformation from Wannier to band basis. 
In both cases, the pairing temperature behaves qualitatively similar, see Fig.~\ref{fig:fRG_phasediagram} b). 

\textit{Discussion.}~~
Our combined HF, RPA, and FRG study with \textit{ab-initio} initial conditions reveals a rich phase diagram of tWSe$_2$ at $\nu=-1$ and $\theta=3.65^\circ$ as a function of displacement field and interaction strength, which depends on gates and substrates in experiment. 
We find an IVC-SDW dome around the displacement field where a VHS crosses the Fermi level. 
Its extent strongly depends on the interaction strength.
Adjacent to the IVC-SDW phase, we find a robust electronically driven superconducting instability with mixed $d$- and $p$-wave singlet-triplet pairing symmetry. 
The superconducting region extends down to zero displacement field, while lines of almost constant critical temperature~$T_c$ follow the IVC-SDW transition. 
We also find a pronounced asymmetry in the displacement-field dependence of the superconducting $T_c$ around the VHS, which is driven not only by the asymmetry of the DOS but also by the structure of the projected initial interactions. 

A key observation of this work, which goes beyond the application to WSe$_{2}$, is that details of the theoretical modeling have a critical impact on the phase diagram. 
A projected screened Coulomb interaction produces a substantially larger and more stable pairing phase than a projected Hubbard interaction or a momentum-independent constant interaction in band space. 

Interestingly, our theoretical phase diagram using the Coulomb interaction exhibits a stable superconducting dome down to zero displacement field and is in qualitative agreement with the experimental phase diagram in~\cite{xiaBandwidthtunedMottTransition2026} at $\nu=-1$, where the superconducting region also extends to zero displacement field at $\theta=3.6^\circ$ and remains sizable next to a correlated insulating phase for increasing angle. 
The superconducting region in another experiment shrinks with decreasing angle down to $\theta \sim 3.8^\circ$ and disappears before reaching zero displacement field~\cite{Guo.2026} consistent with the pairing behavior we get from a Hubbard interaction and in agreement with FRG data presented in~\cite{Guo.2026}. 
Based on our analysis, we argue that the theoretical differences come from different modeling, while the difference in experimental findings may be attributed to different twist angles and sample properties. 
Overall, the experimentally observed phases of broken symmetry are consistent with our description via Fermi-liquid instabilities up to intermediate interactions approaching the order of the band width. 
However, we note that our methodology cannot address other observations such as a strange metal regime \cite{xiaBandwidthtunedMottTransition2026} which may be described more conveniently via strong-coupling treatments in future studies.

From the perspective of moir\'e systems more broadly, our results point to the general principle that theoretical treatments of interaction-driven superconductivity in topological moir\'e bands benefit from using Coulomb or sufficiently extended Hubbard interactions and form factors derived from the continuum (or \textit{ab-initio}) eigenstates.

\begin{acknowledgments}
	We thank Jiabin Yu, Nico Gneist, Lennart Klebl, and Jeyong Park for discussions. D.C. gratefully acknowledges the support provided by the Leverhulme Trust and the support from the UKRI Horizon Europe Guarantee (Grant no. EP/Z002419/1 for a European Research Council Consolidator Grant to S. A. Parameswaran). Y.J. was supported by the European Research Council (ERC) under the European Union’s Horizon 2020 research and innovation program (Grant Agreement No. 101020833), as well as by the IKUR Strategy under the collaboration agreement between Ikerbasque Foundation and DIPC on behalf of the Department of Education of the Basque Government. Q. W. was supported by the National Key R\&D Program of China (2024YFA1408400, 2023YFA1607400).
    B.A.B. was supported by the Gordon and Betty Moore Foundation (grant GBMF8685) towards the Princeton theory program, the Gordon and Betty Moore Foundation’s EPiQS Initiative (grant GBMF11070), the Global Collaborative Network Grant at Princeton University, the Simons Investigator Grant No. 404513, the NSF MRSEC (grant DMR-2011750), the Simons Collaboration on New Frontiers in Superconductivity (grant SFI-MPS-NFS-00006741-01), the Princeton Catalysis Initiative (PCI), the Schmidt Transformative Technology Fund at Princeton University and the NSF through the AI Research Institutes (program award DMR-2433348). 
    M.M.S. was funded by DFG within Project-ID 277146847, SFB 1238 (project C02), and the DFG Heisenberg programme (Project-ID 452976698).
    LC was supported by the Simons Foundation SFI-MPS-NFS-00006741-11, the Deutsche Forschungsgemeinschaft (DFG, German Research Foundation) – TRR 360 – 492547816, and the European Union (ERC-2023-STG, Project 101115758 - QuantEmerge). Views and opinions expressed are, however, those of the authors only and do not necessarily reflect those of the European Union or the European Research Council Executive Agency. Neither the European Union nor the granting authority can be held responsible for them.
\end{acknowledgments}

%TC:ignore
\renewcommand{\addcontentsline}[3]{}
\bibliographystyle{apsrev4-2}
\bibliography{bibliography.tex}
\let\addcontentsline\oldaddcontentsline

\renewcommand{\thetable}{S\arabic{table}}
\renewcommand{\thefigure}{S\arabic{figure}}
\renewcommand{\theequation}{S\arabic{section}.\arabic{equation}}
\onecolumngrid
\pagebreak
\thispagestyle{empty}
\newpage
\begin{center}
	\textbf{\large Supplementary Information for ''\titlePaper``}\\[.2cm]
\end{center}

\appendix
\renewcommand{\thesection}{\Roman{section}}
\tableofcontents
\let\oldaddcontentsline\addcontentsline
\newpage
\section{Twisted \ch{WSe2}}
\label{sec:moire_geometry}

We use the following basis vectors for the moir\'e unit cell:
\begin{equation}
    \mathbf{a}_{M,1}=a_M (\frac{\sqrt{3}}{2}, -\frac{1}{2}),\
    \mathbf{a}_{M,2}=a_M (0, 1), \
    \bb_{M,1}= \frac{4\pi}{\sqrt{3} a_M}(1, 0),\
    \bb_{M,2}= \frac{4\pi}{\sqrt{3} a_M}(\frac{1}{2}, \frac{\sqrt{3}}{2}),
\end{equation}
where $a_M=a_0/2\sin(\frac{\theta}{2})$ is the moir\'e lattice constant.

We first consider monolayer \ch{WSe2}.
At the $K$ valley, the top valence band is mainly composed of the $d_{x^2-y^2}, d_{xy}$ orbitals of W and exhibits spin-valley locking due to strong SOC: the top valence band is spin up at the $K$ valley and spin down at the $-K$ valley.
We initially omit the spin index.
DFT gives the $C_3$ eigenvalue of the top valence band at $K$ as $e^{i2\pi/3}$.
Thus, we can choose the Wannier basis as $d_{x^2-y^2} + id_{xy}$ (denoted as $d$ for simplicity).

We define $\vK_l$ as the $\vK$ point of layer $l$, i.e.,
\begin{equation}
    \mathbf{K}_b=\frac{4\pi}{3a_0} (\cos(\frac{\theta}{2}), \sin(\frac{\theta}{2}))^T,\quad
    \mathbf{K}_t=\frac{4\pi}{3a_0} (\cos(\frac{\theta}{2}), -\sin(\frac{\theta}{2})),\quad
\end{equation}
The vectors $\vq_i$ connect the $\vK$ points of the two layers:
\begin{equation}
    \vq_1=\vK_b-\vK_t=\frac{8\pi}{3a_0} \sin(\frac{\theta}{2}) (0,1)^T,\quad \vq_2=C_3\vq_1,\quad \vq_3=C_3^2\vq_1
\end{equation}

We use $c_{\vrr,\sigma,l}^\dag$ to denote the electron creation operator, where $\sigma =\pm 1$ labels the $\pm K$ valley, $l$ labels the layer, and $\vrr$ denotes the position.
The real-space operator can be expressed in terms of momentum-space operators as
\begin{align}
     & c^\dagger_{\vrr,\sigma,l} = \frac{1}{\sqrt{\Omega_{\tTot}}} \sum_\vk \sum_{\vQ\in \mathcal{Q}_l^\sigma} e^{-i(\vk-\vQ)\cdot \vrr} c_{\vk,\sigma,\vQ}^\dagger\nonumber \\
     & c_{\vk,\sigma,\vQ}^\dag = \frac{1}{\sqrt{\Omega_{\tTot}}}\int d^2\vrr c_{\vrr,\sigma,l}^\dag e^{i(\vk-\vQ)\cdot\vrr},\quad \vQ\in \mathcal{Q}_{l}^\sigma
\end{align}
where $\mathcal{Q}_l^\sigma=\{\vG_M + \sigma (-1)^l \vq_1 |\vG_M \in \mathbb{Z}\bm{b}_{M,1}+\mathbb{Z}\bm{b}_{M,2}\}$ and $\Omega_{\tTot}=N\Omega_0$ is the total sample area for $N$ unit cells of area $\Omega_0$, with $(-1)^{t,b} = \pm 1$.

\section{Symmetry}
\label{app:Sym}
AA-stacked t\ch{WSe2} has the symmetries $C_{3}, C_{2y}, \mathcal{T}$ together with moir\'e translations:
\begin{align}
    C_3 c_{\vrr,\sigma,l}^\dagger C_3^{-1}                                 & = c_{C_3 \vrr,\sigma,l}^\dagger e^{i\sigma \pi/3} \label{eq:sym1} \\
    (C_{2y}\mathcal{T}) c_{\vrr,\sigma,l}^\dagger (C_{2y}\mathcal{T})^{-1} & = c_{C_{2y} \vrr,\sigma,\bar{l}}^\dagger \label{eq:sym2} \\
    \mathcal{T} c_{\vrr,\sigma,l}^\dagger \mathcal{T}^{-1}                 & = c_{\vrr,-\sigma,l}^\dagger (-\sigma) \label{eq:sym3} \\
    T_{\vR_M} c_{\vrr,\sigma,l}^\dagger T_{\vR_M}^{-1}                     & = c_{\vrr+\vR_M,\sigma,l}^\dagger e^{-i\sigma \vR_M\cdot\vK_l}
\end{align}
Here, $C_{2y}T$ exchanges the two layers, and $T_{\vR_M}$ denotes a moir\'e translation.
In addition, in the absence of an electric field, the first-harmonic model possesses an effective inversion symmetry defined as
\begin{align}
    \mathcal{I}c_{\vrr,\sigma,l}^\dag \mathcal{I}^{-1} = c_{-\vrr,\sigma,\bar{l}}^\dag
\end{align}

\section{Hamiltonian}
In the first-harmonic approximation, the Hamiltonian can be written as
\begin{align}
    H = \sum_{\vk,\vQ,\vQ'}h^\sigma_{\vQ,\vQ'}(\vk) c_{\vk,\sigma,\vQ}^\dag c_{\vk,\sigma,\vQ'}
\end{align}
where
\begin{align}
    \label{eq:h_QQ_k}
    h^\sigma_{\vQ,\vQ'}(\vk)
    = &
    \bigg[
        -\frac{\hbar^2}{2m^*} (\vk-\vQ)^2
        +
        \frac{D}{2}
        \sigma(-1)^\vQ
        \bigg]
    \delta_{\vQ\vQ'}
    \\
      &
    +
    V_0
    \sum_{i=1,2,3}
    \left[
        e^{-\sigma(-1)^\vQ i\phi_V}
        \delta_{\vQ+\bm{g}_i,\vQ'}
        +
        e^{\sigma(-1)^\vQ i\phi_V}
        \delta_{\vQ-\bm{g}_i,\vQ'}
    \right] \\
      &
    +
    \sum_{i=1,2,3}
    \left[
        w\delta_{\vQ+\vq_i,\vQ'} +\text{h.c.}
        \right]
\end{align}
where $(-1)^\vQ=\pm1$ for $\vQ\in \{\vG_M\pm \vq_1\}$. $V_0,\phi_V$  characterize the intralayer potential, and $w$ characterizes the interlayer coupling.
We define $\bm{g}_i = C_{3z}^{i-1}\bm{b}_{M,1}$.
$D$ denotes the displacement field.
We use $m^*=0.43m_0$, where $m_0$ is the bare electron mass.

We define the eigenvalues and eigenvectors of the single-particle Hamiltonian through
\begin{align}
    h_{\vQ,\vQ'}^\sigma(\vk)
    \mathbf{U}_{\vQ',n}^\sigma(\vk) 
    = 
    \epsilon^\sigma_{\vk,n}
    \mathbf{U}_{\vQ,n}^{\sigma}(\vk)
    \label{eq:app:eigeneq_continuous_model}
\end{align}
The electron operators in the band basis are then
\begin{align}
    \gamma_{\vk,\sigma,n}
    =
    \sum_{\vQ}
    \left(\mathbf{U}^{\sigma}_{\vQ,n}(\vk)\right)^\ast
    c_{\vk,\sigma,\vQ}
\end{align}
We also use $|\vk,\sigma,n \rangle = \gamma_{\vk,\sigma,n}^\dag|0\rangle$ to denote the corresponding Bloch states.

\section{Band structures and band topology at $\theta=3.65^\circ$} \label{sec:band_structure_and_topology}

\subsection{First-harmonic model}

\begin{figure}[ht]
    \centering
    \includegraphics[width=0.6\textwidth]{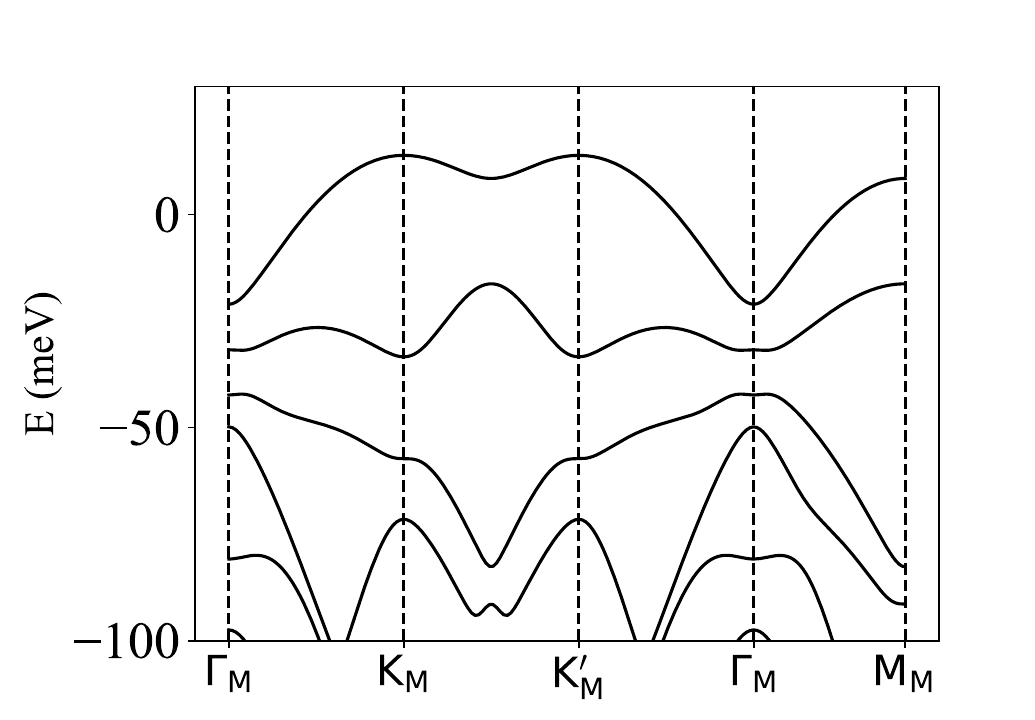}
    \caption{Band dispersions of the system with the parameters given in~\cref{eq:para_LF}. 
    }
    \label{fig:app:disp}
\end{figure}

We use the following parameters for the first-harmonic model \cite{devakulMagicTwistedTransition2021}:
\begin{align}
     & a_0 = 3.317\,\text{\AA} \nonumber \\
     & V_0 = 9\text{meV},\quad w = -18.0\text{meV},\quad
    \phi_V = -128^\circ
    \label{eq:para_LF}
\end{align}
We note that the sign of $\phi_V$ used here is opposite to that in \cite{devakulMagicTwistedTransition2021}.
This is because we assume that the top (bottom) layer is rotated by $-\theta/2$ ($\theta/2$), which is the reverse of the convention in \cite{PhysRevB.109.205122}.
Finally, the following change in the parameter
\begin{align}
    w \rightarrow w e^{i\phi_w }
\end{align}
does not change the spectrum of the system.
The additional phase $\phi_w$ can be removed by the gauge transformation $c^\dag_{\vrr,\sigma,t} \rightarrow c^\dag_{\vrr,\sigma,t}e^{-i\phi_w}$.
We take $\phi_w =0 $ and $w<0$ to match the convention adopted in our DFT calculations.
The dispersion of the model (\cref{eq:para_LF}) is shown in~\cref{fig:app:disp}.

Including the effective inversion symmetry, the magnetic symmetry group for each valley is $\#$164.89 $P\bar{3}m'1$.
The symmetry eigenvalues and the Chern numbers of the top five bands are given in~\cref{tab:irrep_all_lf}.

\begin{table}[ht]
    \centering
    \begin{tabular}{  c|c| c | c  }
        Band index @ $\Gamma_M$ & $C_{3z}$      & $\mathcal{I}$ & IRREP                                       \\
        \hline
        1                       & $e^{i\pi}$    & 1             & {$\bar{\Gamma}_4$} (3Tri) (1Tri2Hon) (2Hon) \\
        2                       & $e^{i\pi/3}$  & -1            & {$\bar{\Gamma}_9$}  (3Tri) (1Tri2Hon)       \\
        3                       & $e^{-i\pi/3}$ & 1             & {$\bar{\Gamma}_5$} (3Tri)                   \\
        4                       & $e^{i\pi }$   & -1            & $\bar{\Gamma}_7$(1Tri2Hon) (2Hon)           \\
    \end{tabular}
    \quad \quad
    \begin{tabular}{  c|c|c  }
        Band index@ $K_M$ & $C_{3z}$       & IRREP                                   \\
        \hline
        1                 & $e^{i\pi/3}$   & {$\bar{K}_6$}  (3Tri) (1Tri2Hon) (2Hon) \\
        2                 & $e^{-i\pi/3 }$ & {$\bar{K}_5$} (3Tri) (1Tri2Hon) (2Hon)  \\
        3                 & $e^{i\pi}$     & {$\bar{K}_4$} (3Tri)                    \\
        4                 & $e^{i\pi/3}$   & $\bar{K}_6 $ (1Tri2Hon)                 \\
    \end{tabular}
    \quad \quad
    \begin{tabular}{  c|c|c  }
        Band index @$M_M$ & $\mathcal{I}$ & IRREP                                  \\
        \hline
        1                 & $-1$          & {$\bar{M}_2$} (3Tri) (1Tri2Hon) (2Hon) \\
        2                 & $1$           & {$\bar{M}_3$} (3Tri) (1Tri2Hon)(2Hon)  \\
        3                 & $-1$          & $\bar{M}_2$ (1Tri2Hon)                 \\
        4                 & $1$           & {$\bar{M}_3$}(3Tri)                    \\
    \end{tabular}
    \quad \quad
    \begin{tabular}{  c|c }
        Band index & Chern number \\
        \hline 1   & $1$          \\
        2          & $1$          \\
        3          & $1$          \\
        4          & $1$          \\
    \end{tabular}

    \caption{Irreps at the high-symmetry points and Chern numbers of the top four bands at zero electric field.
        (3Tri) labels the irreps that form $^2\bar{E}_u@1a \oplus \bar{E}_g@1a \oplus ^1\bar{E}_g@1a $. (1Tri2Hon) labels the irreps that form  $^2\bar{E}_u@1a \oplus \bar{E}@2d $. (2Hon) labels the irreps that form EBR $\bar{E}@2d $.
    }
    \label{tab:irrep_all_lf}
\end{table}

\FloatBarrier
\subsection{DFT-derived continuum model}
\label{sec:dft_continuum}

\subsubsection{Structural relaxation}

The source structures were prepared using the machine-learning force-field workflow described in Appendix A of Ref.~\cite{zhang_universal_2024}.
First-principles molecular-dynamics data for untwisted bilayer configurations with different relative layer translations were generated using VASP.
An equivariant force field was trained with Allegro~\cite{musaelian_Learning_2023}, validated against large-angle DFT-relaxed structures, and used to relax the twisted supercells within the Atomic Simulation Environment.
The source geometries were subsequently refined for ten ionic steps using VASP 6.3.0~\cite{kresse1996efficient}, with PBE, a 300 eV plane-wave cutoff, $\Gamma$-point sampling, and the zero-damping DFT-D3 dispersion correction~\cite{grimme_D3_2010}.
The maximum residual atomic forces after these refinements were approximately 0.013 and 0.020 eV/\AA\ for the $3.48^\circ$ and $3.89^\circ$ structures, respectively.

\subsubsection{First-principles calculations}

Electronic-structure calculations were performed using OpenMX version 3.9.9~\cite{OpenMX39Manual} with the Perdew--Burke--Ernzerhof exchange-correlation functional~\cite{perdew1996generalized}.
We used the \texttt{W\_PBE19} and \texttt{Se\_PBE19} pseudopotentials from the OpenMX 2019 database, together with the \texttt{W7.0-s3p2d2} and \texttt{Se7.0-s3p2d1} pseudoatomic-orbital bases, both with a radial cutoff of 7.0 Bohr.
The real-space integration-grid cutoff was 300 Ry, and Brillouin-zone sampling was restricted to the moir\'e $\Gamma$ point.

The source supercells labelled $3.48^\circ$ and $3.89^\circ$ contained 1626 and 1302 atoms, respectively, with an out-of-plane cell height of approximately 24.86 \AA, including the bilayer.
Following the two-step procedure in Appendix A of Ref.~\cite{zhang_universal_2024}, we first converged the charge density without spin polarization or spin--orbit coupling, using an electronic convergence threshold of $10^{-7}$ Hartree.
The spin--orbit Hamiltonian was then evaluated in a single-iteration noncollinear calculation using this density and the same atomic structure, without further charge self-consistency.

\subsubsection{Construction of the continuum model}

The first-principles Hamiltonian and overlap matrices were projected onto the truncated atomic-plane-wave (TAPW) basis~\cite{miao_truncated_2023}.
After orthogonalization and elimination of high-energy states, we denote the low-energy matrix for source calculation $i=1$ ($3.48^\circ$) or $i=2$ ($3.89^\circ$) by $h_{\mathrm{DFT}}^{(i)}(\mathbf k)$, where $\mathbf k$ is the moir\'e Brillouin-zone momentum and $\theta_i$ is the angle argument used by the source continuum model.
Its continuum representation $h_{\mathrm{cont}}^{(i)}$ is expanded in the analytic operator basis of Ref.~\cite{zhang_universal_2024} [Eqs.~(8)--(9) therein]:
\begin{equation}
 h_{\mathrm{cont}}^{(i)}(\mathbf k)
 =\sum_\alpha r_\alpha^{(i)}Y_\alpha(\mathbf k;\theta_i)
 \simeq h_{\mathrm{DFT}}^{(i)}(\mathbf k),
 \label{eq:dft_operator_expansion}
\end{equation}
Here $r_\alpha^{(i)}$ is the coefficient of the retained operator $Y_\alpha$, labelled by the composite index $\alpha=(M_x,M_y,l,l',n_1,n_2)$:
\begin{equation}
 [Y_\alpha(\mathbf k;\theta)]_{\mathbf Q\mathbf Q'}
 =\delta_{l_{\mathbf Q},l}\delta_{l_{\mathbf Q'},l'}
 (\mathbf k-\mathbf Q)_x^{M_x}(\mathbf k-\mathbf Q)_y^{M_y}
 \delta_{\mathbf Q,\mathbf Q'+\mathbf p_\theta}.
 \label{eq:dft_operator_basis}
\end{equation}
The layer indices $l,l'\in\{b,t\}$ denote the bottom and top layers.
The vectors $\mathbf Q,\mathbf Q'$ label the plane-wave channels introduced in Appendix~\ref{sec:moire_geometry}, and $l_{\mathbf Q}$ is the layer associated with channel $\mathbf Q$.
The integers $M_x,M_y\geq0$ specify the momentum powers, while $n_1,n_2\in\mathbb Z$ specify the moir\'e harmonic.
The momentum transfer is $\mathbf p_\theta=\mathbf K_l(\theta)-\mathbf K_{l'}(\theta)+n_1\mathbf b_{M,1}(\theta)+n_2\mathbf b_{M,2}(\theta)$.
Here $\mathbf K_l(\theta)$ is the rotated monolayer $K$-valley momentum of layer $l$.
The vectors $\mathbf b_{M,1},\mathbf b_{M,2}$ are the moir\'e reciprocal basis vectors.
Both were defined in Appendix~\ref{sec:moire_geometry}.

At the reference momentum, Gram--Schmidt orthonormalization gives $\widetilde Y_\beta^{(i)}=\sum_\alpha Y_\alpha^{(i)}z_{\alpha\beta}^{(i)}$, with $\operatorname{Tr}[(\widetilde Y_\beta^{(i)})^\dagger\widetilde Y_\gamma^{(i)}]=\delta_{\beta\gamma}$.
The indices $\beta,\gamma$ label the orthonormalized operators, and $z_{\alpha\beta}^{(i)}$ is the change-of-basis matrix.
The coefficients $\widetilde r_\beta^{(i)}$ in the orthonormalized basis and $r_\alpha^{(i)}$ in the original analytic basis follow by matrix projection [Eqs.~(11)--(14) of Ref.~\cite{zhang_universal_2024}]:
\begin{equation}
 \widetilde r_\beta^{(i)}
 =\operatorname{Tr}\!\left[(\widetilde Y_\beta^{(i)})^\dagger
 h_{\mathrm{DFT}}^{(i)}\right],
 \qquad
 r_\alpha^{(i)}=\sum_\beta z_{\alpha\beta}^{(i)}\widetilde r_\beta^{(i)}.
 \label{eq:dft_trace_projection}
\end{equation}
For several reference momenta, the matrices in this procedure are replaced by their direct sums.
This determines the analytic coefficients without fitting band energies.
The retained expansion includes higher moir\'e harmonics and higher-order momentum terms beyond the first-harmonic model in Eq.~\eqref{eq:para_LF}.

The two source parameter sets are labelled $3.48^\circ$ and $3.89^\circ$, according to their DFT supercells.
Let $h^{(1)}(\theta,\bar{\mathbf k})$ and $h^{(2)}(\theta,\bar{\mathbf k})$ denote the Hamiltonians constructed from the two source parameter sets, respectively, where $\bar{\mathbf k}=\mathbf k/q_\theta$.
The momentum scale is $q_\theta=(8\pi/3a_0)\sin(\theta/2)$, with the monolayer lattice constant $a_0=3.28524$ \AA\ used in this continuum parameterization.
The vectors $\mathbf k$, $\mathbf Q$ and $\mathbf p_\theta$ scale with $q_\theta$ at fixed dimensionless momentum and channel labels; Eq.~\eqref{eq:dft_operator_basis} therefore gives a factor $q_\theta^{M_x+M_y}$ for each momentum-polynomial term.
For the target angle, both source Hamiltonians are evaluated at $3.65^\circ$ in the same ordered basis and then averaged:
\begin{equation}
    h_{3.65^\circ}(\bar{\mathbf k})
    =\frac{1}{2}\Bigl(
    h^{(1)}(3.65^\circ,\bar{\mathbf k})
    +h^{(2)}(3.65^\circ,\bar{\mathbf k})
    \Bigr).
    \label{eq:dft_target_model}
\end{equation}
At $3.65^\circ$, the curves labelled DFT in Fig.~\ref{fig:disp_DFT_3tri} are the resulting DFT-derived continuum bands, rather than bands from a direct first-principles calculation at that angle.
The three-orbital Wannier bands are shown separately as TB (3Tri).

\subsubsection{Symmetry and topology of the DFT-derived bands}

For the top $2$ bands at the $\Gamma$ point, the $C_{3z}$ eigenvalues are the same as those of the previous model.
For the top $4$ bands at the $K$ point, the $C_{3z}$ eigenvalues of the third and fourth bands are switched compared to the previous model (\ref{eq:para_LF}).

Since the DFT model does not have inversion symmetry, we now quantify the inversion symmetry breaking in our model.
For each band with dispersion $\epsilon_{\vk,n}$ and eigenvector $U_{\vQ,n}(\vk)$, we define
\begin{align}
     & \Delta E_{inv,n} = 
     \text{max}_{\vk} |\epsilon_{\vk,n}-\epsilon_{-\vk,n}|, \quad \quad \nonumber \\
     &
    \Delta U_{inv,n}^{max} = \text{max}_{\vk}\bigg[1- |U_{\vQ,n}^* \mathcal{I}_{\vQ',\vQ}
    U_{\vQ,n}(\vk)|^2 \bigg], \quad \Delta U_{inv,n}^{mean} = \text{mean}_{\vk}\bigg[1- |U_{\vQ,n}^* \mathcal{I}_{\vQ',\vQ} U_{\vQ,n}(\vk)|^2 \bigg]
\end{align}
where $\mathcal{I}_{\vQ,\vQ'} = \delta_{\vQ,-\vQ'}$.
The first term quantifies the deviation from an inversion-symmetric dispersion and $\Delta E_{inv,n}=0$ for an inversion-symmetric dispersion.
The second term quantifies the deviation from inversion-symmetric Bloch wavefunctions.
In the inversion-symmetric case, we should have $\langle -\vk, n |I|\vk,n\rangle = 1$ and $\Delta U_{inv,n} = 0$.
For the top four bands , we obtain
\begin{align}
     & \Delta E_{inv,n=1,2,3,4} = 4\text{meV},\quad 10.8\text{meV},\quad 10.3 \text{meV},\quad 7.2\text{meV}\nonumber \\
     & \Delta U^{max}_{inv,n=1,2,3,4} =0.16,\quad 0.45,\quad 0.48,\quad 0.87 \nonumber \\
     & \Delta U^{mean}_{inv,n=1,2,3,4} =0.05,\quad 0.08,\quad 0.12,\quad 0.23
\end{align}
We conclude that inversion symmetry is almost perfect for the first band.
For the top four bands, inversion symmetry remains a good approximate symmetry.
Without inversion symmetry, the corresponding magnetic group for each valley is $P32'1$($\#150.27$).
The irreducible representations of the top 4 bands are given in~\cref{tab:irrep_dft_model}.
Since inversion symmetry remains approximately valid, we also calculate the expectation value of the inversion operation at the high-symmetry points.
The inversion symmetry eigenvalues shown in \cref{tab:irrep_dft_model} are taken to be $+1$ or $-1$ depending on whether the expectation value of the inversion operation is closer to $+1$ or $-1$.

\begin{table}[ht]
    \centering
    \begin{tabular}{  c|c| c | c  }
        Band index @ $\Gamma$ & $C_{3z}$      & $\mathcal{I}$ & IRREP                                       \\
        \hline
        1                     & $e^{i\pi}$    & 1             & {$\bar{\Gamma}_7$} (3Tri) (1Tri2Hon) (2Hon) \\
        2                     & $e^{i\pi/3}$  & -1            & {$\bar{\Gamma}_6$}  (T) (3Tri) (1Tri2Hon)   \\
        3                     & $e^{-i\pi/3}$ & 1             & {$\bar{\Gamma}_5$} (3Tri)                   \\
        4                     & $e^{i\pi }$   & -1            & $\bar{\Gamma}_4$(1Tri2Hon) (2Hon)
    \end{tabular}
    \quad \quad
    \begin{tabular}{  c|c|c  }
        Band index@ $K$ & $C_{3z}$       & IRREP                                   \\
        \hline
        1               & $e^{i\pi/3}$   & {$\bar{K}_6$}  (3Tri) (1Tri2Hon) (2Hon) \\
        2               & $e^{-i\pi/3 }$ & {$\bar{K}_5$} (3Tri) (1Tri2Hon) (2Hon)  \\
        3               & $e^{i\pi/3}$   & {$\bar{K}_6$} (3Tri)                    \\
        4               & $e^{i\pi}$     & $\bar{K}_4 $ (1Tri2Hon)
    \end{tabular}
    \quad \quad
    \begin{tabular}{  c|c|c  }
        Band index @$M$ & $\mathcal{I}$ & IRREP                                  \\
        \hline
        1               & $-1$          & {$\bar{M}_2$} (3Tri) (1Tri2Hon) (2Hon) \\
        2               & $1$           & {$\bar{M}_3$} (3Tri) (1Tri2Hon)        \\
        3               & $-1$          & $\bar{M}_2$ (1Tri2Hon) (2Hon)          \\
        4               & $1$           & {$\bar{M}_3$}(3Tri)
    \end{tabular}
    \quad \quad
    \begin{tabular}{  c|c }
        Band index & Chern number \\
        \hline 1   & $1$          \\
        2          & $1$          \\
        3          & $-1$         \\
        4          & $3$          \\
    \end{tabular}
    \caption{Irreps at the high-symmetry points and Chern numbers of the top five bands of the DFT model at zero electric field.
        (3Tri) labels the irreps that form $^2\bar{E}_g@1a \oplus \bar{E}_u@1a \oplus ^1\bar{E}_g@1a $. (1Tri2Hon) labels the irreps that form  $^2\bar{E}_g@1a \oplus \bar{E}@2d $. (2Hon) labels the irreps that form EBR $\bar{E}@2d $.
    }
    \label{tab:irrep_dft_model}
\end{table}

\subsection{Wannier orbitals}
From the irreps in~\cref{tab:irrep_dft_model}, we can construct the following three types of models:
\begin{itemize}
    \item A 3Tri model with three Wannier orbitals at the $1a$ positions.
    \item A 1Tri2Hon model with one Wannier orbital at the $1a$ position and two Wannier orbitals at $2d$ positions (honeycomb sites).
    \item A 2Hon model with two Wannier orbitals at the $2d$ positions.
\end{itemize}
We now discuss the construction of each model.

\section{Tight-binding models}

We aim to construct an exponentially localized Wannier orbital that has a large overlap with the top band.
We focus on the DFT model in this section.

\subsection{3Tri model}
\label{app:DFT3Tri}
We first discuss the 3Tri model, whose band representation is given in \cref{tab:irrep_dft_model}.
The three orbitals correspond to $s,p_x-i\sigma p_y,p_x+i\sigma p_y$ orbitals, respectively, in valley $\sigma$.

\subsubsection{Trial wavefunctions}
\label{sec:app:wan_trial_3tri}
We use the following trial wavefunctions to describe the Wannier orbitals $\alpha=1,2,3$:
\begin{align}
    \psi^\dag_{\rm{3Tri},\sigma,\alpha, \vR}
    = & \sum_l\sum_{\vQ\in\mathcal{Q}_{l}^\sigma }A_{\alpha,\sigma, l}(\vR) f_{\alpha,\sigma,l}(\vrr-\vR) e^{-(\vrr-\vR)^2/(2\lambda_{\alpha,l}^2)}c_{\vrr,\sigma,l}^\dag
\end{align}
Here, we introduce the following functions characterizing different Wannier orbitals
\begin{align}
     & f_{1,\sigma,l}(\vrr) =  \frac{1}{\sqrt{\pi\lambda_{1,l}^2}} \nonumber \\
     & f_{2,\sigma,l}(\vrr) = \frac{ i\sigma r_x + r_y}{\sqrt{\pi\lambda_{2,l}^4}} \nonumber \\
     & f_{3,\sigma,l}(\vrr) = \frac{ i\sigma r_x -r_y}{\sqrt{\pi\lambda_{3,l}^4}}
\end{align}
which correspond to the $s,p_x-i\sigma p_y, p_x+i\sigma p_y$ orbitals, respectively.
We now discuss the symmetry properties.

Under moir\'e translational symmetry, we find
\begin{align}
    T_{\vR'} \psi_{\rm{3Tri},\sigma,\alpha,\vR}^\dag T_{\vR'}^{-1} = & \sum_l \int d^2\vrr A_{\alpha,\sigma, l}(\vR) f_{\alpha,\sigma,l}(\vrr-\vR) e^{-(\vrr-\vR)^2/(2\lambda_{\alpha,l}^2)} \frac{1}{\sqrt{\Omega_{\tTot}}} \sum_{\vk,\vQ \in \mathcal{Q}_{l}^\sigma} e^{-i(\vk-\vQ)\cdot\vrr -i\vk\cdot\vR'} c_{\vk,\sigma,\vQ}^\dag \nonumber \\
    =                                                                & \sum_l \int d^2\vrr A_{\alpha,\sigma, l}(\vR) f_{\alpha,\sigma,l}(\vrr-\vR) e^{-(\vrr-\vR)^2/(2\lambda_{\alpha,l}^2)} \frac{1}{\sqrt{\Omega_{\tTot}}} \sum_{\vk,\vQ \in \mathcal{Q}_{l}^\sigma} e^{-i(\vk-\vQ)\cdot(\vrr+\vR) -i\vQ\cdot\vR'} c_{\vk,\sigma,\vQ}^\dag \nonumber \\
    =                                                                & \sum_l e^{-i\kappa_l\cdot\vR'} \int d^2\vrr A_{\alpha,\sigma, l}(\vR) f_{\alpha,\sigma,l}(\vrr-\vR-\vR') e^{-(\vrr-\vR-\vR')^2/(2\lambda_{\alpha,l}^2)} \frac{1}{\sqrt{\Omega_{\tTot}}} \sum_{\vk,\vQ \in \mathcal{Q}_{l}^\sigma} e^{-i(\vk-\vQ)\cdot(\vrr+\vR)} c_{\vk,\sigma,\vQ}^\dag \nonumber \\
    =                                                                & \sum_l e^{-i\kappa_l\cdot\vR'} \int d^2\vrr A_{\alpha,\sigma, l}(\vR) f_{\alpha,\sigma,l}(\vrr-\vR-\vR') e^{-(\vrr-\vR-\vR')^2/(2\lambda_{\alpha,l}^2)} c_{\vrr,\sigma,l}
\end{align}
where we have used $T_{\vR}c_{\vk,\sigma,\vQ}^\dag T_{\vR}^{-1} = e^{-i\vk\cdot\vR}c_{\vk,\sigma,\vQ}$ and introduced $\bm{\kappa}_l = (-)^l\bm{q}_1$.
Therefore, we can choose
\begin{align}
    A_{\alpha,\sigma,l}(\vR) = A_{\alpha,\sigma,l}e^{-i\sigma \vR\cdot \sigma\bm{\kappa}_l}
\end{align}
which leads to
\begin{align}
    T_{\vR'}
    \psi_{\rm{3Tri},\alpha,\sigma,\vR}^\dag
    T_{\vR'}^{-1}
    =
    \psi_{\rm{3Tri},\alpha,\sigma,\vR+\vR'}^\dag
\end{align}

Under the $C_{3z}$ transformation, we obtain
\begin{align}
      & C_{3z}
    \psi_{\rm{3Tri},\alpha, \sigma,\vR}^\dag
    C_{3z}^{-1}
    =
    e^{i\sigma \pi/3}
    \sum_l \int d^2\vrr
    A_{\alpha,\sigma,l}
    e^{-i\sigma\vR\cdot\bm{\kappa}_l}
    f_{\alpha,l,\sigma}(\vrr-\vR)
    e^{-(\vrr-\vR)^2/(2\lambda_{\alpha,l}^2)}
    c_{C_{3z}\vrr,\sigma,l}^\dag
    \nonumber \\
    = &
    e^{i\sigma \pi/3}
    \sum_l
    e^{i\sigma(C_{3z}\vR-\vR)\cdot \bm{\kappa}_l}
    \int d^2\vrr'
    A_{\alpha, \sigma,l}
    e^{-i\sigma C_{3z}\vR\cdot\bm{\kappa}_l}
    f_{\alpha,\sigma,l}(C_{3z}^{-1}\vrr-\vR)
    e^{-(\vrr'-C_{3z}\vR)^2/(2\lambda_{\alpha,l}^2)}
    c_{\vrr',\sigma,l}^\dag    \nonumber \\
    = &
    e^{i\sigma \pi/3}
    \sum_l \int d^2\vrr'
    A_{\alpha, \sigma,l}
    e^{-i\sigma C_{3z}\vR\cdot\bm{\kappa}_l}
    f_{\alpha,\sigma,l}(C_{3z}^{-1}\vrr-\vR)
    e^{-(\vrr'-C_{3z}\vR)^2/(2\lambda_{\alpha,l}^2)}
    c_{\vrr',\sigma,l}^\dag
\end{align}
We note that
\begin{align}
     & f_{1,\sigma,l}(C_{3z}^{-1}\vrr) =f_{1,\sigma,l}(\vrr) \nonumber \\
     & f_{2,\sigma,l}(C_{3z}^{-1}\vrr) =  e^{i\sigma 2\pi/3} f_{2,\sigma,l}(\vrr)  \nonumber \\
     & f_{3,\sigma,l}(C_{3z}^{-1}\vrr) =  e^{-i\sigma 2\pi/3} f_{3,\sigma,l}(\vrr)
\end{align}
Then we find
\begin{align}
     & C_{3z}\psi_{\rm{3Tri},1, \sigma,\vR}^\dag C_{3z}^{-1} = \psi_{\rm{3Tri},1,\sigma,C_{3z}\vR}^\dag e^{i\sigma \pi/3 } \nonumber \\
     & C_{3z}\psi_{\rm{3Tri},2, \sigma,\vR}^\dag C_{3z}^{-1} = \psi_{\rm{3Tri},2,\sigma,C_{3z}\vR}^\dag e^{i\sigma \pi} \nonumber \\
     & C_{3z}\psi_{\rm{3Tri},3, \sigma,\vR}^\dag C_{3z}^{-1} = \psi_{\rm{3Tri},3,\sigma,C_{3z}\vR}^\dag e^{-i\sigma\pi/3} \label{app:eqn:C3z_action_trial_3tri}
\end{align}

Under the effective intra-valley inversion symmetry, the trial wavefunctions transform as
\begin{align}
      &
    \mathcal{I}
    \psi_{1,\sigma,\vR}^\dag
    \mathcal{I}c^{-1}
    =
    \sum_l \int d^2\vrr
    A_{1,\sigma,l}
    \frac{
        e^{-i\sigma\vR\cdot\bm{\kappa}_l}
        e^{-(\vrr-\vR)^2/(2\lambda_{1,l}^2)}
    }{
        \sqrt{2\pi\lambda^2}
    }
    c_{-\vrr,\sigma,\bar{l}}^\dag
    \nonumber \\
    = &
    \sum_l \int d^2\vrr
    A_{1,\sigma,\bar{l}}
    \frac{
        e^{+i\sigma\vR\cdot\bm{\kappa}_l}
        e^{-(-\vrr-\vR)^2/(2\lambda_{1,\bar{l}}^2)}
    }{
        \sqrt{2\pi\lambda_{1,\bar{l}}^2}
    }
    c_{\vrr,\sigma,l}^\dag
    \nonumber \\
    = &
    \sum_l \int d^2\vrr
    A_{1,\sigma,\bar{l}}
    \frac{
        e^{-i\sigma(-\vR)\cdot\bm{\kappa}_l}
        e^{-(\vrr-(-\vR))^2/(2\lambda_{1,\bar{l}}^2)}
    }{
        \sqrt{2\pi\lambda_{1,\bar{l}}^2}
    }
    c_{\vrr,\sigma,l}^\dag
\end{align}
We take $A_{1,\sigma,\bar{l}} = A_{1,\sigma,l} =(-1)^lA_{1,\sigma}$ and
$\lambda_{1,\bar{l}} = \lambda_{1, l} = \lambda_{1}$, which gives
\begin{align}
    \mathcal{I}
    \psi_{\rm{3Tri},1,\sigma,\vR}^\dag
    \mathcal{I}^{-1}
    =
    -\psi_{\rm{3Tri},1,\sigma,-\vR}
\end{align}

For the second orbital, we find
\begin{align}
      &
    \mathcal{I}
    \psi_{\rm{3Tri},2,\sigma,\vR}^\dag
    \mathcal{I}^{-1}
    =
    \sum_l \int d^2\vrr
    A_{2,\sigma,l}
    \frac{  i\sigma \bigg[\vrr_x-\vR_x-i\sigma(\vrr_y-\vR_y)\bigg] e^{-i\sigma\vR\cdot\bm{\kappa}_l}e^{-(\vrr-\vR)^2/(2\lambda_{2,l}^2)}}{\sqrt{2\pi\lambda_{2,l}^4}}  c_{-\vrr,\sigma,\bar{l}}^\dag  \nonumber \\
    = & \sum_l \int d^2\vrr A_{2,\sigma,\bar{l}}\frac{ i\sigma  \bigg[-\vrr_x-\vR_x-i\sigma(-\vrr_y-\vR_y)\bigg] e^{i\sigma\vR\cdot\bm{\kappa}_l}e^{-(-\vrr-\vR)^2/(2\lambda_{2,\bar{l}}^2)}}{\sqrt{2\pi\lambda_{2,\bar{l}}^4}}  c_{\vrr,\sigma,{l}}^\dag \nonumber \\
    = & -\sum_l \int d^2\vrr A_{2,\sigma,\bar{l}} \frac{ i\sigma  \bigg[\vrr_x-(-\vR_x)-i\sigma(\vrr_y-(-\vR_y))\bigg] e^{i\sigma\vR\cdot\bm{\kappa}_l}e^{-(\vrr-(-\vR))^2/(2\lambda_{2,\bar{l}}^2)}}{\sqrt{2\pi\lambda_{2,\bar{l}}^4}}  c_{\vrr,\sigma,{l}}^\dag
\end{align}
We therefore take
\begin{align}
    A_{2,\sigma,l} =(-1)^l A_{2,\sigma},\quad \lambda_{2,\bar{l}} = \lambda_{2,l} = \lambda_{2}
\end{align}
which gives
\begin{align}
    \mathcal{I} \psi_{\rm{3Tri},2,\sigma,\vR}^\dag \mathcal{I}^{-1} = \psi_{\rm{3Tri},2,\sigma,-\vR}
\end{align}
For the third orbital, we observe
\begin{align}
      &
    \mathcal{I}
    \psi_{\rm{3Tri},3,\sigma,\vR}^\dag
    \mathcal{I}^{-1}
    =
    \sum_l \int d^2\vrr
    A_{3,\sigma,l}
    \frac{
        i\sigma
        \bigg[
            \vrr_x
            -\vR_x
            +i\sigma(\vrr_y-\vR_y)
            \bigg]
        e^{-i\sigma\vR\cdot\bm{\kappa}_l}
        e^{-(\vrr-\vR)^2/(2\lambda_{3,l}^2)}
    }{
        \sqrt{2\pi\lambda_{3,l}^4}
    }
    c_{-\vrr,\sigma,\bar{l}}^\dag
    \nonumber \\
    = &
    \sum_l \int d^2\vrr
    A_{3,\sigma,\bar{l}}
    \frac{
        i\sigma
        \bigg[
            -\vrr_x
            -\vR_x
            +i\sigma(-\vrr_y-\vR_y)
            \bigg]
        e^{i\sigma\vR\cdot\bm{\kappa}_l}
        e^{-(-\vrr-\vR)^2/(2\lambda_{3,\bar{l}}^2)}
    }{
        \sqrt{2\pi\lambda_{3,\bar{l}}^4}
    }
    c_{\vrr,\sigma,{l}}^\dag
\end{align}
We take $A_{3,\sigma,l} =(-1)^l A_{3,\sigma}$ and $\lambda_{3,\bar{l}}=\lambda_{3,l} = \lambda_{3}$, which gives
\begin{align}
    \mathcal{I} \psi_{\rm{3Tri},3,\sigma,\vR}^\dag \mathcal{I}^{-1} = \psi_{\rm{3Tri},3,\sigma,-\vR}
\end{align}

In general, we note that the effective inversion symmetry fixes the prefactors of the Wannier orbitals in the two layers to be
\begin{align}
    A_{\alpha, \sigma, l} = (-1)^l A_{\alpha,\sigma} ,\quad \lambda_{\alpha,l} = \lambda_{\alpha,\bar{l}} = \lambda_{\alpha}
    \label{eq:prefactor_inversion_symmetry}
\end{align}

Under $C_{2y}\mathcal{T}$ symmetry, we observe
\begin{align}
    & 
    C_{2y}
    \mathcal{T}
    \psi_{\rm{3Tri},\alpha, \sigma,\vR}^\dag
    (C_{2y}\mathcal{T})^{-1} 
    = 
    \sum_l 
    \int d^2\vrr (-1)^l 
    A_{\alpha, \sigma}
    e^{-i\sigma\vR\cdot\bm{\kappa}_l}
    e^{-(\vrr-\vR)^2/(2\lambda_{\alpha,l}^2)}
    f_{\alpha,\sigma,l}(\vrr-\vR)^*  c_{C_{2y}\vrr,\sigma,\bar{l}}^\dag  
    \nonumber \\
    = & 
    \sum_l 
    \int d^2\vrr (-1)^l 
    A^*_{\alpha, \sigma}
    e^{i\sigma (C_{2y}\vR\cdot\bm{\kappa}_{\bar{l}} - \vR\cdot\bm{\kappa}_l)}
    e^{-i\sigma C_{2y}\vR\cdot\bm{\kappa}_{\bar{l}}}
    e^{-(\vrr-C_{2y}\vR)^2/(2\lambda_{\alpha,l}^2)}
    f_{\alpha,\sigma,l}
    (C_{2y}^{-1}(\vrr-C_{2y}\vR))^*
    c_{\vrr,\sigma,\bar{l}}^\dag 
    \nonumber \\
    = & 
    \sum_l 
    \int d^2\vrr(-1)^l  
    A^*_{\alpha, \sigma}
    e^{i\sigma (C_{2y}\vR\cdot\bm{\kappa}_{\bar{l}} - \vR\cdot\bm{\kappa}_l)}
    e^{-i\sigma C_{2y}\vR\cdot\bm{\kappa}_{\bar{l}}}
    e^{-(\vrr-C_{2y}\vR)^2/(2\lambda_{\alpha,l}^2)}
    f_{\alpha,\sigma,l}(\vrr-C_{2y}\vR)
    c_{\vrr,\sigma,\bar{l}}^\dag
\end{align}
We note that
\begin{align}
     & f_{1,\sigma,l}(C_{2y}^{-1}(\vrr-C_{2y}\vR))^* = f_{1,\sigma,l}(\vrr-C_{2y}\vR) \nonumber \\
     & f_{2,\sigma,l}(C_{2y}^{-1}(\vrr-C_{2y}\vR))^* = f_{2,\sigma,l}(\vrr-C_{2y}\vR)\nonumber \\
     & f_{3,\sigma,l}(C_{2y}^{-1}(\vrr-C_{2y}\vR))^* = f_{3,\sigma,l}(\vrr-C_{2y}\vR)
\end{align}
and we take $A^*_{\alpha, \sigma}  \in \mathbb{R}$. This leads to
\begin{align}
     & C_{2y}\mathcal{T}\psi_{\rm{3Tri},a,\sigma,\vR}^\dag (C_{2y}\mathcal{T})^{-1} =- \psi_{\rm{3Tri},a,\sigma,C_{2y}\vR}^\dag  ,\quad a=1,2,3 \label{app:eqn:C2yT_action_trial_3tri}
\end{align}

From the time-reversal transformation, we obtain
\begin{align}
    \mathcal{T}\psi_{\rm{3Tri},\alpha, \sigma,\vR}^\dag \mathcal{T}^{-1} = & \sum_l \int d^2\vrr A^*_{\alpha, \sigma}e^{i\sigma\vR\cdot\bm{\kappa}_l}e^{-(\vrr-\vR)^2/(2\lambda_\alpha^2)}f^*_{\alpha,\sigma,l}(\vrr-\vR) c_{\vrr,-\sigma,l}^\dag(-\sigma)
\end{align}
We can then take $A_{1, \sigma} =A_{1, -\sigma}^* =A_1,\quad A_{2/3,\sigma} = A_{2/3,-\sigma}^* = A_{2/3}  $, which gives
\begin{align}
    \mathcal{T}\psi_{\rm{3Tri},\alpha, \sigma,\vR}^\dag \mathcal{T}^{-1}= -\sigma \psi_{\rm{3Tri},\alpha, -\sigma,\vR}^\dag
    \label{app:eqn:T_action_trial_3tri}
\end{align}

In summary, our choices of trial wavefunctions for the 3Tri model are
\begin{align}
     & \psi_{\rm{3Tri},1,\sigma,\vR}^\dag = \sum_l \int d^2\vrr \frac{(-1)^l  e^{-i\sigma\vR\cdot\bm{\kappa}_l}e^{-(\vrr-\vR)^2/(2\lambda_1^2)}}{\sqrt{2\pi\lambda_1^2}}  c_{\vrr,\sigma,l}^\dag \nonumber \\
     & \psi_{\rm{3Tri},2,\sigma,\vR}^\dag = \sum_l \int d^2\vrr \frac{ (-1)^l \bigg[i\sigma(\vrr_x-\vR_x)+(\vrr_y-\vR_y)\bigg] e^{-i\sigma\vR\cdot\bm{\kappa}_l}e^{-(\vrr-\vR)^2/(2\lambda_2^2)}}{\sqrt{2\pi\lambda_2^4}}  c_{\vrr,\sigma,l}^\dag \nonumber \\
     & \psi_{\rm{3Tri},3,\sigma,\vR}^\dag = \sum_l \int d^2\vrr \frac{ (-1)^l \bigg[i\sigma(\vrr_x-\vR_x)-(\vrr_y-\vR_y)\bigg] e^{-i\sigma\vR\cdot\bm{\kappa}_l}e^{-(\vrr-\vR)^2/(2\lambda_3^2)}}{\sqrt{2\pi\lambda_3^4}}  c_{\vrr,\sigma,l}^\dag
    \label{eq:trial_wave_3tri}
\end{align}
where we take $A_1=A_2=A_3=1/\sqrt{2}$ so that each wavefunction has norm $1$.

\subsubsection{Wannier model}
We use Wannier90~\cite{marzari1997maximally, souza2001maximally, marzari2012maximally, pizzi2020wannier90} to construct the Wannier model from the trial wavefunctions introduced in~\cref{eq:trial_wave_3tri} with $\lambda_1=\lambda_2=\lambda_3=0.4a_M$.
We use
\begin{align}
    f_{\vR,\sigma},\quad c_{\vR,1,\sigma},\quad c_{\vR,2,\sigma}
\end{align}
to denote the electron operators after Wannierization corresponding to $\psi_{\rm{3Tri},1,\sigma,\vR},\psi_{\rm{3Tri},2,\sigma,\vR},\psi_{\rm{3Tri},3,\sigma,\vR}$, respectively.

Using Wannier90, we obtain three Wannier orbitals at the $1a$ position with spreads
\begin{align}
    f:0.405a_M,\quad c_1:0.522a_M,\quad c_2:0.568a_M
\end{align}

\begin{figure}[ht]
    \centering
    \includegraphics[width=0.5\textwidth]{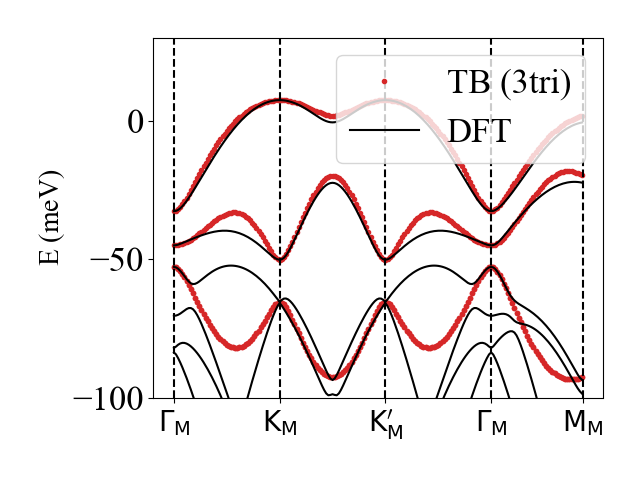}
    \caption{Comparison of the DFT-derived continuum bands at $\theta=3.65^\circ$ (black lines) with those of the effective three-orbital tight-binding model, TB (3Tri) (red points).
    The continuum Hamiltonian is constructed according to Eq.~\eqref{eq:dft_target_model}.}
    \label{fig:disp_DFT_3tri}
\end{figure}
The tight-binding model can now be written as
\begin{align}
    H_0 = \sum_\vk d_{\vk,\sigma,n}^\dag h^\sigma_{nm}(\vk) d_{\vk,\sigma,m},\quad
    d_{\vk,\sigma,m} =
    \begin{bmatrix}
       f_{\vk,\sigma} & c_{\vk,\sigma,1} & c_{\vk,\sigma,2}
   \end{bmatrix}
    _m
\end{align}
The effective tight-binding Hamiltonian has the matrix elements up to nearest neighbors
\begin{align}
     & h_{\vk,11}^+ = \epsilon_f + 2t_{ff}[\cos(k_1) + \cos(k_2) + \cos(k_1+k_2)]\nonumber \\
     & h^+_{\vk,22} = \epsilon_{c,1} + 2t_{cc,1} [\cos(k_1) + \cos(k_2) + \cos(k_1+k_2)] \nonumber \\
     & h^+_{\vk,33} = \epsilon_{c,2}+ 2t_{cc,2} [\cos(k_1) + \cos(k_2) + \cos(k_1+k_2)] \nonumber \\
     & h_{\vk,12}^+(\vk) = h_{\vk,21}^{+,*}(\vk)= \sum_{n=1,2,3}\left[t_{fc,1}e^{-i\frac{2\pi(n-1)}{3}}e^{iC_{3,z}^{n-1}\bm{a}_2 \cdot \vk}
    -t_{fc,1}'e^{-i\frac{2\pi(n-1)}{3}}e^{-iC_{3,z}^{n-1}\bm{a}_2 \cdot \vk}\right]
    \nonumber \\
     & h_{\vk,13}^+(\vk) = h_{\vk,21}^{+,*}(\vk)= \sum_{n=1,2,3}\left[-t_{fc,2}e^{i\frac{2\pi(n-1)}{3}}e^{iC_{3,z}^{n-1}\bm{a}_2 \cdot \vk}
    +t_{fc,2}'e^{i\frac{2\pi(n-1)}{3}}e^{-iC_{3,z}^{n-1}\bm{a}_2 \cdot \vk}\right]
    \nonumber \\
     & h_{\vk,23}^+(\vk) = h_{\vk,32}^{+,*}(\vk)  = \sum_{n=1,,2,3}
    \left[ -t_{cc,12}e^{-i\frac{2\pi(n-1)}{3}}e^{iC_{3,z}^{n-1}\bm{a}_2 \cdot \vk}
    -t_{cc,12}'e^{-i\frac{2\pi(n-1)}{3}}e^{-iC_{3,z}^{n-1}\bm{a}_2 \cdot \vk}
    \right]
\end{align}
The parameters obtained from Wannier90 are
\begin{align}
     & \epsilon_f = -13.4\text{meV}, \quad t_{ff} = -6.0\text{meV}\nonumber \\
     & \epsilon_{c,1} = - 61.0\text{meV},\quad t_{cc,1} = 4.2\text{mev}\nonumber \\
     & \epsilon_{c,2} = -54.46\text{meV}, \quad t_{cc,2}=-0.1\text{meV}\nonumber \\
     & t_{fc,1} =5.8\text{meV} ,\quad t_{fc,1}' = 7.1\text{meV} \nonumber \\
     & t_{fc,3} =5.4\text{meV} ,\quad t_{fc,4}' = 4.3\text{meV} \nonumber \\
     & t_{cc,12} =10.5\text{meV} ,\quad t_{cc,12}' = 7.7\text{meV}
\end{align}
Because inversion symmetry is absent, the magnitudes of the nearest-neighbor hoppings along the six directions can differ.
However, such anisotropy is weak (less than $0.1$ meV) for intra-orbital hopping and has been ignored. 
We can also obtain a better fit to the dispersion by requiring the tight-binding model to exactly reproduce the energies of the top three bands at the high-symmetry points $\Gamma_M,K_M,K_M^\prime$.
This gives the following updated parameters:
\begin{align}
     & \epsilon_f = -10.0\text{meV}, \quad t_f = 5.8\text{meV}\nonumber \\
     & \epsilon_{c,1} = - 54.4\text{meV},\quad t_{cc,1} = 3.6\text{mev}\nonumber \\
     & \epsilon_{c,2} = -51.0\text{meV}, \quad t_{cc,2}=0.3\text{meV}
\end{align}
In~\cref{fig:disp_DFT_3tri}, we compare the DFT band structure with that of the effective tight-binding model.
The Chern numbers of the top three bands in the tight-binding model are $1,1,-2$ (from top to bottom).
The overlaps between the wavefunctions of the top three bands obtained from the effective tight-binding model and from DFT are given in~\cref{tab:tbp_3tri_ov_orbital_weight}. For our numerical Hartree-Fock, RPA and FRG calculations, we use a more precise tight-binding model including the full Wannierization without truncation after nearest neighbors.

\begin{table}[ht]
    \centering
    \begin{tabular}{c|c|c | c| }
        Band index                                              & 1    & 2    & 3    \\
        \hline Overlap between the DFT and tight-binding models & 1.00 & 0.94 & 0.41 \\
        \hline Weight of $f$ orbital                            & 0.85 & 0.05 & 0.10 \\
        Weight of $c_1$ orbital                                 & 0.12 & 0.39 & 0.49 \\
        Weight of $c_2$ orbital                                 & 0.02 & 0.56 & 0.41
    \end{tabular}
    \caption{Wavefunction overlaps between the DFT and 3Tri tight-binding models.
        Orbital weights of the 3Tri tight-binding model.
    }
    \label{tab:tbp_3tri_ov_orbital_weight}
\end{table}

We next discuss the interactions.
The dominant interactions are on-site repulsions:
\begin{align}
    H_{onsite} = \frac{1}{2}\sum_{\vR, nm,\sigma\sigma'} U_{nm}: d_{\vR,\sigma,n}^\dag d_{\vR,\sigma,n} d_{\vR,\sigma',m}^\dag d_{\vR,\sigma',n}:
\end{align}
where
\begin{align}
    U_{nm,\sigma\sigma} = \begin{bmatrix}
                              0               & 49.5\text{meV}  & 46.1\text{meV} \\
                              49.5 \text{meV} & 0               & 51.9\text{meV} \\
                              46.1\text{meV}  & 51.9 \text{meV} & 0
                          \end{bmatrix}
    ,\quad
    U_{nm,(\sigma,-\sigma)} = \begin{bmatrix}
                                  116.9 \text{meV} & 91.6 \text{meV} & 83.8 \text{meV} \\
                                  91.6\text{meV}   & 83.5\text{meV}  & 77.7\text{meV}  \\
                                  83.8 \text{meV}  & 77.7\text{meV}  & 73.6\text{meV}
                              \end{bmatrix}
\end{align}
In addition, we have the following generic on-site interactions:
\begin{align}
    H_{int,2} = \frac{1}{2}\sum_{\vR,n_1m_1n_2m_2,\sigma}U'^\sigma_{n_1m_1n_2m_2}:d_{\vR,\sigma,m_1}^\dag d_{\vR,\sigma,n_2}d_{\vR,\bar{\sigma},n_2}^\dag d_{\vR,\bar{\sigma},m_2}:
\end{align}
where
\begin{align}
     & U'^\sigma_{1212} =U'^\sigma_{2121} = 42.2\text{meV}\nonumber \\
     & U'^\sigma_{1313} =U'^\sigma_{3131} = 37.7\text{meV}\nonumber \\
     & U'^\sigma_{2323} =U'^\sigma_{3232} = 25.9\text{meV}\nonumber \\
     & U'^\sigma_{1321} =U'^\sigma_{2113} =U'^\sigma_{2113} =U'^\sigma_{3112}=-39.6\text{meV}
\end{align}
The nearest-neighbor interaction reads
\begin{align}
    H_{NN} = \frac{U_2}{2}\sum_{\vR,n,m,\sigma,\sigma'}\sum_{\bm{e}\in NN} :d_{\vR,\sigma,m}^\dag d_{\vR,\sigma,m}d_{\vR+\bm{e},\sigma',n}^\dag d_{\vR+\bm{e},\sigma',n}:
\end{align}
where $U_2= 28.2$ meV.
The next-nearest-neighbor Coulomb repulsion is $6.5$ meV and has been omitted.
All other interactions that are not of the density–density type are smaller than \(15\,\mathrm{meV}\).

\subsection{1Tri2Hon model}
\label{app:DFT1tri2hon}
As shown in \cref{tab:irrep_dft_model}, the low-energy bands can form $^2\bar{E}_u@1a \oplus \bar{E}@2d $.
We can construct a three-orbital interacting model with one Wannier orbital at the $1a$ position (triangular sites) and one Wannier orbital at each $2d$ position (honeycomb sites).

\subsubsection{Trial wavefunctions}
\label{sec:app:wan_trial_1tri2hon}
We now discuss the trial wavefunctions used for Wannierization.
For the orbital at the $1a$ position, we use the trial wavefunction derived for the $f$ electron in the 3Tri model ($\psi_{\rm{3Tri},1,\sigma,\vR}$ in~\cref{eq:trial_wave_3tri}).

The two Wannier orbitals at the $2d$ positions are located at
\begin{align}
    \vrr_+ = \frac{\bm{a}_1-\bm{a}_2}{3},\quad \vrr_- = C_6\vrr_+
\end{align}
The trial wavefunction for the electrons at the $2d$ positions can be written as
\begin{align}
    \psi^\dag_{\rm{1Tri2Hon},\vR,\alpha,\sigma}= \sum_l \int d^2\vrr A_{\alpha\sigma,l}(\vR) \frac{ e^{-(\vrr-\vrr_\alpha -\vR)^2/(2\lambda_{\alpha,\sigma,l}^2)} }{\sqrt{\pi\lambda_{\alpha,\sigma,l}^2} }c_{\vrr,\sigma,l}^\dag
\end{align}
where $\alpha=1,2$ with $\vrr_{1/2} = \vrr_{\pm }$.

We now discuss the symmetry properties.
Under translation, we use
\begin{align}
    T_\vR c_{\vrr,\sigma,l}^\dag T_\vR^\dag = e^{-i\sigma \kappa_l \cdot\vR} c_{\vrr+\vR,\sigma,l}^\dag
\end{align}
and find
\begin{align}
    T_{\vR'}
    \psi^\dag_{\rm{1Tri2Hon},\vR,\alpha,\sigma}
    T_{\vR'}^\dag
    =
    \sum_l \int d^2\vrr
    e^{-i\sigma \kappa_l \cdot\vR'}
    A_{\alpha\sigma,l}(\vR)
    \frac{1}{\sqrt{\pi\lambda_{\alpha,\sigma,l}^2}}
    e^{-\frac{
                (\vrr-\vrr_\alpha-\vR-\vR')^2
            }{
                (2\lambda_{\alpha,\sigma,l}^2)}}
    c_{\vrr,\sigma,l}
    ^\dag
\end{align}
We take
\begin{align}
    A_{\alpha\sigma l}(\vR) =e^{-i\sigma \kappa_l\cdot\vR}A_{\alpha\sigma l }
\end{align}
which leads to
\begin{align}
    T_{\vR'} \psi_{\rm{1Tri2Hon},\vR,\alpha,\sigma}  T_{\vR'}^\dag  = \psi_{\rm{1Tri2Hon},\vR+\vR',\alpha,\sigma}
\end{align}

Under $C_{3z}$, we find
\begin{align}
      & C_{3z} \psi^\dag_{\rm{1Tri2Hon},\vR,\alpha\sigma} C_{3z}^{-1} = \sum_l \int d^2\vrr A_{\alpha\sigma,l}e^{-i\sigma \kappa_l\cdot\vR}
    \frac{1}{\sqrt{\pi\lambda_{\alpha,\sigma,l}^2}}e^{-(\vrr-\vrr_\alpha-\vR)^2/(2\lambda_{\alpha,\sigma,l}^2)}c_{C_{3z}\vrr,\sigma,l}^\dag e^{i\sigma \pi /3 } \nonumber \\
    = & \sum_l \int d^2\vrr A_{\alpha\sigma,l}e^{-i\sigma \kappa_l\cdot\vR}
    \frac{1}{\sqrt{\pi\lambda_{\alpha,\sigma,l}^2}} e^{-(\vrr- C_{3z}(\vR-\vrr_\alpha))^2/(2\lambda_{\alpha,\sigma,l}^2)}c_{\vrr,\sigma,l}e^{i\sigma \pi/3} \nonumber \\
    = & \sum_l \int d^2\vrr A_{\alpha\sigma,l}
    e^{-i\sigma \kappa_l\cdot(C_{3z}(\vR+\vrr_\alpha)-\vrr_\alpha)}
    e^{-i\sigma \kappa_l\cdot\bigg(
            \vR - C_{3z}(\vR+\vrr_\alpha)+\vrr_\alpha)
            \bigg)}  \frac{1}{\sqrt{\pi\lambda_{\alpha,\sigma,l}^2}}e^{-\bigg(\vrr-(C_{3z}(\vR+\vrr_\alpha)\bigg)^2/(2\lambda_{\alpha,\sigma,l}^2) }c_{\vrr,\sigma,l}e^{i\sigma \pi/3}
\end{align}
We note that
\begin{align}
    e^{-i\sigma \kappa_l\cdot  (\vR- (C_{3z}(\vR+\vrr_\alpha)-\vrr_\alpha))}
    =
    e^{i\sigma \kappa_l\cdot (C_{3z}\vrr_\alpha -\vrr_\alpha) }
    =
    e^{i\sigma (-1)^l (-1)^{\alpha } \frac{2\pi}{3}}
\end{align}
with $\alpha=1,2$. We can choose a layer-polarized Wannier function with
\begin{align}
    A_{\alpha\sigma,l} = A_{\alpha,\sigma} \delta_{(-1)^l, (-1)^\alpha}
\end{align}
which gives
\begin{align}
      & C_{3z}\psi_{\rm{1Tri2Hon},\vR,\alpha\sigma}^\dag C_{3z}^{-1}
    =
    e^{i\sigma 2\pi/3+i\sigma \pi/3}
    \psi_{\rm{1Tri2Hon},C_{3z}(\vR+\vrr_\alpha)-\vrr_\alpha ,\alpha\sigma }
    \nonumber \\
    = &
    e^{-i\pi  }
    \psi^\dag_{\rm{1Tri2Hon},C_{3z}(\vR+\vrr_\alpha)-\vrr_\alpha,\alpha\sigma  }
\end{align}

Under $C_{2y}\mathcal{T}$, we find
\begin{align}
      & C_{2y}\mathcal{T}
    \psi_{\rm{1Tri2Hon},\vR,\alpha\sigma}^\dag (C_{2y}\mathcal{T})^{-1}= \nonumber \\
      & \sum_l \int d^2\vrr A^*_{\alpha\sigma,l} e^{+i\sigma \kappa_l\cdot\vR} \frac{1}{\sqrt{\pi\lambda_{\alpha,\sigma,l}^2}}e^{-(\vrr-\vrr_\alpha-\vR)^2/(2\lambda_{\alpha,\sigma,l}^2)}c_{C_{2y}\vrr,\sigma,\bar{l}}^\dag \nonumber \\
    = & \sum_l \int d^2\vrr A^*_{\alpha\sigma,\bar{l}}e^{-i\sigma \kappa_l\cdot\vR} \frac{1}{\sqrt{\pi\lambda_{\alpha,\sigma,l}^2}}e^{-[\vrr-C_{2y}(\vrr_\alpha-\vR)]^2/(2\lambda_{\alpha,\sigma,l}^2)}c_{\vrr,\sigma,{l}}^\dag \nonumber \\
    = & \sum_l \int d^2\vrr A^*_{{\alpha}\sigma,\bar{l}} e^{-i\sigma \kappa_l\cdot\bigg(C_{2y}(\vrr_\alpha+\vR)-\vrr_{\bar{\alpha}}\bigg)} e^{i\sigma \kappa_l\cdot \bigg[ \bigg(C_{2y}(\vrr_\alpha+\vR)-\vrr_{\bar{\alpha}}\bigg) -\vR \bigg] } \frac{1}{\sqrt{\pi \lambda_{\alpha,\sigma,l}^2}}e^{-\bigg[\vrr- \bigg(C_{2y}(\vrr_\alpha+\vR)-\vrr_{\bar{\alpha}}\bigg) -\vrr_{\bar{\alpha}} \bigg]^2/(2\lambda_{\alpha,\sigma,l}^2)}c_{\vrr,\sigma,l}^\dag \nonumber \\
    = & \sum_l \int d^2\vrr A^*_{{\alpha}\sigma,\bar{l}} e^{-i\sigma \kappa_l\cdot\bigg(C_{2y}(\vrr_\alpha+\vR)-\vrr_{\bar{\alpha}}\bigg)} e^{-i\sigma2\pi/3} \frac{1}{\sqrt{\pi \lambda_{\alpha,\sigma,l}^2}}e^{-\bigg[\vrr- \bigg(C_{2y}(\vrr_\alpha+\vR)-\vrr_{\bar{\alpha}}\bigg) -\vrr_{\bar{\alpha}} \bigg]^2/(2\lambda_{\alpha,\sigma,l}^2)}c_{\vrr,\sigma,l}^\dag
\end{align}
where $\bar{\alpha}=3-\alpha$.
We take
\begin{align}
    A_{\alpha \sigma,l} = \delta_{(-1)^l, -(-1)^\alpha} e^{-i\sigma \pi/3 -i(-1)^l\sigma 2\pi/3},\quad \lambda_{\alpha,\sigma,l} = \lambda_{\sigma}
\end{align}
which gives
\begin{align}
    C_{2y}\mathcal{T}
    \psi_{\rm{1Tri2Hon},\vR,\alpha\sigma}^\dag
    (C_{2y}\mathcal{T})^{-1}
    =
    \psi_{\rm{1Tri2Hon},C_{2y}(\vR+\vrr_\alpha)-\vrr_{\bar{\alpha}},\bar{\alpha}\sigma}
\end{align}

Under inversion, we find
\begin{align}
    \mathcal{I}
    \psi_{\rm{1Tri2Hon},\vR,\alpha\sigma}^\dag
    \mathcal{I}^{-1}
    = & \sum_l \int d^2\vrr A_{\alpha \sigma ,l}e^{-i\sigma \kappa_l\cdot\vR }\frac{1}{\sqrt{\pi\lambda_{\sigma}^2}}e^{-\bigg(\vrr-\vR-\vrr_\alpha\bigg)^2/(2\lambda_{\sigma}^2)} c_{-\vrr,\sigma,\bar{l}}  \nonumber \\
    = & \sum_l \int d^2\vrr A_{\alpha \sigma ,\bar{l}}e^{i\sigma \kappa_l\cdot\vR }\frac{1}{\sqrt{\pi\lambda_{\sigma}^2}}e^{-\bigg(\vrr +\vR+\vrr_\alpha\bigg)^2/(2\lambda_{\sigma}^2)} c_{\vrr,\sigma,{l}}  \nonumber \\
    = & \sum_l \int d^2\vrr A_{\bar{\alpha}\sigma,l} e^{-i\sigma\kappa_l\cdot(\mathcal{I}(\vR+\vrr_\alpha)-\vrr_{\bar{\alpha}})}e^{-(\vrr -\mathcal{I}(\vR+\vrr_\alpha)^2/(2\lambda^2)}c_{\vrr,\sigma,l } \nonumber \\
    = &
    \psi^\dag_{\rm{1Tri2Hon},\mathcal{I}(\vR+\vrr_\alpha)-\vrr_{\bar{\alpha}},\bar{\alpha}\sigma }
\end{align}

Finally, the time-reversal transformation gives
\begin{align}
    \mathcal{T}
    \psi^\dag _{\rm{1Tri2Hon},\vR,\alpha\sigma}
    \mathcal{T}^{-1}
    =
    \sum_l \int d^2\vrr
    A_{\alpha \sigma,l} ^*
    e^{i\sigma \kappa_l\cdot \vR }
    e^{-(\vrr-\vrr_\alpha-\vR)^2/(2\lambda_\sigma^2)}\frac{1}{\sqrt{\pi\lambda_\sigma^2}}
    (-\sigma)
    c_{r,-\sigma,l}^\dag
    (-\sigma)
\end{align}
We take $\lambda_\sigma = \lambda$, which gives
\begin{align}
    \mathcal{T}
    \psi^\dag _{\rm{1Tri2Hon},\vR,\alpha\sigma}
    \mathcal{T}^{-1}
    =
    - \sigma \psi^\dag _{\rm{1Tri2Hon},\vR,\alpha -\sigma}
\end{align}

We also consider the Fourier transform:
\begin{align}
    \psi^\dag_{\rm{1Tri2Hon},\vR,\alpha\sigma} 
    = & 
    \sum_l \int d^2\vrr A_{\alpha\sigma,l} e^{-i\sigma \kappa_l\cdot\vR} \frac{1}{\sqrt{\pi\lambda^2}} e^{-(\vrr-\vR-\vrr_{\alpha})^2/(2\lambda^2)}\frac{1}{\sqrt{\Omega_{\tTot}}} \sum_{\vk,\vQ} c_{\vk,\sigma,\vQ}^\dag e^{-i(\vk-\vQ)\cdot\vrr} \nonumber \\
    =                                            
    & \frac{1}{\sqrt{\Omega_{\tTot}}}\sum_{\vk,\vQ}
    A_{\alpha\sigma,l}e^{-\frac{1}{2}\lambda^2(\vk-\vQ)^2 -i(\vk)\cdot(\vR+\vrr_\alpha) +i\vQ\cdot \vrr_\alpha }c_{\vk,\sigma,\vQ}^\dag \label{eq:trial_1tri2hon}
\end{align}

\subsubsection{Wannier model} The trial Wannier functions we use are given by $\psi_{\rm{3Tri},1,\sigma,\vR}^\dag$ in~\cref{eq:trial_wave_3tri} with $\lambda =0.4a_M$ and $\psi_{\rm{1Tri2Hon},\vR,\alpha\sigma s}^\dag$ given in~\cref{eq:trial_1tri2hon} with $\lambda = 0.4a_M$.
After Wannierization, the spreads of the three Wannier orbitals in the 1Tri2Hon model are
\begin{align}
    f:0.352a_M,\quad c_1:0.382a_M,\quad c_2:0.382a_M
\end{align}
Here, $c_1,c_2$ denote the two orbitals at the $2d$ positions, and $f$ denotes the orbital at the $1a$ position.

The 1Tri2Hon tight-binding Hamiltonian can be written as
\begin{align}
    H_0 = \sum_\vk d_{\vk,\sigma,n}^\dag h^\sigma_{nm}(\vk) d_{\vk,\sigma,m},\quad
    d_{\vk,\sigma,m} = \begin{bmatrix}
                           f_{\vk,\sigma} & c_{\vk,\sigma,1} & c_{\vk,\sigma,2}
                       \end{bmatrix}
    _m
\end{align}
Keeping only nearest-neighbor hopping, we find
\begin{align}
     & h^+_{11}(\vk) = \epsilon_f
    \nonumber \\
     & h^+_{22}(\vk) = \epsilon_{c} \nonumber \\
     & h^+_{33}(\vk) = \epsilon_{c}
    \nonumber \\
     & h^+_{12}(\vk) = h_{21}^{+,*}=t_{fc}e^{i\phi_1} \sum_{j=1,2,3} e^{i4\pi(j-1)/3}e^{i\vk\cdot C_{3z}^{j-1}\vrr_1} \nonumber \\
     & h^+_{13}(\vk) =t_{fc}e^{i(-\phi_1+\pi/3)} \sum_{j=1,2,3} e^{-i2\pi(j-1)/3}e^{i\vk\cdot C_{3z}^{j-1}\vrr_2} \nonumber \\
     & h^+_{23}(\vk) =h^{+,*}_{32}(\vk) = t_{cc}e^{i\phi_2} \sum_{j=1,2,3} e^{i\vk\cdot C_{3z}^{j-1}(\vrr_2-\vrr_1) }
\end{align}
where
\begin{align}
     & \vrr_1= (\frac{1}{2\sqrt{3}} ,-\frac{1}{2})a_M,\quad \vrr_2 =\frac{1}{\sqrt{3}}(1,0)a_M \nonumber \\
     & \epsilon_f= -34.7\text{meV},\quad \epsilon_c = -43.9\text{meV} \nonumber \\
     & t_{fc} = 14.0,\quad \phi_1 =0.01 \pi  \nonumber \\
     & t_{cc} = 2.56, \quad \phi_2 =-0.78\pi
\end{align}

\begin{figure}[ht]
    \centering
    \includegraphics[width=0.7\textwidth]{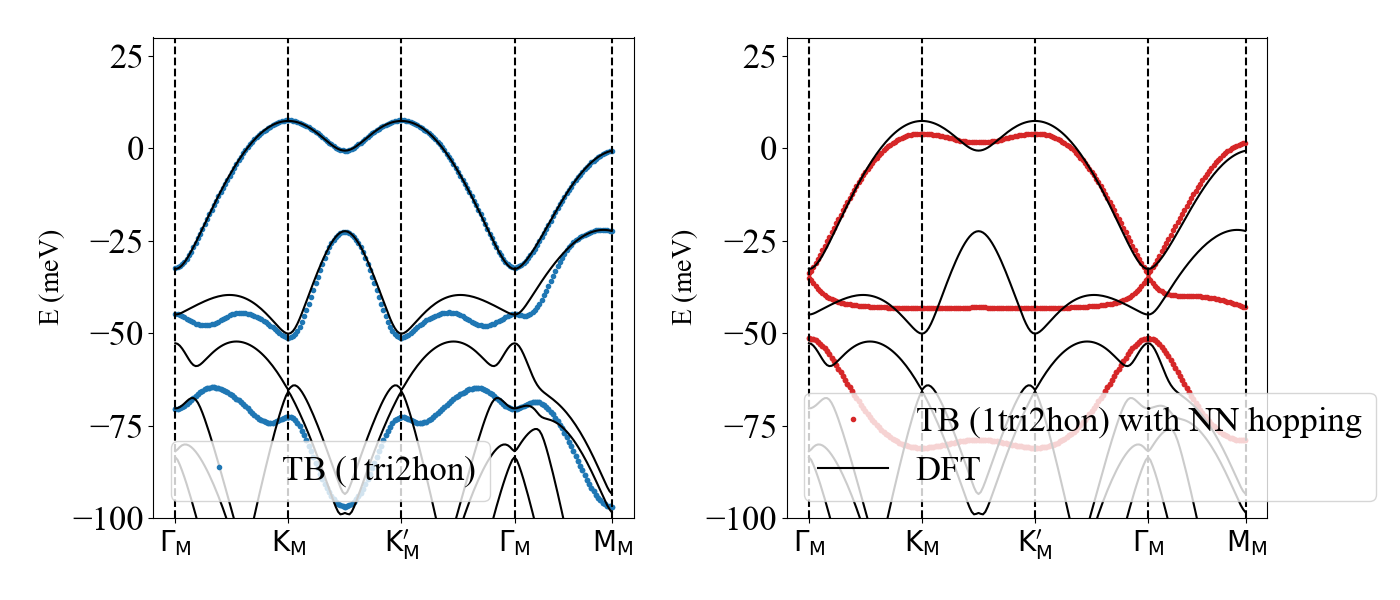}
    \caption{Comparison of the DFT band structure with that of the effective tight-binding model (1Tri2Hon). (Left) Tight-binding model including hoppings over distances shorter than $5a_M$. (Right) Tight-binding model with nearest-neighbor hopping.  }
    \label{fig:disp_DFT_1tri2hon}
\end{figure}

We compare the band structures of the DFT model and the 1Tri2Hon tight-binding model in~\cref{fig:disp_DFT_1tri2hon}.
We also list the overlaps between the wavefunctions obtained from the DFT model and the tight-binding model in~\cref{tab:tbp_tri_honey_ov_orbital_weight}.

We find that the Chern numbers, from the top band to the bottom band, are $0,-1,1$ in our nearest-neighbor tight-binding model.
However, after including all hoppings over distances shorter than $5a_M$, the Chern numbers, from the top band to the bottom band, are $1,1,-2$.
The band structures of the two tight-binding models are also shown in~\cref{fig:disp_DFT_1tri2hon}.
The band gaps at the $\Gamma_M$ point are small in the nearest-neighbor tight-binding model.
Band inversion can occur when longer-range hoppings are included.

\begin{table}[ht]
    \centering
    \begin{tabular}{c|c|c | c| }
        Band index                                              & 1    & 2    & 3    \\
        \hline Overlap between the DFT and tight-binding models & 0.98 & 0.80 & 0.54 \\
        \hline Weight of $f$ orbital                            & 0.56 & 0.0  & 0.44 \\
        Weight of $c_1$ orbital                                 & 0.22 & 0.50 & 0.28 \\
        Weight of $c_2$ orbital                                 & 0.22 & 0.50 & 0.28
    \end{tabular}
    \caption{Wavefunction overlaps between the DFT and 1Tri2Hon tight-binding models.
        Orbital weights of the 1Tri2Hon tight-binding model.
    }
    \label{tab:tbp_tri_honey_ov_orbital_weight}
\end{table}

The dominant interactions are the on-site Hubbard interactions with
\begin{align}
    H_{int,1} = \frac{1}{2}\sum_{\vR,\sigma,n}U_n :d_{\vR,\sigma,n}^\dag d_{\vR,\sigma,n} d_{\vR,-\sigma,n}^\dag d_{\vR,-\sigma,n}:
\end{align}
where
\begin{align}
    U_1 = 160.3\text{meV}, \quad
    U_2 = 144.7\text{meV},\quad
    U_3 = 144.7\text{meV}
\end{align}

The nearest-neighbor and next-nearest-neighbor Coulomb repulsions are
\begin{align}
    H_{int,2}
    =&
    \frac{V_1^{\rm{NN}}}{2}
    \sum_{\langle (\vR,n), (\vR',m)\rangle}
    :
        d_{\vR,\sigma,n}^\dag
        d_{\vR,\sigma,n}
        d_{\vR',\sigma',m}^\dag
        d_{\vR',\sigma',m}
    :
    \nonumber \\
    &
    +
    \frac{V_2^{\rm{NN}}}{2}
    \sum_{\langle\langle (\vR,n), (\vR',m)\rangle\rangle }
    :
        d_{\vR,\sigma,n}^\dag
        d_{\vR,\sigma,n}
        d_{\vR',\sigma',m}^\dag
        d_{\vR',\sigma',m}
    :
    \nonumber \\
    &
    +
    \frac{V_3^{\rm{NN}}}{2}
    \sum_{\langle\langle (\vR,n), (\vR',m)\rangle\rangle }
    :
        d_{\vR,\sigma,n}^\dag
        d_{\vR,\sigma,n}
        d_{\vR',\sigma',m}^\dag
        d_{\vR',\sigma',m}
    :
\end{align}
with $V_1^{\rm{NN}}= 60.7$ meV, $V_2^{\rm{NN}}=22.5$ meV, and $V_3^{\rm{NN}} =16.1$ meV.
All other interaction strengths are smaller than $15$ meV.
The Coulomb repulsions also depend on the orbitals.
However, the orbital dependence is smaller than $2$ meV and has been omitted.

\subsection{2Hon model}
\label{app:DFT2hon}
As shown in \cref{tab:irrep_dft_model}, we can also construct a 2Hon model using only Wannier orbitals at the honeycomb sites.

\subsubsection{Wannier model}
We use the trial wavefunctions introduced in \cref{eq:trial_1tri2hon}, with $\lambda = 0.4a_M$, retaining only the two orbitals centered at the honeycomb sites.
After Wannierization, the spreads of the Wannier orbitals are
\begin{align}
    c_1:0.510a_M,\quad c_2: 0.510 a_M
\end{align}

The tight-binding model can now be written as
\begin{align}
    H_0 = \sum_\vk d_{\vk,\sigma,n}^\dag h^\sigma_{nm}(\vk) d_{\vk,\sigma,m},\quad
    d_{\vk,\sigma,m} = \begin{bmatrix}
                           c_{\vk,\sigma,1} & c_{\vk,\sigma,2}
                       \end{bmatrix}
    _m
\end{align}
Keeping only nearest-neighbor and next-nearest-neighbor hoppings, we find
\begin{align}
     & h^+_{11}(\vk) = \epsilon_c + \sum_{j=1,...,6}e^{i\vk\cdot C_{6z}^{j-1}\bm{a}_{M,2} + i\phi_2(-1)^{j-1}} t_2
    \nonumber \\
     & h^+_{22}(\vk) = \epsilon_{c} +\sum_{j=1,...,6}e^{i\vk\cdot C_{6z}^{j-1}\bm{a}_{M,2} - i\phi_2(-1)^{j-1}} t_2 \nonumber \\
     & h^+_{12}(\vk) = h^{+,*}_{21}(\vk)= t_1 e^{i\phi_1} \sum_{j=1,2,3} e^{i\vk\cdot(\vrr_2-\vrr_1)}
\end{align}
where
\begin{align}
     & \vrr_1= (\frac{1}{2\sqrt{3}} ,-\frac{1}{2})a_M,\quad \vrr_2 =\frac{1}{\sqrt{3}}(1,0)a_M \nonumber \\
     & \epsilon_c= -13.3\text{ meV},\quad t_1 =6.2\text{ meV},\quad  t_2= 5.8\text{ meV}  \nonumber \\
     & \phi_1 =2\pi/3 ,\quad \phi_2 =-2\pi/3
\end{align}

\begin{figure}[ht]
    \centering
    \includegraphics[width=0.7\textwidth]{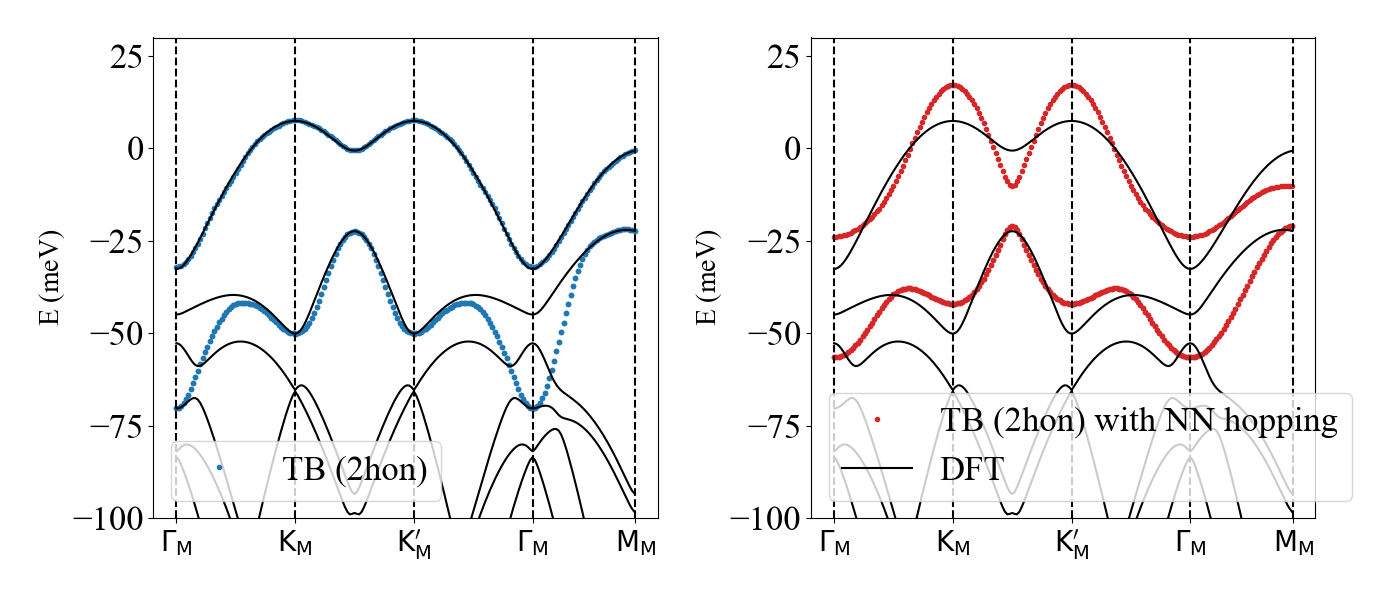}
    \caption{Comparison of the DFT band structure with that of the effective tight-binding model (2Hon). (Left) Tight-binding model including hoppings over distances shorter than $5a_M$. (Right) Tight-binding model with nearest-neighbor hopping.  }
    \label{fig:disp_DFT_2hon}
\end{figure}

We compare the dispersions of the DFT model and the tight-binding model in~\cref{fig:disp_DFT_2hon}.
The overlaps between the wavefunctions of the DFT model and the tight-binding model are given in~\cref{tab:tbp_hony_ov_orbital_weight}.
The Chern numbers calculated from the tight-binding model are $+1,-1$ for the top and bottom bands, respectively.

The interactions of the 2Hon model are
\begin{align}
    H_{\tInt} = & \frac{U}{2}\sum_{\vR,n}
    :
    d_{\vR,\sigma,n}^\dag
    d_{\vR,\sigma,n}
    d_{\vR,-\sigma,n}^\dag
    d_{\vR,-\sigma,n}
    :
    \nonumber \\
                &
    +
    \frac{V_1^{\rm{NN}}}{2}
    \sum_{\langle (\vR,n), (\vR',m)\rangle}
    :
    d_{\vR,\sigma,n}^\dag
    d_{\vR,\sigma,n}
    d_{\vR',\sigma',m}^\dag
    d_{\vR',\sigma',m}
    :
    \nonumber \\
                &
    +
    \frac{V_2^{\rm{NN}}}{2}
    \sum_{\langle\langle  (\vR,n), (\vR',m)\rangle\rangle }
    :
    d_{\vR,\sigma,n}^\dag
    d_{\vR,\sigma,n}
    d_{\vR',\sigma',m}^\dag
    d_{\vR',\sigma',m}
    :
    \nonumber \\
                &
    +
    \frac{V_3^{\rm{NN}}}{2}
    \sum_{\langle\langle  (\vR,n), (\vR',m)\rangle\rangle }
    :
    d_{\vR,\sigma,n}^\dag
    d_{\vR,\sigma,n}
    d_{\vR',\sigma',m}^\dag
    d_{\vR',\sigma',m}
    :
\end{align}
where
\begin{align}
    U = 109.3\text{meV},\quad V_1^{\rm{NN}} = 57.0\text{meV},\quad V_2^{\rm{NN}} = 25.6\text{meV},\quad V_3^{\rm{NN}} = 17.1 \text{meV},\quad
\end{align}
All other interactions are smaller than $15$ meV and have been omitted.

\begin{table}[ht]
    \centering
    \begin{tabular}{c|c|c  }
        Band index                                              & 1    & 2    \\
        \hline Overlap between the DFT and tight-binding models & 0.96 & 0.81 \\
        \hline Weight of $c_1$ orbital                          & 0.5  & 0.5  \\
        Weight of $c_2$ orbital                                 & 0.5  & 0.5
    \end{tabular}
    \caption{Wavefunction overlaps between the DFT and 2Hon tight-binding models.
        Orbital weights of the 2Hon tight-binding model.
    }
    \label{tab:tbp_hony_ov_orbital_weight}
\end{table}

\subsection{Effect of the electric field}
\begin{figure}[ht]
    \centering
    \includegraphics[width=1.0\textwidth]{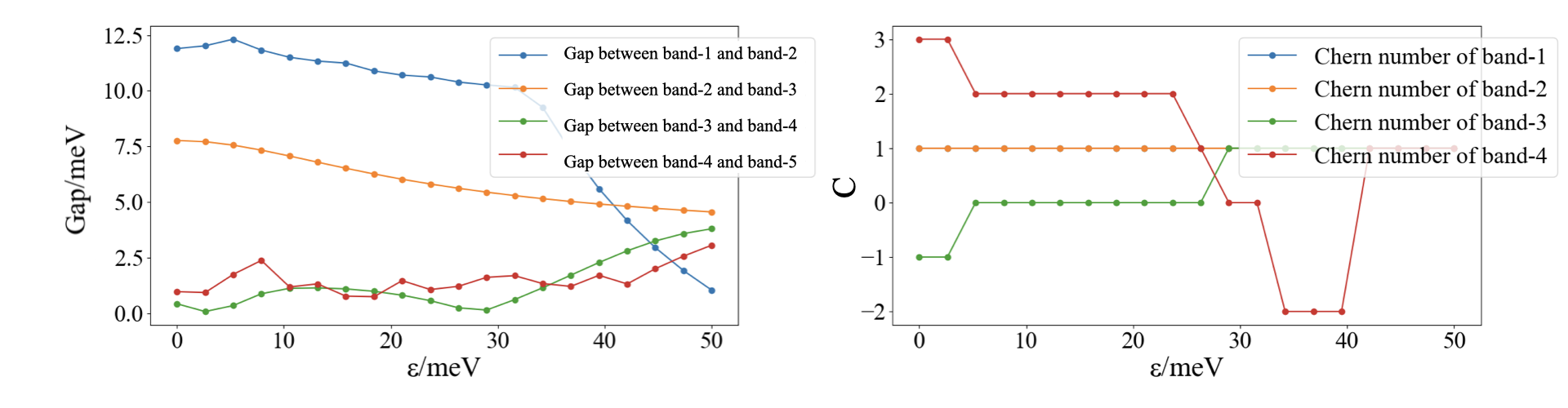}
    \caption{Gaps between the top 5 bands (left) and Chern numbers of the top 4 bands (right) of the DFT model in an electric field. The Chern numbers of first band and the second band are identical within the plotted region. }
    \label{fig:Gap_chern_number_ele_field_DFT}
\end{figure}

We now explore the effect of the electric field.
We first take the DFT model and consider a finite displacement field (see \cref{eq:h_QQ_k}).
We then plot the direct gap and the Chern numbers for the top 4 bands in~\cref{fig:Gap_chern_number_ele_field_DFT}.
Several gaps close as the electric field increases, accompanied by changes in the band Chern numbers.

We now check whether the existing Wannier orbitals still describe the top band of the model at finite electric fields.
For a given set of Wannier orbitals and the corresponding Bloch states
\begin{align}
    d_{\vk,\sigma,n}^\dag|0\rangle  = |d,\vk,\sigma,n\rangle \, ,
\end{align}
we introduce the projection operator
\begin{align}
    P_{\vk,wan} = \sum_{n,\sigma}|d, \vk,\sigma,n\rangle \langle d, \vk,\sigma,n|
\end{align}
We then calculate the overlap between the $n$-th band and the Wannier states, defined as
\begin{align}
    OV_{n} = \frac{1}{N}\sum_\vk \langle \vk,\sigma,n| P_{\vk,wan} |\vk,\sigma,n\rangle
\end{align}
where $|\vk,\sigma,n\rangle$ denotes the $n$-th band.

We also obtain the electric-field term in the Wannier basis by direct projection.
The corresponding Hamiltonian now reads
\begin{align}
    H_{D} = 
    D
    \sum_{\vk,nm} 
    \epsilon_{nm}^{D,\sigma}(\vk)
    d_{\vk,\sigma,n}^\dag
    d_{\vk,\sigma,m}
    \label{eq:displacement_field_dependency}
\end{align}

\subsubsection{3Tri model}

The overlaps of the top two bands for the 3Tri model are given in~\cref{tab:ov_3tri_ele_field}.
The existing Wannier orbitals still describe the first band well, which is the most relevant to low-energy physics.

\begin{table}[ht]
    \centering
    \begin{tabular}{c|c|c|c|c|c}
        $D/$meV & 10   & 20   & 30   & 40   & 50   \\
        \hline
        $OV_1$         & 0.99 & 0.96 & 0.94 & 0.93 & 0.92 \\
        $OV_2$         & 0.97 & 0.93 & 0.83 & 0.68 & 0.55 \\
    \end{tabular}
    \caption{Overlaps between the top two bands and the Wannier states for the 3Tri model.}
    \label{tab:ov_3tri_ele_field}
\end{table}

\begin{figure}[ht]
    \centering
    \includegraphics[width=1.0\linewidth]{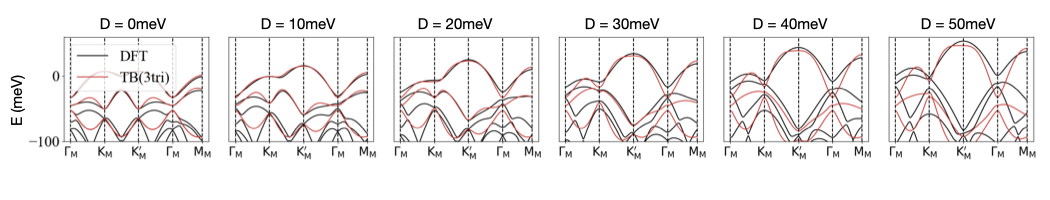}
    \caption{Comparison of the DFT model and the tight-binding model (3Tri) at different electric fields in valley $+$.
    }
    \label{fig:dft_ele_3tri_disp}
\end{figure}

The electric-field term in the Wannier basis for the 3Tri model reads
\begin{align}
    &
    \epsilon^{D,\sigma=+}_{11}(\vk)
    =
    \sum_{j=1,...,6}it_{11}^D
    e^{i\pi(j-1)}
    e^{i\vk\cdot C_{6z}^{j-1}\bm{a}_{M,2}}
    \nonumber \\
    &
    \epsilon^{D,\sigma=+}_{22}(\vk)
    =
    \sum_{j=1,...,6}it_{22}^D e^{i\pi(j-1)} e^{i\vk\cdot C_{6z}^{j-1}\bm{a}_{M,2}} 
    \nonumber \\
    & 
    \epsilon^{D,\sigma=+}_{33}(\vk) 
    = 
    \sum_{j=1,...,6}
    it_{33}^D 
    e^{i\pi(j-1)} 
    e^{i\vk\cdot C_{6z}^{j-1}\bm{a}_{M,2}} 
    \nonumber \\
    & \epsilon^{D,\sigma=+}_{12}(\vk) 
    = 
    [\epsilon^{D,\sigma=+}_{21}(\vk)]^* 
    = 
    \sum_{j=1,2,3}(-i)
    t_{12,1}^{D}
    e^{i 4(j-1)\pi/3}
    e^{i\vk\cdot C_{3z}^{j-1}\bm{a}_{M,2}}
    +
    \sum_{j=1,2,3}
    (-i)t_{12,2}^{D}
    e^{i 4(j-1)\pi/3}
    e^{-i\vk\cdot C_{3z}^{j-1}\bm{a}_{M,2}}
    \nonumber \\
    & \epsilon^{D,\sigma=+}_{13}(\vk) 
    = 
    [\epsilon^{D,\sigma=+}_{31}(\vk)]^* 
    = 
    \sum_{j=1,2,3}
    it_{13,1}^{D}
    e^{-i 4(j-1)\pi/3}
    e^{i\vk\cdot C_{3z}^{j-1}\bm{a}_{M,2}}
    +
    \sum_{j=1,2,3}
    it_{13,2}^{D}e^{-i 4(j-1)\pi/3}
    e^{-i\vk\cdot C_{3z}^{j-1}\bm{a}_{M,2}} 
    \nonumber \\
    & \epsilon^{D,\sigma=+}_{23}(\vk) 
    = 
    [\epsilon^{D,\sigma=+}_{32}(\vk)]^* 
    = 
    \sum_{j=1,2,3}
    it_{23,1}^{D}
    e^{-i 2(j-1)\pi/3}
    e^{i\vk\cdot C_{3z}^{j-1}\bm{a}_{M,2}}
    +
    \sum_{j=1,2,3}
    (-i)t_{23,2}^{D}e^{i 4(j-1)\pi/3}
    e^{-i\vk\cdot C_{3z}^{j-1}\bm{a}_{M,2}}
\end{align}
where
\begin{align}
     &
    t_{11}^{D} =  0.15\text{ meV},\quad
    t_{22}^{D} =  0.16\text{ meV}, \quad
    t_{33}^{D} =  0.16\text{ meV} \nonumber \\
     &
    t_{12,1}^{D} = 0.17\text{ meV},\quad
    t_{12,2}^{D} = 0.14\text{ meV}
    \nonumber \\
     &
    t_{23,1}^{D} = 0.18\text{ meV},\quad
    t_{23,2}^{D} = 0.14\text{ meV}
\end{align}
In~\cref{fig:dft_ele_3tri_disp}, we compare the band structures of the effective tight-binding model and the DFT model at different electric fields.

\subsubsection{1Tri2Hon model}

The overlaps of the top two bands for the 1Tri2Hon model are given in~\cref{tab:ov_1tri2hon_ele_field}.
The existing Wannier orbitals still describe the first band well, which is the most relevant to low-energy physics.

The electric-field term in the Wannier basis for the 1Tri2Hon model reads
\begin{align}
     & 
     \epsilon_{11}^{D,+}(\vk)=0,
     \quad \epsilon_{22}^{D,+}(\vk)  = -t_{11}^D ,
     \quad \epsilon_{33}^{D,+}(\vk) =t_{11}^D 
     \nonumber \\
     & 
     \epsilon_{12}^{D,+ }(\vk)=
     [\epsilon_{21}^{D,+ }(\vk)]^* = 
     \sum_{j=1,2,3} 
     t_{12}^D  
     e^{i\frac{-1-2j}{3}\pi} 
     e^{i\vk\cdot C_{3z}^{j-1}\bm{r}_1}  
     \nonumber \\
     & 
     \epsilon_{13}^{D,+ }(\vk)=
     [\epsilon_{31}^{D,+ }(\vk)]^* = 
     \sum_{j=1,2,3} 
     t_{12}^D  
     e^{i\frac{3-2j}{3}\pi} 
     e^{i\vk\cdot C_{3z}^{j-1}\bm{r}_2}  
     \nonumber \\
     & \epsilon_{23}^{D,+ }(\vk)=
     [\epsilon_{32}^{D,+ }(\vk)]^* = 0
\end{align}
where we retain terms only up to nearest neighbors, with
\begin{align}
    t_{11}^D = 0.84\text{ meV},
    \quad 
    t^D_{12} = 0.21\text{ meV}
\end{align}

\begin{figure}[ht]
    \centering
    \includegraphics[width=1.0\linewidth]{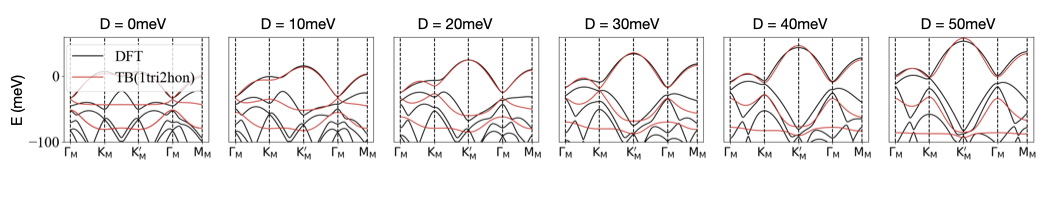}
    \caption{Comparison of the DFT model and the tight-binding model (1Tri2Hon) at different electric fields in valley $+$.
    }
    \label{fig:dft_ele_1tri2hon_disp}
\end{figure}

In~\cref{fig:dft_ele_1tri2hon_disp}, we compare the band structures of the effective tight-binding model and the DFT model at different electric fields.

\begin{table}[ht]
    \centering
    \begin{tabular}{c|c|c|c|c|c}
        $D/$meV & 10   & 20   & 30   & 40   & 50   \\
        \hline
        $OV_1$         & 1.00 & 0.99 & 0.98 & 0.97 & 0.97 \\
        $OV_2$         & 0.94 & 0.90 & 0.82 & 0.71 & 0.62 \\
    \end{tabular}
    \caption{Overlaps between the top two bands and the Wannier states for the 1Tri2Hon model.}
    \label{tab:ov_1tri2hon_ele_field}
\end{table}

\subsubsection{2Hon model}
The overlaps of the top two bands for the 2Hon model are given in~\cref{tab:ov_2hon_ele_field}.
The existing Wannier orbitals still describe the first band well, which is the most relevant to low-energy physics.

The electric-field term in the Wannier basis for the 2Hon model reads
\begin{align}
    & 
    \epsilon_{11}^{D,+}(\vk)=-t_{11}^D,
    \quad 
    \epsilon_{22}^{D,+}(\vk)  = t_{11}^D,
    \quad
    \epsilon_{12}^{D,+}(\vk) = 
    [\epsilon_{21}^{D,+}]^*=0
\end{align}
where we retain terms only up to nearest neighbors, with
\begin{align}
    t_{11}^D = 0.86\text{ meV}
\end{align}

\begin{figure}[ht]
    \centering
    \includegraphics[width=1.0\linewidth]{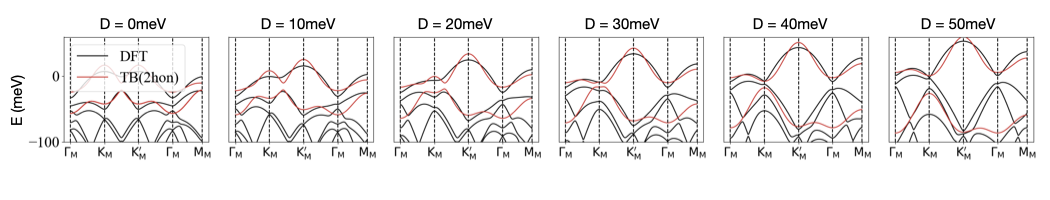}
    \caption{Comparison of the DFT model and the tight-binding model (2Hon) at different electric fields in valley $+$.
    }
    \label{fig:dft_ele_2hon_disp}
\end{figure}

In~\cref{fig:dft_ele_2hon_disp}, we compare the band structures of the effective tight-binding model and the DFT model at different electric fields.

\begin{table}[ht]
    \centering
    \begin{tabular}{c|c|c|c|c|c}
        $D/$meV & 10   & 20   & 30   & 40   & 50   \\
        \hline
        $OV_1$         & 0.99 & 0.97 & 0.95 & 0.94 & 0.94 \\
        $OV_2$         & 0.85 & 0.78 & 0.64 & 0.45 & 0.30 \\
    \end{tabular}
    \caption{Overlaps between the top two bands and the Wannier states for the 2Hon model.}
    \label{tab:ov_2hon_ele_field}
\end{table}

\FloatBarrier
\newpage

\section{Hartree-Fock calculations}\label{app:sec:HF}

In this \siSection{}, we provide details of the Hartree-Fock calculations performed in this work.
We begin by introducing a unified notation for the various Wannier models obtained in this work.
We then describe how states that spontaneously break moir\'e translation symmetry can be treated within a supercell Hartree-Fock formalism.
We discuss the properties of the Hartree-Fock ground states of interest and define the observables, or equivalently order parameters, that we use to characterize them.
Finally, we present the results of our Hartree-Fock calculations for the 3Tri wannier model obtained both from the continuum model of Ref.~\cite{devakulMagicTwistedTransition2021} and from our DFT model.

\subsection{Wannier model Hamiltonian}\label{app:sec:HF:Wannier_model}

For a given Wannier model, the single-particle Hamiltonian expressed in the Wannier basis is given by
\begin{align}
    H_0
    =
    \sum_{\substack{\vk \in \tMBZ \\ n,m,\sigma}}
    h^\sigma_{nm}\left( \vk \right)
    \cre{d}{\vk,\sigma,n}
    \des{d}{\vk,\sigma,m}. \label{app:eqn:single_particle_wannier_hamiltonian_general}
\end{align}
We employ the following Fourier transform convention to relate the momentum-space and real-space Wannier operators
\begin{align}
    \cre{d}{\vR,\sigma,n}
    & =
    \frac{1}{\sqrt{N}}
    \sum_{\vk\in\tMBZ}
    \cre{d}{\vk,\sigma,n}
    e^{-i\vk\cdot\left( \vR+\vrr_n \right)},
    \label{app:eqn:ft_conv_general_k_to_r} \\
    \cre{d}{\vk,\sigma,n}
    & =
    \frac{1}{\sqrt{N}}
    \sum_{\vR}
    \cre{d}{\vR,\sigma,n}
    e^{+i\vk\cdot\left( \vR+\vrr_n \right)}.
    \label{app:eqn:ft_conv_general_r_to_k}
\end{align}
Here, $\vR$ labels the moir\'e unit cell while $\vR+\vrr_n$ denotes the physical center of Wannier orbital $n$ in that unit cell.
In real space, the single-particle Hamiltonian therefore reads
\begin{align}
    H_0 = \sum_{\substack{\vR,\Delta\vR \\ n,m,\sigma}} h^\sigma_{nm}\left( \Delta\vR \right)
    \cre{d}{\vR+\Delta\vR,\sigma,n}
    \des{d}{\vR,\sigma,m},
\end{align}
where
\begin{align}
    h^\sigma_{nm}\left( \vk \right)
    & = 
    \sum_{\Delta\vR} 
    h^\sigma_{nm}\left( \Delta\vR \right) 
    e^{-i\vk\cdot\left( \Delta\vR+\vrr_n-\vrr_m \right)}, 
    \\
    h^\sigma_{nm}\left( \Delta\vR \right) 
    & = 
    \frac{1}{N} 
    \sum_{\vk\in\tMBZ} 
    h^\sigma_{nm}\left( \vk \right) 
    e^{+i\vk\cdot\left( \Delta\vR+\vrr_n-\vrr_m \right)}.
\end{align}

To obtain the interaction Hamiltonian in the Wannier basis, we rewrite the original plane-wave-basis electron operators within one of the Wannier orbital bases introduced in this work,
\begin{align}
    \cre{c}{\vk,\sigma,\vQ}
    \approx
    \sum_n
    \mathbf{L}_{n,\vQ}^{\sigma,*}\left( \vk \right)
    \cre{d}{\vk,\sigma,n}.
\end{align}
The density operator then becomes
\begin{align}
    \rho\left( \vq,\vG \right)
    & =
    \sum_{\vk,\vQ,\sigma}
    \cre{c}{\vk,\sigma,\vQ}
    \des{c}{\vk+\vq,\sigma,\vQ-\vG}
    \nonumber \\
    & \approx
    \sum_{\substack{\vk,\vQ,\sigma \\ n,m}}
    \mathbf{L}_{n,\vQ}^{\sigma,*}\left( \vk \right)
    \mathbf{L}_{m,\vQ-\vG}^{\sigma}\left( \vk+\vq \right)
    \cre{d}{\vk,\sigma,n}
    \des{d}{\vk+\vq,\sigma,m}.
\end{align}
We consider the Coulomb repulsion 
\begin{align}
    \label{app:eqn:continuum_coulomb_interaction}
    H_{\tInt} = 
    \frac{1}{2\Omega_{\tTot}} 
    \sum_{\vq\in\tMBZ} 
    \sum_{\vG\in\mathcal{Q}} 
    V\left( \vq+\vG \right) 
    \normord{\rho\left( \vq,\vG \right)} 
    \normord{\rho\left( -\vq,-\vG \right)},
\end{align}
which has been normal ordered with respect to the Slater determinant state with all the valence bands fully filled. In \cref{app:eqn:continuum_coulomb_interaction}, $V\left( \vq+\vG \right)$ is the Fourier transform of the Coulomb interaction potential arising in a dual-gated setup with the sample located between two metallic gates separated by a distance $d$,
\begin{align}
	V \left( \vrr \right) 
    & = 
    \frac{e^2}{4 \pi \epsilon \epsilon_0 d } 
    \sum_{n=-\infty}^{\infty} 
    \frac{\left( -1 \right)^n}{\sqrt{\left( \frac{r}{d} \right)^2 + n^2}}
    = 
    \int \frac{\dd[2]{q}}{(2 \pi)^2} 
    V \left( \vq \right) 
    e^{- i \vq \cdot \vrr}, 
    \nonumber \\
	V \left( \vq \right)  
    & = 
    \frac{e^2}{4 \epsilon \epsilon_0}
    \frac{\tanh \left( qd/2 \right)}{q/ 2}.
	\label{app:eqn:Coulomb_interaction}
\end{align}
Rewritten in the Wannier basis, the interaction Hamiltonian reads
\begin{align}
    H_{\tInt} 
    = 
    \frac{1}{2N} 
    \sum_{\substack{\vk,\vk',\vq,\sigma,\sigma' \\ n,m,n',m'}} 
    V_{\sigma\sigma'}^{nm;n'm'}\left( \vq;\vk,\vk'\right)
    \normord{\cre{d}{\vk,\sigma,n}
    \des{d}{\vk+\vq,\sigma,m}}
    \normord{\cre{d}{\vk'+\vq,\sigma',m'}
    \des{d}{\vk',\sigma',n'}},
\end{align}
where the momentum-space interaction tensor expressed in the Wannier basis is
\begin{align} \label{eq:app:interaction_orbital_basis}
    V_{\sigma\sigma'}^{nm;n'm'}\left( \vq;\vk,\vk' \right) 
    = \frac{N}{\Omega_{\tTot}} 
    \sum_{\substack{\vG,\vQ,\vQ'\in\mathcal{Q}}} 
    V\left( \vq+\vG \right) 
    \mathbf{L}_{n,\vQ}^{\sigma,*}\left( \vk \right) 
    \mathbf{L}_{m,\vQ-\vG}^{\sigma}\left( \vk+\vq \right) 
    \mathbf{L}_{m',\vQ'}^{\sigma',*}\left( \vk'+\vq \right) 
    \mathbf{L}_{n',\vQ'+\vG}^{\sigma'}\left( \vk' \right).
\end{align}
Using the Fourier transform of the Wannier operators introduced in \cref{app:eqn:ft_conv_general_k_to_r,app:eqn:ft_conv_general_r_to_k}, the interaction Hamiltonian can equivalently be written in real space as
\begin{align}
    H_{\tInt} = 
    \frac{1}{2} 
    \sum_{\substack{\vR,\Delta\vR,\Delta\vrr,\Delta\vrr' 
    \\ 
    \sigma,\sigma',n,m,n',m'}} 
    V_{\sigma\sigma'}^{nm;n'm'}
    \left( 
        \Delta\vR,\Delta\vrr,\Delta\vrr' \right) 
        \normord{
            \cre{d}{\vR+\Delta\vrr,\sigma,n}
            \des{d}{\vR,\sigma,m}
            }
        \normord{
            \cre{d}{\vR+\Delta\vR,\sigma',m'}
            \des{d}{\vR+\Delta\vR+\Delta\vrr',\sigma',n'}
            },
\end{align}
where
\begin{align}
    V_{\sigma\sigma'}^{nm;n'm'}\left( \Delta\vR,\Delta\vrr,\Delta\vrr' \right) 
    = 
    \frac{1}{N^3}
    \sum_{\vk,\vk',\vq} 
    V_{\sigma\sigma'}^{nm;n'm'}
    \left( \vq;\vk,\vk' \right) 
    e^{
        +i\vk\cdot\left( \Delta\vrr+\vrr_n-\vrr_m \right)
        +i\vk'\cdot\left( -\Delta\vrr'+\vrr_{m'}-\vrr_{n'} \right)
        +i\vq\cdot\left( \Delta\vR-\vrr_m+\vrr_{m'} \right)
        }.
\end{align}

In our Hartree-Fock calculations, we retain all hopping terms in the single-particle Hamiltonian but truncate the interaction Hamiltonian in real space.
This truncation is justified by two complementary effects.
First, the exponential localization of the Wannier orbitals causes the interaction tensor $V_{\sigma\sigma'}^{nm;n'm'}\left( \Delta\vR,\Delta\vrr,\Delta\vrr' \right)$ to decay rapidly with increasing $\abs{\Delta\vrr}$ or $\abs{\Delta\vrr'}$.
Second, screening by the metallic gates suppresses the long-range Coulomb interaction, causing the interaction tensor to decay rapidly with increasing $\abs{\Delta\vR}$.

\subsection{Real-space Hartree-Fock calculations with supercells}\label{app:sec:HF:real_space_suopercell}

\newcommand{\vRsc}{\boldsymbol{\mathcal{R}}}
\newcommand{\vrsc}{\boldsymbol{\bm{r}}}
\newcommand{\vTau}{\boldsymbol{\tau}}
\newcommand{\tMF}{\text{MF}}
\newcommand{\tHF}{\text{HF}}
\newcommand{\tSBZ}{\text{SBZ}}
\newcommand{\vEll}{\boldsymbol{\ell}}
\newcommand{\tSC}{\text{sc}}

To allow for the possibility of translational symmetry breaking, we perform Hartree-Fock calculations using supercells.
We consider a system with periodic boundary conditions containing $N_i$ supercells of size $n_i$ along the moir\'e lattice vectors $\vec{a}_{M_i}$, with $i=1,2$.
We define
\begin{align}
    N_{\tSC}=N_1N_2,\qquad n_{\tSC}=n_1n_2,\qquad N=N_{\tSC}n_{\tSC},
\end{align}
such that $N_{\tSC}$ is the number of supercells, $n_{\tSC}$ is the number of primitive moir\'e unit cells within each supercell, and $N$ is the total number of primitive moir\'e unit cells.

We label the supercells by
\begin{align}
    \vRsc = m_1n_1\vec{a}_{M_1}+m_2n_2\vec{a}_{M_2},\qquad m_i=0,1,\ldots,N_i-1,
\end{align}
and the primitive-cell offset within each supercell by
\begin{align}
    \vTau = \ell_1\vec{a}_{M_1}+\ell_2\vec{a}_{M_2},\qquad \ell_i=0,1,\ldots,n_i-1.
\end{align}
We denote the corresponding Wannier operators by $\cre{d}{\vRsc,\vTau,\sigma,n}$, with
\begin{align}
    \cre{d}{\vRsc,\vTau,\sigma,n} \equiv \cre{d}{\vRsc+\vTau,\sigma,n}.
\end{align}
Thus, the physical center of this Wannier orbital is $\vRsc+\vTau+\vrr_n$.

Within the supercell Hartree-Fock calculation, we assume that translation symmetry under the supercell Bravais lattice is preserved.
The corresponding Slater determinants are therefore parametrized by the real-space supercell density matrix
\begin{align}
    \rho_{\vTau \sigma n;\vTau'\sigma'n'}\left( \Delta\vRsc \right) = \expec{\normord{\cre{d}{\vRsc+\Delta\vRsc,\vTau,\sigma,n} \des{d}{\vRsc,\vTau',\sigma',n'}}},\qquad \qq{for any }\vRsc.
\end{align}

\subsubsection{Momentum-space density matrix and Brillouin-zone folding}\label{app:sec:HF:real_space_suopercell:momentum_space_rho}

We now formulate the same density matrix in momentum space.
Let $\vec{b}_{M_i}$ denote the reciprocal lattice vectors of the primitive moir\'e lattice, $\vec{a}_{M_i}\cdot\vec{b}_{M_j}=2\pi\delta_{ij}$.
The reciprocal lattice vectors of the supercell are
\begin{align}
    \vec{b}^{\,\tSC}_{M_i} = \frac{1}{n_i}\vec{b}_{M_i}.
\end{align}
Correspondingly, the primitive moir\'e Brillouin zone folds into a reduced supercell Brillouin zone, which we denote by $\tSBZ$.
We introduce the $n_{\tSC}$ folding vectors
\begin{align}
    \vK_{\vEll } = \frac{\ell_1}{n_1}\vec{b}_{M_1}+\frac{\ell_2}{n_2}\vec{b}_{M_2},\qquad \ell_i=0,1,\ldots,n_i-1.
\end{align}
Every momentum $\vk\in\tMBZ$ can then be written uniquely, modulo a primitive reciprocal lattice vector, as
\begin{align}
    \vk = \vk_{\tSC}+\vK_{\vEll },\qquad \vk_{\tSC}\in\tSBZ.
\end{align}

Consistently with the Fourier-transform convention introduced above, we define
\begin{align}
    \cre{d}{\vRsc,\vTau,\sigma,n} = \frac{1}{\sqrt{N_{\tSC}}}\sum_{\vk_{\tSC}\in\tSBZ} \cre{d}{\vk_{\tSC},\vTau,\sigma,n} e^{-i\vk_{\tSC}\cdot\left( \vRsc+\vTau+\vrr_n \right)}.
\end{align}
The momentum-space supercell density matrix is therefore
\begin{align}
    \rho_{\vTau \sigma n;\vTau'\sigma'n'}\left( \vk_{\tSC} \right) & = \expec{\normord{\cre{d}{\vk_{\tSC},\vTau,\sigma,n} \des{d}{\vk_{\tSC},\vTau',\sigma',n'}}} = \sum_{\Delta\vRsc}\rho_{\vTau \sigma n;\vTau'\sigma'n'}\left( \Delta\vRsc \right)e^{+i\vk_{\tSC}\cdot\left( \Delta\vRsc+\vTau-\vTau'+\vrr_n-\vrr_{n'} \right)}.
\end{align}

The relation between the supercell momentum operators and the original primitive-cell momentum operators is
\begin{align}
    \cre{d}{\vk_{\tSC},\vTau,\sigma,n} = \frac{1}{\sqrt{n_{\tSC}}}\sum_{\vEll }e^{-i\vK_{\vEll }\cdot\left( \vTau+\vrr_n \right)}\cre{d}{\vk_{\tSC}+\vK_{\vEll },\sigma,n},
\end{align}
or, equivalently,
\begin{align}
    \cre{d}{\vk_{\tSC}+\vK_{\vEll },\sigma,n} = \frac{1}{\sqrt{n_{\tSC}}}\sum_{\vTau}e^{+i\vK_{\vEll }\cdot\left( \vTau+\vrr_n \right)}\cre{d}{\vk_{\tSC},\vTau,\sigma,n}.
\end{align}
It is consequently useful to introduce the density matrix directly in the folded primitive-momentum basis,
\begin{align}
    \rho^{\mathrm{fold}}_{\vEll \sigma n;\vEll '\sigma'n'}\left( \vk_{\tSC} \right) \equiv \expec{\normord{\cre{d}{\vk_{\tSC}+\vK_{\vEll },\sigma,n}\des{d}{\vk_{\tSC}+\vK_{\vEll '},\sigma',n'}}}.
\end{align}
In terms of the supercell density matrix,
\begin{align}
    \rho^{\mathrm{fold}}_{\vEll \sigma n;\vEll '\sigma'n'}\left( \vk_{\tSC} \right) = \frac{1}{n_{\tSC}}\sum_{\vTau,\vTau'} e^{+i\vK_{\vEll }\cdot\left( \vTau+\vrr_n \right)}\rho_{\vTau \sigma n;\vTau'\sigma'n'}\left( \vk_{\tSC} \right)e^{-i\vK_{\vEll '}\cdot\left( \vTau'+\vrr_{n'} \right)}.
\end{align}
Equivalently, directly in terms of the real-space density matrix,
\begin{align}
    \rho^{\mathrm{fold}}_{\vEll \sigma n;\vEll '\sigma'n'}\left( \vk_{\tSC} \right) = \frac{1}{n_{\tSC}}
    \sum_{\substack{\Delta\vRsc,\vTau,\vTau'}} 
    \rho_{\vTau \sigma n;\vTau'\sigma'n'}
    \left( \Delta\vRsc \right)
    e^{+i\vk_{\tSC}\cdot\Delta\vRsc}
    e^{
        +i\left( \vk_{\tSC}+\vK_{\vEll } \right)
        \cdot
        \left( \vTau+\vrr_n \right)
    }
    e^{-i\left( \vk_{\tSC}+\vK_{\vEll '} \right)\cdot\left( \vTau'+\vrr_{n'} \right)}.
\end{align}
Thus, supercell translation symmetry implies conservation of the reduced momentum $\vk_{\tSC}$, but it does not imply conservation of the original momentum in the primitive moir\'e Brillouin zone.
In particular, the density matrix can couple different folds,
\begin{align}
    \vk_{\tSC}+\vK_{\vEll }\longleftrightarrow\vk_{\tSC}+\vK_{\vEll '},
\end{align}
with $\vEll \neq\vEll '$.
These off-diagonal matrix elements describe translation-symmetry breaking within the supercell.
If the primitive moir\'e translation symmetry is unbroken, the density matrix is diagonal in the folding indices $\vEll$ and $\vEll '$.

\subsubsection{Hamiltonian in the supercell basis}\label{app:sec:HF:real_space_suopercell:hamiltonian}

For later convenience, we rewrite both the kinetic and interaction Hamiltonians directly in the supercell basis.
The kinetic Hamiltonian is
\begin{align}
    H_0 = 
    \sum_{\substack{\vRsc,\Delta\vRsc \\ \vTau,n,\sigma,\vTau',n',\sigma'}} 
    h_{\vTau \sigma n;\vTau'\sigma'n'}\left( \Delta\vRsc \right)
    \cre{d}{\vRsc+\Delta\vRsc,\vTau,\sigma,n}
    \des{d}{\vRsc,\vTau',\sigma',n'},
\end{align}
where
\begin{align}
    h_{\vTau \sigma n;\vTau'\sigma'n'}\left( \Delta\vRsc \right) 
    = 
    h^\sigma_{nn'}\left( \Delta\vRsc+\vTau-\vTau' \right)\delta_{\sigma\sigma'}.
\end{align}

Similarly, the interaction Hamiltonian can be written as
\begin{align}
    H_{\tInt} = 
    \frac{1}{2}
    \sum_{\substack{\vRsc,\Delta\vRsc,\Delta\vrsc,\Delta\vrsc' \\ \vTau_1, n_1, \sigma_1,\vTau_2, n_2, \sigma_2 \\ \vTau_3, n_3, \sigma_3,\vTau_4, n_4, \sigma_4}} & V_{\vTau_1\sigma_1n_1,\vTau_2\sigma_2n_2;\vTau_3\sigma_3n_3,\vTau_4\sigma_4n_4}\left( \Delta\vRsc,\Delta\vrsc,\Delta\vrsc' \right) \nonumber \\
     & \times\normord{\cre{d}{\vRsc+\Delta\vrsc,\vTau_1,\sigma_1,n_1}\des{d}{\vRsc,\vTau_2,\sigma_2,n_2}}\normord{\cre{d}{\vRsc+\Delta\vRsc,\vTau_3,\sigma_3,n_3}\des{d}{\vRsc+\Delta\vRsc+\Delta\vrsc',\vTau_4,\sigma_4,n_4}}.
\end{align}
The supercell matrix elements are related to the primitive-cell interaction tensor according to
\begin{align}
    & V_{\vTau_1\sigma_1n_1,\vTau_2\sigma_2n_2;\vTau_3\sigma_3n_3,\vTau_4\sigma_4n_4}\left( \Delta\vRsc,\Delta\vrsc,\Delta\vrsc' \right)
    =
    \delta_{\sigma_1\sigma_2}\delta_{\sigma_3\sigma_4}
    V_{\sigma_1\sigma_3}^{n_1n_2;n_4n_3}\left( \Delta\vRsc+\vTau_3-\vTau_2,\Delta\vrsc+\vTau_1-\vTau_2,\Delta\vrsc'+\vTau_4-\vTau_3 \right).
\end{align}
The reversal $n_4n_3$ in the second pair follows from our convention for the primitive-cell interaction tensor, $V_{nm;n'm'}$, for which the corresponding second bilinear is $\cre{d}{m'}\des{d}{n'}$.
The kinetic Hamiltonian can additionally contain the effects of a perpendicularly applied displacement field, which we keep implicit in the expressions above.

\subsubsection{Hartree-Fock decoupling}\label{app:sec:HF:real_space_suopercell:HF_decoupling}

We now introduce the composite index
\begin{align}
    i \equiv \left( \vTau_i,\sigma_i,n_i \right),
\end{align}
and denote the position of orbital $i$ within the supercell by
\begin{align}
    \vrr_i^{\tSC}\equiv\vTau_i+\vrr_{n_i}.
\end{align}
In terms of these composite indices, the Hamiltonian takes the generic form
\begin{align}
    H_0       & = 
    \sum_{\vRsc,\Delta\vRsc} 
    h_{ij}\left( \Delta\vRsc \right)
    \cre{d}{\vRsc+\Delta\vRsc,i}
    \des{d}{\vRsc,j}, \\
    H_{\tInt} & = 
    \frac{1}{2}
    \sum_{\substack{\vRsc,\Delta\vRsc \\ \Delta\vrsc_1,\Delta\vrsc_2}} 
    V_{ijkl}\left( \Delta\vRsc,\Delta\vrsc_1,\Delta\vrsc_2 \right)
    \normord{
        \cre{d}{\vRsc+\Delta\vrsc_1,i}
        \des{d}{\vRsc,j}
    }
    \normord{
        \cre{d}{\vRsc+\Delta\vRsc,k}
        \des{d}{\vRsc+\Delta\vRsc+\Delta\vrsc_2,l}
    },
\end{align}
where repeated composite indices are summed implicitly.

The real-space density matrix is
\begin{align}
    \rho_{ij}\left( \Delta\vRsc \right)=\expec{\normord{\cre{d}{\vRsc+\Delta\vRsc,i}\des{d}{\vRsc,j}}}.
\end{align}
Up to the constant Hartree-Fock double-counting contribution, the mean-field decoupling of the interaction is
\begin{align}
    H_{\tInt}^{\tMF} = & 
    \frac{1}{2}
    \sum_{\substack{\vRsc,\Delta\vRsc \\ \Delta\vrsc_1,\Delta\vrsc_2}}
    V_{ijkl}\left( \Delta\vRsc,\Delta\vrsc_1,\Delta\vrsc_2 \right) 
    \left( \rho_{ij}\left( \Delta\vrsc_1 \right)
    \cre{d}{\vRsc+\Delta\vRsc,k}
    \des{d}{\vRsc+\Delta\vRsc+\Delta\vrsc_2,l} 
    +
    \rho_{kl}\left( -\Delta\vrsc_2 \right)
    \cre{d}{\vRsc+\Delta\vrsc_1,i}
    \des{d}{\vRsc,j} \right.
    \nonumber \\
    & \left. 
    -
    \rho_{il}\left( \Delta\vrsc_1-\Delta\vRsc-\Delta\vrsc_2 \right)
    \cre{d}{\vRsc+\Delta\vRsc,k}
    \des{d}{\vRsc,j} 
    -\rho_{kj}\left( \Delta\vRsc \right)
    \cre{d}{\vRsc+\Delta\vrsc_1,i}
    \des{d}{\vRsc+\Delta\vRsc+\Delta\vrsc_2,l}\right).
\end{align}
After relabeling the summation variables, this can be brought to the form
\begin{align}
    H_{\tInt}^{\tMF} = 
    \sum_{\vRsc,\Delta\vrsc_2} 
    h^{\tMF}_{kl}\left( \Delta\vrsc_2 \right)
    \cre{d}{\vRsc+\Delta\vrsc_2,k}
    \des{d}{\vRsc,l},
\end{align}
where
\begin{align}
    h^{\tMF}_{kl}\left( \Delta\vrsc_2 \right) = 
    \sum_{\Delta\vrsc_1}
    \mathcal{T}_{klij}\left( \Delta\vrsc_2,\Delta\vrsc_1 \right)\rho_{ij}\left( \Delta\vrsc_1 \right).
\end{align}
The Hartree-Fock tensor is determined entirely by the interaction matrix elements and is given by
\begin{align}
    \mathcal{T}_{klij}\left( \Delta\vrsc_2,\Delta\vrsc_1 \right) = \frac{1}{2}\sum_{\Delta\vRsc}
    \bigg( & V_{ijkl}\left( \Delta\vRsc,\Delta\vrsc_1,-\Delta\vrsc_2 \right)+V_{klij}\left( \Delta\vRsc,\Delta\vrsc_2,-\Delta\vrsc_1 \right) \nonumber \\
           & -V_{ilkj}\left( \Delta\vrsc_2,\Delta\vRsc,\Delta\vRsc-\Delta\vrsc_1-\Delta\vrsc_2 \right)-V_{kjil}\left( \Delta\vrsc_1,\Delta\vRsc,\Delta\vRsc-\Delta\vrsc_1-\Delta\vrsc_2 \right)\bigg).
\end{align}

Finally, the Hartree-Fock Hamiltonian is block diagonal in the reduced supercell momentum,
\begin{align}
    H_{\tHF} = \sum_{\vk_{\tSC}\in\tSBZ}h^{\tHF}_{ij}\left( \vk_{\tSC} \right)\cre{d}{\vk_{\tSC},i}\des{d}{\vk_{\tSC},j}+E_{\mathrm{dc}},
\end{align}
where
\begin{align}
    h^{\tHF}_{ij}\left( \vk_{\tSC} \right) = \sum_{\Delta\vRsc}\left[h_{ij}\left( \Delta\vRsc \right)+h^{\tMF}_{ij}\left( \Delta\vRsc \right)\right]e^{-i\vk_{\tSC}\cdot\left( \Delta\vRsc+\vrr_i^{\tSC}-\vrr_j^{\tSC} \right)}.
\end{align}
The corresponding momentum-space density matrix satisfies
\begin{align}
    \rho_{ij}\left( \vk_{\tSC} \right) = \sum_{\Delta\vRsc}\rho_{ij}\left( \Delta\vRsc \right)e^{+i\vk_{\tSC}\cdot\left( \Delta\vRsc+\vrr_i^{\tSC}-\vrr_j^{\tSC} \right)}.
\end{align}

\subsubsection{Order parameters}\label{app:sec:HF:real_space_suopercell:ops}

\begin{figure}[!t]
    \centering
    \includegraphics[width=\textwidth]{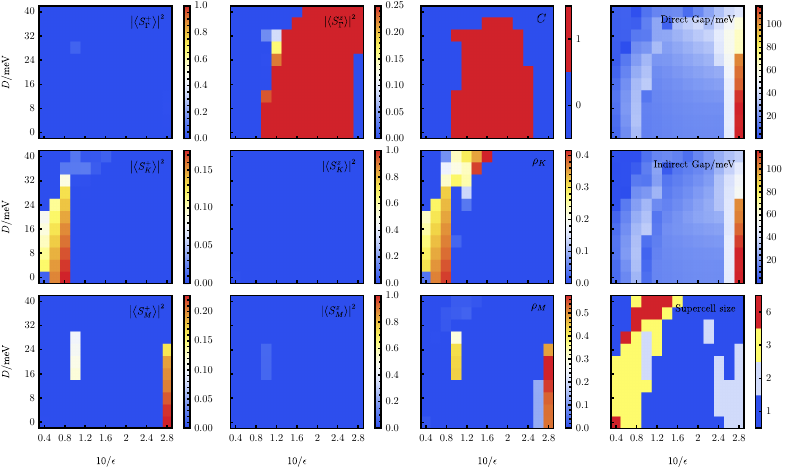}
    \caption{Hartree-Fock phase diagram of the 3Tri model obtained from the continuum model of Ref.~\cite{devakulMagicTwistedTransition2021}. 
    The phase diagram is shown as a function of the displacement field $D$ and the interaction strength, parametrized by the dielectric constant $10/\epsilon$.
    Each panel shows one of the order parameters defined in \cref{app:sec:HF:real_space_suopercell:ops}. 
    We additionally show the direct and indirect gaps, together with the linear size of the supercell for which the lowest-energy Hartree-Fock solution was obtained.
    }
    \label{app:fig:HF_LF}
\end{figure}

\begin{figure}[!t]
    \centering
    \includegraphics[width=\textwidth]{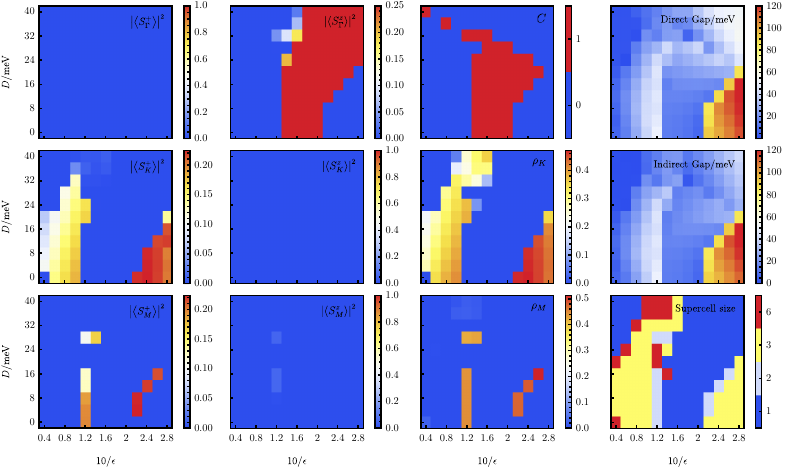}
    \caption{Hartree-Fock phase diagram of the 3Tri model obtained from our DFT calculation. The layout is the same as in \cref{app:fig:HF_LF}.
    }
    \label{app:fig:HF_DFT}
\end{figure}

For a given Hartree-Fock ground state, specified by the supercell density matrix $\rho_{\vTau \sigma n;\vTau'\sigma'n'}\left( \Delta\vRsc \right)$ and the corresponding Hartree-Fock Hamiltonian, we compute several observables to characterize its symmetry-breaking pattern and topology. One such observable is the Chern number $C$ of the occupied bands, which we determine from the Wilson loop of the occupied Hartree-Fock bands in the supercell Brillouin zone.

We also characterize translation-symmetry breaking through the momentum-off-diagonal components of the density matrix associated with specific momentum transfers. In particular, we focus on momentum transfers corresponding to the $K_M$ and $M_M$ points of the primitive moir\'e Brillouin zone and define
\begin{align}
    \rho_{K} & \equiv \frac{1}{N_{\tSC}} \sum_{\vk_{\tSC}} \sum_{\substack{n,\sigma,\vEll \\ n',\sigma', \vEll'}} \abs{\rho^{\mathrm{fold}}_{\vEll \sigma n;\vEll '\sigma'n'}\left( \vk_{\tSC} \right)}^2 \left( \delta_{\vK_{\vEll}, \vK_{\vEll'} + \vK_{K_M}} + \delta_{\vK_{\vEll}, \vK_{\vEll'} - \vK_{K_M}} \right), \\
    \rho_{M} & \equiv \frac{1}{N_{\tSC}} \sum_{i=0}^2 \sum_{\vk_{\tSC}} \sum_{\substack{n,\sigma,\vEll \\ n',\sigma', \vEll'}} \abs{\rho^{\mathrm{fold}}_{\vEll \sigma n;\vEll '\sigma'n'}\left( \vk_{\tSC} \right)}^2 \left( \delta_{\vK_{\vEll}, \vK_{\vEll'} + C^i_{3z} \vK_{M_M}} + \delta_{\vK_{\vEll}, \vK_{\vEll'} - C^i_{3z}\vK_{M_M}} \right),
\end{align}
where $\vK_{K_M} = \frac{1}{3}\left( \vec{b}_{M_1} + \vec{b}_{M_2} \right)$ and $\vK_{M_M} = \frac{1}{2} \vec{b}_{M_1}$ denote the momenta of one of the two $K_M$ points and one of the three $M_M$ points, respectively. The Kronecker delta functions select density-matrix elements whose momentum difference is equal to the specified momentum transfer.

To gain further insight into the nature of the Hartree-Fock ground states, we additionally compute the expectation values of the spin operators at selected high-symmetry momenta of the primitive moir\'e Brillouin zone. Letting $s^z$ and $s^{+}$ denote the $2\times2$ spin matrices corresponding to the $z$ component and raising operator of the spin, respectively, we define
\begin{equation}
    S^{z,+}_{\vq} \equiv \frac{1}{N_{\tSC}} \sum_{\vk_{\tSC}} \sum_{\substack{n,\sigma,\vEll \\ \sigma', \vEll'}} \rho^{\mathrm{fold}}_{\vEll \sigma n;\vEll ' \sigma' n}\left( \vk_{\tSC} \right) s^{z,+}_{\sigma\sigma'} \delta_{\vK_{\vEll}, \vK_{\vEll'} + \vq}.
\end{equation}
We focus on the $\Gamma_M$, $K_M$, and $M_M$ momenta and define the corresponding spin-order parameters as
\begin{align}
    \abs{\expec{S^{z,+}_{\Gamma}}}^2 \equiv S^{z,+}_{\vec{0}}, \quad
    \abs{\expec{S^{z,+}_{K}}}^2 \equiv \sum_{\vq} \abs{S^{z,+}_{\vq}}^2 \left( \delta_{\vq, \vK_{K_M}} + \delta_{\vq, -\vK_{K_M}} \right), \nonumber \\
    \abs{\expec{S^{z,+}_{M}}}^2 \equiv \sum_{\vq} \sum_{i=0}^2 \abs{S^{z,+}_{\vq}}^2 \left( \delta_{\vq, C^i_{3z} \vK_{M_M}} + \delta_{\vq, - C^i_{3z}\vK_{M_M}} \right).
\end{align}

\subsection{Numerical results}\label{app:sec:HF:numerical_results}

We now present the Hartree-Fock results for twisted \ch{WSe2} at filling $\nu = -1$. We consider systems of size $36 \times 36$ and allow for $1 \times 1$, $2 \times 2$, $3 \times 3$, and $6 \times 6$ supercells. We vary both the displacement field $D$ and the interaction strength, the latter by changing the dielectric constant $\epsilon$ of the surrounding dielectric environment. For each set of parameters and each supercell size, we initialize the Hartree-Fock calculation from five different random density matrices and identify the converged solution with the lowest energy as the Hartree-Fock ground state. To accelerate convergence, we employ several standard algorithms, including RCA~\cite{CAN00}, DIIS~\cite{PUL80,PUL82}, and EDIIS~\cite{KUD02,GAR12}, whose implementations are described in the corresponding references.

We perform Hartree-Fock calculations for the 3Tri models obtained both from the continuum model of Ref.~\cite{devakulMagicTwistedTransition2021} and from our DFT calculation, which we refer to below as the DFT model. The resulting phase diagrams are shown in \cref{app:fig:HF_LF,app:fig:HF_DFT}. The two models yield qualitatively similar phase diagrams, although the precise locations of the phase boundaries and other quantitative details differ.

At weak interactions and large displacement fields, both models stabilize a metallic state consistent with strong layer polarization. Upon reducing the displacement field, we find a state that breaks moir\'e translation symmetry at the $K_M$ momentum, as indicated by a large value of the $\rho_K$ order parameter. The simultaneously large value of $\abs{\expec{S^{+}_{K}}}^2$ reveals inter-valley coherence in which the two valleys are coupled by a momentum transfer at $K_M$. This state is consistent with the $120^{\circ}$ antiferromagnetic order identified in previous works~\cite{devakulMagicTwistedTransition2021,ZAN21,WAN24d,FOU24,BRI26} and has zero Chern number. 

Upon further increasing the interaction strength, the ground state becomes valley polarized while preserving moir\'e translation symmetry. Over most of the corresponding region of parameter space, this valley-polarized state has Chern number $C = 1$, although a small region with $C = 0$ appears at large interaction strengths and large displacement fields. At still stronger interactions, the system undergoes another transition into a zero-Chern-number state that breaks moir\'e translation symmetry at the $K_M$ momentum. Finally, small regions in which translation symmetry is instead broken at the $M_M$ momentum appear between the $120^{\circ}$ antiferromagnetic and valley-polarized states at relatively weak interactions.

\section{Functional Renormalization Group} \label{sec:app:FRG}

The functional renormalization group (FRG) is a powerful theoretical tool for studying competing orders in interacting many-body systems, particularly in the context of correlated electron materials \cite{salmhoferFermionicRenormalizationGroup2001,metznerFunctionalRenormalizationGroup2012,kopietzIntroductionFunctionalRenormalization2010,platt_functional_2013,dupuis_nonperturbative_2021,honerkampRecentDevelopmentsFunctional2022}.
FRG provides a systematic way to analyze the dependence (or flow) of functionals, such as the effective interactions, on an energy scale, allowing us to identify instabilities towards various ordered phases, such as superconductivity, magnetism, or charge order. 

It is based on an exact differential equation for the effective action \cite{metznerFunctionalRenormalizationGroup2012}
\begin{align}\label{eq:app:wetterich_equation}
    \frac{d\Gamma^{\Lambda}[\psi, \bar \psi]}{d\Lambda} = 
    -\big(\bar \psi, \dot{\mathcal{Q}}^{\Lambda,0} \psi\big)
    - 
    \frac{1}{2} 
    \text{Tr} 
    \left[ 
        \dot{\mathbf{Q}}^{\Lambda}_{0} \left(\mathbf{\Gamma}^{(2),\Lambda}[\psi, \bar \psi]\right)^{-1}
    \right] ,
\end{align}
where $\Gamma^\Lambda$ is the effective action at scale $\Lambda$, 
$\psi$ and $\bar{\psi}$ are fermionic fields, $\mathcal{G}^{\Lambda}_{0}$ is the regularized bare propagator  with inverse $\mathcal{Q}^{\Lambda}_{0} = (\mathcal{G}^{\Lambda}_{0})^{-1}$, $\dot{\mathbf{Q}}^{\Lambda}_{0} = \mathrm{diag}\big(\dot{\mathcal{Q}}^{\Lambda}_{0}, -(\dot{\mathcal{Q}}^{\Lambda}_{0})^{t}\big)$, and
$\mathbf{\Gamma}^{(2),\Lambda}$ denotes the matrix of second functional derivatives of the effective action with respect to the fermionic fields.
The dot indicates the derivative with respect to the flow parameter $\Lambda$. 
The scale dependence is introduced via a regularization, implemented here on the level of the bare propagator $\mathcal{G}_{0}\rightarrow \mathcal{G}^{\Lambda}_{0}$. The regularization is necessary to tame UV or IR divergences in a calculation of $\Gamma$. 
It can be given by, e.g., the temperature \cite{honerkampTemperatureflowRenormalizationGroup2001}, an infrared cutoff around the Fermi surface (as in Wilsonian RG) \cite{kopietzExactRenormalizationGroup2001,corellCorrelationFunctionsThreedimensional2018}, or a frequency cutoff for low energies \cite{hilleQuantitativeFunctionalRenormalization2020}. 
The full effective action, which includes all (quantum) fluctuations, can, in principle, be obtained as the solution of the differential equation Eq.~\eqref{eq:app:wetterich_equation} at the end of the flow.

In the FRG framework, we start with a microscopic model of the system at the ultraviolet cutoff scale $\Lambda_{\rm{UV}}$. 
Starting from there, we apply a vertex expansion and a so-called level-two truncation (see below) to obtain a flow equation for the static one-particle irreducible (1PI) two-particle vertex, which can be interpreted as the (anti-symmetrized) effective interaction \cite{shankar_renormalization-group_1994, dupuis_nonperturbative_2021, rosten_fundamentals_2012, berges_non-perturbative_2002, platt_functional_2013, metzner_fermi_1998, ma_introduction_1973, polonyi_lectures_2003, honerkamp_iron_2010, honerkamp_recent_2022, metznerFunctionalRenormalizationGroup2012}. 
The FRG flow equations describe how the effective interaction evolves as we integrate out high-energy degrees of freedom, effectively lowering the energy scale towards the infrared cutoff scale~$\Lambda_{\rm{IR}}$.
By solving the FRG flow equations, we can track the growth of effective interactions and identify the leading instability, which corresponds to the dominant ordering tendency in the system. Restricted to a single channel, this reduces to an RPA-like  resummation. 
However, the FRG method goes beyond such RPA-like treatments because it includes all interaction channels (particle-particle, direct particle-hole and crossed particle-hole) and their inter-dependencies. Therefore, it 
can capture competing orders on equal footing, making it particularly suitable for studying systems with multiple intertwined phases and superconductivity from an electronic pairing mechanism.

For this work, we employ the temperature-flow FRG scheme \cite{honerkampTemperatureflowRenormalizationGroup2001}, where the temperature $T$ serves as the flow parameter, allowing us to directly access an estimate of the critical temperature of the leading instability.
This is done by regularizing the bare propagator as
\begin{equation}
    \mathcal{G}^{T}_{0,n}(\mathbf{k}, i\omega) 
    = 
    \frac{T^{1/2}}{i\omega - E_n(\mathbf{k})}\quad . \label{eq:appregulator}
\end{equation}
Here, $E_n(\vk)$ denotes the $n$th energy band. 
The factor $T^{1/2}$ results from a rescaling of the fermionic fields that renders the interaction term of the action temperature independent. The entire temperature dependence is thus carried by the propagator, such that $T$ can be used as the flow parameter, and the flow starts from the bare interaction at high temperatures~\cite{honerkampTemperatureflowRenormalizationGroup2001,salmhoferFermionicRenormalizationGroup2001}.
Throughout this work, we operationally identify the critical temperature~$T_c$ with the temperature at which the FRG flow diverges, which provides an estimate of the corresponding ordering scale rather than the physical transition temperature. 
Since order-parameter fluctuations are not included, this scale should be interpreted, for example in the superconducting channel, as an estimate of the pair-formation temperature, while phase coherence and the associated reduction of the transition temperature are beyond the present treatment.

The effective action is now expanded in terms of one-particle irreducible $n$-particle vertices $\Gamma^{(2n),\Lambda}$ as~\cite{metznerFunctionalRenormalizationGroup2012}
\begin{align}
    \Gamma^\Lambda[\psi, \bar{\psi}] 
    &= 
    \sum_{n=1}^\infty 
    \frac{(-1)^n}{(n!)^2} 
    \int_{k'_1} \dots \int_{k'_n} 
    \int_{k_1} \dots \int_{k_n} \nonumber \\
    &\quad \times 
    \Gamma^{(2n),\Lambda}(k'_1, \dots, k'_n; k_1, \dots, k_n)
    \bar{\psi}(k'_1) 
    \dots 
    \bar{\psi}(k'_n) 
    \psi(k_n) 
    \dots 
    \psi(k_1) \, ,
\end{align}
up to a field-independent constant, where the primed arguments $k'_1, \dots, k'_n$ refer to the outgoing and the unprimed arguments $k_1, \dots, k_n$ to the incoming fermions. 
We use the multi-index $k=(i\omega, \mathbf{k}, \sigma,\dots)$ to denote Matsubara frequency, momentum, spin, and other fermion flavor indices of the fermionic fields.
The integration $\int_k=(T/V_{\rm{BZ}})\sum_{i\omega}\int_\mathbf{k} \sum_{\sigma,\ldots}$ then implies the sum over those indices as well. Here, $V_{\rm{BZ}} = (2\pi)^2/\Omega_0$ is the area of the Brillouin zone and $\Omega_0$ is the unit cell area. 
Inserting the expansion into the flow equation and comparing coefficients of the same order in the fermionic fields yields an infinite hierarchy of coupled differential equations for the $n$-particle vertices $\Gamma^{(2n),\Lambda}$.
The $n=1$ term is the inverse full propagator, $\Gamma^{(2),\Lambda} = (\mathcal{G}^{\Lambda}_{0})^{-1} - \Sigma^\Lambda$, and thus contains the self-energy $\Sigma^\Lambda$, which we neglect in our calculations as explained later.
The $n=2$ vertex $\Gamma^{(4),\Lambda}(k'_1, k'_2; k_1, k_2)$ corresponds to the two-particle effective interaction, which is the main focus of our analysis as it captures the leading instabilities towards ordered phases.
Within the static approximation discussed below, its frequency dependence is neglected, and we denote the resulting interaction vertex by
\begin{equation}
    V^\Lambda(k_1, k_2, k_3, k_4) \equiv \Gamma^{(4),\Lambda}(k_1, k_2; k_4, k_3) \, ,
\end{equation}
where the multi-indices now comprise only momentum, spin, and other flavor indices.
All terms with $n>2$ are neglected (``level-two truncation''), which is a common approximation in FRG studies and has been found to capture the essential physics of many correlated systems \cite{dupuis_nonperturbative_2021,platt_functional_2013,honerkampRecentDevelopmentsFunctional2022,metznerFunctionalRenormalizationGroup2012}. 
This truncation is controlled for weak coupling, where the neglected higher-order vertices are generated only at higher orders in the interaction, and is commonly used as an approximation at moderate interaction strengths, which is the regime we are interested in for twisted WSe$_2$.

The flow equation for the two-particle effective interaction $V^\Lambda(k_1,\dots,k_4)$ is then given by
\begin{align}
    \partial_\Lambda V^\Lambda(k_1, \dots, k_4) &= 
    \frac{1}{2} 
    \int_p \int_{p'} 
    \int_q \int_{q'}
    \mathcal{G}^{\Lambda}_{0}(p,p') 
    \mathcal{S}^{\Lambda}_{0}(q,q') 
    \nonumber\\
    &\quad \times \left[ 
    V^\Lambda(k_1, k_2, p, q) 
    V^\Lambda(p', q', k_3, k_4)
    \right. \nonumber \\
    &\qquad-\left(
    V^\Lambda(k_1, q', k_3, p)
    V^\Lambda(p', k_2, q, k_4) 
    + 
    (p\leftrightarrow q, p'\leftrightarrow q')\right)\nonumber\\
    &\qquad +\left.\left(  V^\Lambda(k_2, q', k_3, p) V^\Lambda(p', k_1, q, k_4) + (p\leftrightarrow q, p'\leftrightarrow q')\right) \right]\nonumber \\
    &\quad - \int_p \int_{p'} \mathcal{S}_0^\Lambda(p,p') \Gamma^{(6),\Lambda}(k_1, k_2, p, p', k_3,k_4) \,,
    \label{eq:app:flow_equation}
\end{align}

where we have defined the single-scale propagator $\mathcal S_0^\Lambda=(d/d\Lambda)\mathcal G_0^\Lambda|_{\Sigma^\Lambda \mathrm{fixed}}$. 
Another simplifying approximation in our approach is the neglect of self-energy corrections and of the external frequency dependence \cite{husemannFrequencyDependentVertex2012,metznerFunctionalRenormalizationGroup2012}. 
Together with the level-two truncation, this yields the standard one-loop flow equation, in which the single-scale propagator does not contain interaction corrections.
The static approximation is motivated by previous studies of related correlated electron models, which have shown that static FRG calculations capture the leading Fermi-surface instabilities and the qualitative structure of the resulting phase diagrams, while frequency-dependent extensions can lead to quantitative modifications of the associated pseudocritical scales~\cite{husemannFrequencyDependentVertex2012,PhysRevB.96.235110,vilardiDynamicalFunctionalRenormalization2020,metznerFunctionalRenormalizationGroup2012,SciPostPhys.6.1.009,hilleQuantitativeFunctionalRenormalization2020,yirgaFrequencydependentFunctionalRenormalization2021,andergassenFunctionalRenormalizationGroup2004}.

Including full frequency dependence would significantly increase the computational complexity, while it is not expected to substantially alter the qualitative physics or the identification of the leading instability, making the static approximation a pragmatic choice for our study.
The neglect of self-energy corrections during the flow allows us to focus on the evolution of the effective interactions without the added complexity of renormalizing the single-particle properties \cite{hilleQuantitativeFunctionalRenormalization2020,uebelackerSelfenergyFeedbackFrequencydependent2012,eberleinSelfenergyEffectsFunctional2015,kuglerMultiloopFunctionalRenormalization2018}. Self-energy corrections enter the flow of the vertex only at third order in the interaction, which suggests that the band renormalization due to the self-energy can be neglected for a qualitative determination of instabilities in the weak to moderate coupling regime we study.

In summary, truncating at the four-point vertex and neglecting self-energy corrections in the single-scale propagator yields three contributions to the flow of the effective interaction $V^\Lambda$, which can be expressed as
\begin{equation}
    \partial_\Lambda V^\Lambda
    =
    \tau^{\text{PP}, \Lambda}
    + 
    \tau^{\text{PH}, \Lambda}
    + 
    \tau^{\text{PHC}, \Lambda}
    \quad , \label{eq:app:vertex_decomposition_flow}
\end{equation}
where $\tau^{\text{PP}, \Lambda}$, $\tau^{\text{PH}, \Lambda}$, and $\tau^{\text{PHC}, \Lambda}$ represent the contributions from the particle-particle, direct particle-hole, and crossed/exchange particle-hole channels, respectively. 
They are given by 
\begin{align}
    \tau^{\text{PP}, \Lambda} &= 
    - 
    \frac{1}{2} 
    \sum_{\mu_1 \dots \mu_4}
    \int_q 
    \frac{d}{d\Lambda}
    \left[
            \mathcal{G}_{0,\mu_1, \mu_3}^{\Lambda}(i\omega, \vq + \vk_1 + \vk_2)
    \mathcal{G}_{0,\mu_2, \mu_4}^{\Lambda}(-i\omega, -\vq)
    \right]\nonumber \\
    &\quad \times 
    V_{\sigma_1, \sigma_2, \mu_1, \mu_2}^\Lambda(\vk_1, \vk_2, \vq + \vk_1 + \vk_2)
    V_{\mu_3, \mu_4, \sigma_3, \sigma_4}^\Lambda(\vq + \vk_1 + \vk_2, -\vq, \vk_3)\, ,
    \\
    \tau^{\text{PH}, \Lambda} &= 
    \sum_{\mu_1 \dots \mu_4}
    \int_q 
    \frac{d}{d\Lambda}
    \left[
    \mathcal{G}_{0,\mu_1, \mu_3}^{\Lambda}(i\omega, \vq + \vk_1 - \vk_3)
    \mathcal{G}_{0,\mu_2, \mu_4}^{\Lambda}(i\omega, \vq)
    \right]\nonumber \\
    &\quad \times 
    V_{\sigma_1, \mu_2, \sigma_3, \mu_1}^\Lambda(\vk_1, \vq , \vk_3)
    V_{\mu_3, \sigma_2, \mu_4, \sigma_4}^\Lambda(\vq + \vk_1 - \vk_3, \vk_2, \vq)\, ,\\
    \tau^{\text{PHC}, \Lambda} &= -\sum_{\mu_1 \dots \mu_4}\int_q \frac{d}{d\Lambda}
    \left[
        \mathcal{G}_{0,\mu_1, \mu_3}^{\Lambda}(i\omega, \vq + \vk_1 - \vk_4)
        \mathcal{G}_{0,\mu_2, \mu_4}^{\Lambda}(i\omega, \vq)
    \right]\nonumber \\
    &\quad \times 
    V_{\sigma_1, \mu_2, \mu_1, \sigma_4}^\Lambda(\vk_1, \vq,  \vq+ \vk_1 - \vk_4)
    V_{\mu_3, \sigma_2, \sigma_3, \mu_4}^\Lambda(\vq + \vk_1 - \vk_4, \vk_2, \vk_3)\, , 
\end{align}
where we used momentum conservation $\vk_1 + \vk_2 = \vk_3 + \vk_4$ to eliminate the fourth momentum argument of the vertices, $\sigma_i$ and $\mu_i$ denote spin/orbital (or band) indices, and $\Lambda = T$.

A high momentum resolution is important for the analysis of non-local instabilities, such as the ones towards incommensurate charge and spin orders, or unconventional superconductivity. Since a straightforward implementation of the flow equation in Eq.~\eqref{eq:app:vertex_decomposition_flow} is very costly for a high momentum resolution, we use the truncated unity FRG (TUFRG) approach \cite{schoberTruncatedUnityFunctionalRenormalization2018,oConsistentCombinationTruncatedunity2024,lichtensteinHighperformanceFunctionalRenormalization2017,husemannEfficientParametrizationVertex2009}, which keeps the full dependence on the transfer momentum of each channel, while projecting the dependence on the remaining fermionic momenta onto a finite basis of form factors (see below). This significantly reduces the computational complexity while retaining the essential physics. 

To this end, we decompose $V^\Lambda$ in the TUFRG formulation according to its contributions from the three channels
\begin{align}\label{eq:app:vertex_decomposition_bosonic_representation}
    V^{\Lambda}_{\sigma_1 \dots \sigma_4}(\vk_1, \vk_2, \vk_3)
    &= 
    V^{\Lambda_0}_{\sigma_1 \dots \sigma_4}(\vk_1, \vk_2, \vk_3) \nonumber\\
    &\quad + \phi^{\Lambda, \text{PP}}_{\sigma_1 \dots \sigma_4}(
    \mathbf{q}_{\rm{PP}}=\vk_1 + \vk_2; 
    \vk_{\rm{PP}} = -\vk_2, 
    \vk_{\rm{PP}}' = -\vk_4) 
    \nonumber \\
    &\quad + 
    \phi^{\Lambda, \text{PH}}_{\sigma_1 \dots \sigma_4}(
    \mathbf{q}_{\rm{PH}} =\vk_1 - \vk_3; 
    \vk_{\rm{PH}} = \vk_3, 
    \vk_{\rm{PH}}' = \vk_2)
    \nonumber \\
    &\quad + 
    \phi^{\Lambda, \text{PHC}}_{\sigma_1 \dots \sigma_4}(
    \mathbf{q}_{\rm{PHC}} =\vk_1 - \vk_4; 
    \vk_{\rm{PHC}} = \vk_4, 
    \vk_{\rm{PHC}}' = \vk_2) \, ,
\end{align}
with the bare interaction $V^{\Lambda_0}$ and the $\phi^{\Lambda, \rm{PP}}$, $\phi^{\Lambda, \rm{PH}}$, and $\phi^{\Lambda, \rm{PHC}}$ generated during the flow in the particle-particle, direct particle-hole, and crossed particle-hole channels, respectively, which vanish at the initial scale $\Lambda_0$. 
Each channel $\phi^{\Lambda, X}$ depends strongly on a particular combination of external momenta, the transfer momentum $\mathbf{q}_X$ in Eq.~\eqref{eq:app:vertex_decomposition_bosonic_representation}, which we keep explicit, while the weaker dependence on $\vk_X$ and $\vk_X'$ is expanded in form factors.

Specifically, the particle-particle channel depends strongly on the total momentum $\vq_{\mathrm{PP}}=\vk_1+\vk_2$, and the particle-hole channels depend strongly on the direct $\vq_{\mathrm{PH}}=\vk_1-\vk_3$ or exchanged $\vq_{\mathrm{PHC}}=\vk_1-\vk_4$ momentum transfer. 
The weaker momentum dependences of these channels, described by the fermionic momenta $\vk_X,\vk_X'$ in Eq.~\eqref{eq:app:vertex_decomposition_bosonic_representation}, are then expressed in a form factor basis
\begin{equation}
    \phi^{\Lambda, X}_{\sigma_1 \dots \sigma_4}(\mathbf{q}_X; \vk_X, \vk_X') 
    = 
    \sum_{l,l'} 
    \phi^{\Lambda,X, (l,l')}_{({\sigma_1 \dots \sigma_4})_X}(\mathbf{q}_X)
    f_l^{\phantom{\ast}}(\vk_X) 
    f_{l'}^\ast(\vk_X') 
    \label{eq:app:ff_expansion}
\end{equation}
with $X\in\{\text{PP, PH, PHC}\}$. 
For each channel, we reorder the valley indices $\sigma_i$ as
\begin{align}
   & ({\sigma_1 \dots \sigma_4})_{\text{PP}}= \sigma_1 \dots \sigma_4\\
   & ({\sigma_1 \dots \sigma_4})_{\text{PH}}= \sigma_1 \sigma_3 \sigma_4 \sigma_2\\
   & ({\sigma_1 \dots \sigma_4})_{\text{PHC}}= \sigma_1 \sigma_4 \sigma_3 \sigma_2 \quad.
\end{align}
With this ordering, the flow equations take the common form
\begin{align}
    \frac{d}{d\Lambda} 
    \phi^{\Lambda,X, (l,l')}_{\sigma_1 \dots \sigma_4}&(\vq_X) 
    = 
    \sum_{n,m}
    \sum_{\mu_1 \dots \mu_4}
    V^{\Lambda, X, (l, n)}_{\sigma_1, \sigma_2, \mu_1, \mu_2}(\vq_X)
    \dot{\Pi}^{\Lambda, X, (n,m)}_{\mu_1 \dots \mu_4}(\vq_X)
    V^{\Lambda, X, (m,l')}_{\mu_3, \mu_4,\sigma_3, \sigma_4 }(\vq_X)\, ,
    \label{eq:app:TUFRG_general_flow_equations}
\end{align}
where 
\begin{align}
    \dot{\Pi}^{\Lambda, \text{PP/PH},(n,m)}_{\mu_1\dots\mu_4}(\vq_{X})
    &=
    \int_{k_X,k_X'} 
    f_n^{\phantom{\ast}}(\vk_X) 
    f_{m}^\ast(\vk_X)
    \frac{d}{d\Lambda}
    \left[
        \mathcal{G}_{0,\mu_1, \mu_3}^{\Lambda}(i\omega, \vq_X + \vk_X)
        \mathcal{G}_{0,\mu_2, \mu_4}^{\Lambda}(\mp i\omega, \mp\vk_X)
    \right]\label{eq:app:bubble_derivatives}
\end{align}
is the scale derivative of the particle-particle (upper sign) or particle-hole (lower sign) bubble, projected onto the form-factor basis. 
Here, $V^{\Lambda, X, (l, n)}_{\sigma_1, \sigma_2, \mu_1, \mu_2}(\vq_X)$ denotes the full vertex, i.e., the left-hand side of Eq.~\eqref{eq:app:vertex_decomposition_bosonic_representation}, projected onto the momentum and form-factor structure of channel $X$. It contains the bare interaction as well as the projections $\hat{X}[\phi^{\Lambda, X', (l, l')}_{\sigma_1, \sigma_2, \mu_1, \mu_2}(\vq_X)]$ of all three channel couplings $X'$ onto channel $X$.
This is where inter-channel feedback enters the flow.

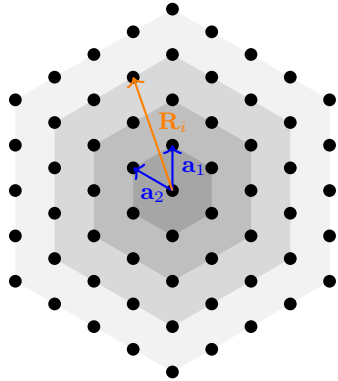
\begin{figure}[th]
    \centering
    \begin{tikzpicture}[scale=0.6, rotate = 90, x={(1cm,0cm)}, y={(0.5cm,0.866025cm)}]

        \fill[gray!10] (4,0) -- (0,4) -- (-4,4) -- (-4,0) -- (0,-4) -- (4,-4) -- cycle;
        \fill[gray!30] (3,0) -- (0,3) -- (-3,3) -- (-3,0) -- (0,-3) -- (3,-3) -- cycle;
        \fill[gray!50] (2,0) -- (0,2) -- (-2,2) -- (-2,0) -- (0,-2) -- (2,-2) -- cycle;
        \fill[gray!70] (1,0) -- (0,1) -- (-1,1) -- (-1,0) -- (0,-1) -- (1,-1) -- cycle;
        \foreach \m in {-5,...,5}{
            \foreach \n in {-5,...,5}{
                \pgfmathsetmacro{\shell}{max(max(abs(\m),abs(\n)),abs(\m+\n))}
                \ifdim \shell pt<4.1pt
                    \fill[black] (\m,\n) circle (4pt);
                \fi
            }
        }
        \draw[->,thick,blue] (0,0) -- (1,0) node[midway,right] {$\mathbf a_1$};
        \draw[->,thick,blue] (0,0) -- (0,1) node[midway,below] {$\mathbf a_2$};
        \draw[->,thick,orange] (0,0) -- (2,1)
            node[pos=0.6,right] {$\mathbf R_{i}$};
    \end{tikzpicture}
    \caption{Real-space triangular lattice generated from primitive basis vectors (blue). The shading indicates shells 1-4. The red arrow labels site $R_i$ which defines the form factor $f_i(\vk)=\rm{exp}(i\vk \cdot \mathbf{R}_i)$.}
    \label{fig:app:triangular_lattice_basis}
\end{figure}

The form factors $f_l(\vk)$ are a set of orthonormal functions defined on the Brillouin zone, which must satisfy the orthonormality and completeness relations
\begin{align}
    \int_{\mathbf{k}} 
    f_l^{\phantom{\ast}}(\vk) 
    f_{l'}^\ast(\vk)
    = \delta_{l,l'} \nonumber \\
    \sum_l 
    f_l^{\phantom{\ast}}(\vk) 
    f_l^\ast(\vk') 
    =
    \delta(\vk - \vk'). \label{eq:app:orthonormality_formfactors}
\end{align}
for an exact rewriting. For the form factor expansion, we choose plane waves $f_l(\vk) = e^{i \mathbf{k} \cdot \mathbf{R}_l}$ labeled by the real-space lattice vectors $\mathbf{R}_l$ of the Wannierized model. 
An equivalent alternative would be lattice harmonics ($s$-, $p$-, $d$-wave, \ldots), i.e., symmetry-adapted linear combinations of these plane waves. 
The crucial approximation of TUFRG is to truncate the sum $\sum_{l,l'}$ at a finite number of form factors.
In our case, we include four shells of real-space lattice vectors, i.e., up to the eighth nearest neighbor, as shown in Fig.~\ref{fig:app:triangular_lattice_basis}, which corresponds to 61 form factors in total. 
We discuss the dependence on the number of form factors further in \cref{sec:app:convergence_check}. 
Let us reiterate that this form factor projection is only performed for parts of the momentum dependence, while we keep the full momentum dependence of the interactions in the total or transfer momenta, that are crucial for accurately capturing the instabilities in the different channels. 
By further keeping a large number of form factors, this approximation is capable of accurately capturing a non-trivial momentum dependence and the non-local nature of the interactions. 
Both are important in our system due to its quantum geometry and topology: after projecting onto the narrow band around the Fermi level, the overlaps of single-particle eigenstates introduce a strong momentum dependence, which corresponds to an extended interaction in real space.

It can also be instructive to view the bare interaction as a function of the momentum combinations specific to the particle-hole or particle-particle channel. To this end, we define
\begin{align}
    V^{\mathrm{PH}}_{\sigma_1 \ldots \sigma_4}(
        \mathbf{q}_\mathrm{PH}; 
        \vk_\mathrm{PH}, 
        \vk'_\mathrm{PH}
        ) 
    &= 
    V_{\sigma_1 \ldots \sigma_4}(
        \mathbf{q}_\mathrm{PH}+\vk_\mathrm{PH},
        \vk'_\mathrm{PH},
        \vk_\mathrm{PH},
        \mathbf{q}_\mathrm{PH}+\vk'_\mathrm{PH}
    )\\
    V^{\mathrm{PP}}_{\sigma_1 \ldots \sigma_4}(
        \mathbf{q}_\mathrm{PP}; 
        \vk_\mathrm{PP}, 
        \vk'_\mathrm{PP}
    ) 
    &= 
    V_{\sigma_1 \ldots \sigma_4}(
        \mathbf{q}_\mathrm{PP}+\vk_\mathrm{PP},
        -\vk_\mathrm{PP},
        \mathbf{q}_\mathrm{PP}+\vk'_\mathrm{PP},
        -\vk'_\mathrm{PP}
    )\,,
    \label{eq:app:channels_bare}
\end{align}
where we omitted the superscript $\Lambda_0$ for brevity.
We define the channel-dependent  reparametrization $(\vq_X,\vk_X,\vk_X')\rightarrow \vk_i$ via the function $F_i^{X}(\vq_X,\vk_X,\vk_X')=\vk_i$.
The projection onto the form-factor basis is given by
\begin{equation}\label{eq:app:ia_FF}
    V^{X,(l,l')}_{\sigma_1\ldots\sigma_4}(\vq_X)
    =
    (1/N_k^2)
    \sum_{\vk_X \vk_X'}
    f_l^\ast(\vk_X) 
    f_{l'}^{\phantom{\ast}}(\vk_X')
    V^X_{\sigma_1\ldots\sigma_4}(\vq_X;\vk_X,\vk_X'). 
\end{equation}
where $N_k$ is the number of momentum points in the Brillouin zone.
We plot the maximum of $V^{X,(l,l')}_{\sigma_1,\sigma_2,\sigma_3, \sigma_4}(\mathbf{q}_X)$ for $l=l'$ in the particle-particle (X=PP) and particle-hole (X=PH) channels in Fig.~\ref{fig:model}(c) of the main text, i.e., the maximum of 
\begin{equation}\label{eq:app:Vii}
    V^{X,(i,i)}_{\sigma_1,\sigma_2,\sigma_3, \sigma_4}(\vq_X)
    =
    (1/N_k^2)
    \sum_{\vk_X \vk_X'}
    e^{i(\vk_X-\vk_X')\cdot \mathbf{R}_i}
    V^X_{\sigma_1,\sigma_2,\sigma_3, \sigma_4}(\vq_X;\vk_X,\vk_X'). 
\end{equation}
The maximum for the $l=l'=1$ component occurs at $\mathbf{q}_X=0$ in both channels for $\sigma_1 = \sigma_3$ and $\sigma_2=\sigma_4$.

\subsection{Band-projected Hamiltonian}\label{sec:gen_hf:band_projected}

In this work, we focus on the leading interaction-driven instabilities of the system at  filling $\nu = -1$, i.e., one hole per moiré unit cell, which indicate the tendency towards symmetry-broken phases. 
To this end, we project the t\ch{WSe2} Hamiltonian into the uppermost valence band of the single-particle Wannier Hamiltonian from \cref{app:eqn:single_particle_wannier_hamiltonian_general} within each valley.
For the band projection, we employ the 3Tri model obtained from our DFT calculations.
We define the band basis of the single-particle Wannier Hamiltonian according to
\begin{equation}
    \sum_{m} 
    h^{\sigma}_{nm} 
    \left( \vk \right) 
    \mathbf{Y}_{m,\alpha}^\sigma \left( \vk \right) 
    = 
    E_{\alpha}^\sigma \left( \vk \right)
    \mathbf{Y}_{n,\alpha}^\sigma \left( \vk \right),
\end{equation}
where $E_{\alpha}^\sigma \left( \vk \right)$ and $\mathbf{Y}_{n,\alpha}^\sigma \left( \vk \right)$ denote the energy and the wave function of the $\alpha$-th valence band of the system at momentum $\vk$, respectively ($\alpha \geq 1$ indexes the valence bands of the system in each valley in order of decreasing energy).
We will be concerned exclusively with the uppermost valence band of the system. We adopt the following gauge-fixing procedure:
\begin{enumerate}
    \item We start by diagonalizing $h^{\sigma}_{nm} \left( \vk \right)$ in valley $\sigma = +$. The phase of $\mathbf{Y}_{n,1}^+ \left( \vk \right)$ is chosen such that the component $\mathbf{Y}_{1,1}^+ \left( \vk \right)$ is real and positive whenever $\abs{\mathbf{Y}_{1,1}^+ \left( \vk \right)} \neq 0$. If $\abs{\mathbf{Y}_{1,1}^+ \left( \vk \right)} = 0$, the phase of the wave function is chosen such that the largest component of $\mathbf{Y}_{n,1}^+ \left( \vk \right)$ is real and positive. This procedure uniquely fixes the phase of $\mathbf{Y}_{n,1}^+ \left( \vk \right)$ throughout the entire Brillouin zone.

    It is important to specify what this gauge-fixing procedure implies for the sewing matrices of the wave functions under the action of the valley-preserving crystalline symmetries of the system. We first note from \cref{app:eqn:C3z_action_trial_3tri,app:eqn:C2yT_action_trial_3tri} that the action of these symmetries on the Wannier-basis operators of the 3Tri model is given by
    \begin{align}
        C_{3z} \cre{d}{\vk,\sigma,n} C_{3z}^{-1} &= \sum_{m} \cre{d}{C_{3z} \vk,\sigma,m} D^{\sigma}_{mn} \left( C_{3z} \right), \\
        \left( C_{2y} \mathcal{T} \right) \cre{d}{\vk,\sigma,n}  \left( C_{2y} \mathcal{T} \right)^{-1} &= \sum_{m}\cre{d}{ C_{2y} \mathcal{T} \vk,\sigma,m} D^{\sigma}_{mn} \left( C_{2y} \mathcal{T} \right),
    \end{align}
    where
    \begin{align}
        D^{\sigma}_{mn} \left( C_{3z} \right) &= e^{i \frac{2\pi}{3} \sigma \left( n - \frac12 \right)} \delta_{mn},\\
        D^{\sigma}_{mn} \left( C_{2y} \mathcal{T} \right) &= -\delta_{mn}.
    \end{align}
    Our gauge-fixing procedure implies the following sewing relations for the wave function of the uppermost valence band in valley $\sigma=+$:
    \begin{align}
        \sum_{m} D^{+}_{nm} 
        \left( C_{3z} \right) 
        \mathbf{Y}_{m,1}^+ \left( \vk \right) 
        &= 
        e^{i \frac{\pi}{3}} 
        \mathbf{Y}_{n,1}^+ \left( C_{3z} \vk \right), \quad \text{if } \mathbf{Y}_{1,1}^+ \left( \vk \right) \neq 0, \\
        \sum_{m} 
        D^{+}_{nm} 
        \left( C_{2y} \mathcal{T} \right) 
        \left(\mathbf{Y}^{+}_{m,1}\left( \vk \right) \right)^\ast 
        &= - \mathbf{Y}_{n,1}^+
        \left( C_{2y} \mathcal{T} \vk \right).
    \end{align}
    The first condition is valid for all $\vk$ points in the moir\'e Brillouin zone except for the $\Gamma_M$ point, where $\mathbf{Y}_{1,1}^+ \left( \vec{0} \right) = 0$ and where the IRREP of the first valence band (as detailed in \cref{tab:irrep_dft_model}) imposes the condition
    \begin{equation}
        \sum_{m} 
        D^{+}_{nm} 
        \left( C_{3z} \right) 
        \mathbf{Y}_{m,1}^+ \left( \vec{0} \right) 
        = 
        e^{i \pi} 
        \mathbf{Y}_{n,1}^+ \left( \vec{0} \right).
    \end{equation}
    The fact that the sewing matrix of the $\mathbf{Y}_{n,1}^+ \left( \vk \right)$ wave function under $C_{3z}$ symmetry cannot be chosen continuously throughout the entire Brillouin zone is a manifestation of the non-trivial topology of the uppermost valence band of twisted \ch{WSe2}, which is a Chern band.

    \item In valley $\sigma = -$, the wave function of the uppermost valence band is obtained from the one in valley $\sigma = +$ using time-reversal symmetry,
    \begin{equation}
        \mathbf{Y}_{m,1}^- \left( - \vk \right) = 
        \sum_{n} 
        D^{+}_{mn} \left( \mathcal{T} \right) 
        \left(\mathbf{Y}^{+}_{n,1} \left( \vk \right)\right)^\ast,
    \end{equation}
    where the matrix representation of the time-reversal operator in the Wannier basis is given by $D^{\sigma}_{mn} \left( \mathcal{T} \right) = -\sigma \delta_{mn}$, as explained in \cref{app:eqn:T_action_trial_3tri}.
\end{enumerate}
Once the gauge-fixed wave functions are obtained, we define the band operators corresponding to the uppermost valence band as
\begin{equation}\label{eq:app:Wannier_orbital_to_top_band}
    \cre{\tilde{\gamma}}{\vk,\sigma} 
    \equiv 
    \sum_{n} 
    \mathbf{Y}_{n,1}^\sigma \left( \vk \right) 
    \cre{d}{\vk,\sigma,n}, \qquad 
    \des{\tilde{\gamma}}{\vk,\sigma} 
    \equiv
    \sum_n
    \left( \mathbf{Y}_{n,1}^\sigma \left( \vk \right) \right)^\ast
    \des{d}{\vk, \sigma, n}
\end{equation}
in terms of which the projected many-body Hamiltonian reads
\begin{equation}
    \mathcal{H}^{\rm{band}} = 
    \mathcal{H}_{\rm{kin}}^{\rm{band}} 
    + H_{\tInt}^{\rm{band}},
\end{equation}
with
\begin{align}
    \mathcal{H}_{\rm{kin}}^{\rm{band}} 
    &= 
    \sum_{\vk,\sigma} 
    E_{1}^\sigma \left(  \vk \right) 
    \cre{\tilde{\gamma}}{\vk,\sigma} 
    \des{\tilde{\gamma}}{\vk,\sigma}
    , \label{app:eqn:tb_single_part_proj}\\
    H_{\tInt}^{\rm{band}}
    &= 
    \frac{1}{2N_\vq} 
    \sum_{\substack{\sigma,\sigma' \\ \vk, \vk', \vq}} \tilde{V}_{\sigma\sigma'} \left( \vq; \vk, \vk' \right) 
    :: 
        \cre{\tilde{\gamma}}{\vk,\sigma} 
        \des{\tilde{\gamma}}{\vk + \vq,\sigma} 
        \cre{\tilde{\gamma}}{\vk' + \vq,\sigma'} 
        \des{\tilde{\gamma}}{\vk',\sigma'} 
    ::.\label{app:eqn:tb_int_proj}
\end{align}
This interaction is written in the density-density/PH channel, for which we drop the index $X$ on the right hand side.
The corresponding interaction in the PP channel is obtained by the reparametrization \cref{eq:app:channels_bare}.
The focus of our analysis on the top bands is valid as they contain the Fermi level and thus are most relevant for the low-energy physics of the system.
In \cref{app:eqn:tb_int_proj}, the projected interaction tensor is given by
\begin{equation}\label{eq:app:interaction_projected}
    \tilde{V}_{\sigma\sigma'} \left( \vq; \vk, \vk' \right) = \sum_{n,m,n',m'} 
    V_{\sigma\sigma'}^{nm;n'm'} \left( \vq; \vk, \vk' \right) 
    \mathbf{Y}^{\sigma}_{n,1} \left( \vk \right) 
    \left(\mathbf{Y}_{m,1}^\sigma \left( \vk + \vq \right)  \right)^\ast
    \mathbf{Y}^{\sigma'}_{m',1} \left( \vk' + \vq \right) 
    \left(\mathbf{Y}_{n',1}^{\sigma'} \left( \vk' \right)\right)^\ast
    ,
\end{equation}
where $n,m,n',m'$ label the Wannier orbital index. $V_{\sigma\sigma'}^{nm;n'm'} \left( \vq, \vk, \vk' \right)$ is given by \cref{eq:app:interaction_orbital_basis}.
Note that the index order of the Wannier orbitals is not equivalent to the order of momenta where incoming and outgoing momenta were grouped, respectively.
Here, the density operator governs the ordering \cref{eq:app:density_operator}.

Combining eigenvectors to overlap matrices
\begin{equation}
    \Lambda^\sigma_{nm}(\vk,\vq)
    =
    \mathbf{Y}^\sigma_{n,1}(\vk)
    \left(
        \mathbf{Y}^\sigma_{m,1}(\vk+\vq)
    \right)^\ast .
    \label{eq:app:density_operator}
 \end{equation}
 gives
 \begin{equation}
    \rho(\vq)
    =
    \sum_{\vk,\sigma,m,n} 
    \tilde{\gamma}_{\vk,\sigma}^\dagger 
    \Lambda^\sigma_{nm}(\vk,\vq)
    \tilde{\gamma}_{\vk+\vq,\sigma}^{\phantom{\dagger}}
\end{equation}
 and recovers the interaction in the main text
 \begin{align}
    \tilde{V}_{\sigma\sigma'} \left( \vq, \vk, \vk' \right) = \sum_{n,m,n',m'} 
    V_{\sigma\sigma'}^{nm;n'm'} \left( \vq, \vk, \vk' \right)
    \Lambda^\sigma_{nm}(\vk,\vq)
    \Lambda^{\sigma'}_{m'n'}(\vk'+\vq,-\vq).
\end{align}

With the gauge-fixing procedure outlined above, the projected fermion operators transform under the action of the symmetry operators of the system as
\begin{align}
    \mathcal{T} 
        \cre{\tilde{\gamma}}{\vk,\sigma} 
    \mathcal{T}^{-1} 
    &= 
    \sigma \cre{\tilde{\gamma}}{-\vk,-\sigma}, \\
    C_{3z} \cre{\tilde{\gamma}}{\vk,\sigma} C_{3z}^{-1} 
    &= 
    \begin{cases}
        e^{i \frac{\pi}{3}\sigma} 
        \cre{\tilde{\gamma}}{C_{3z} \vk, \sigma}, & \text{if } \vk \neq \vec{0},\\
        e^{i \pi \sigma} \cre{\tilde{\gamma}}{C_{3z} \vk, \sigma}, & \text{if } \vk = \vec{0},
    \end{cases} \\
    \left( C_{2y} \mathcal{T} \right)  
    \cre{\tilde{\gamma}}{\vk,\sigma} \left( C_{2y} \mathcal{T} \right)^{-1} 
    &= 
    -\cre{\tilde{\gamma}}{C_{2y} \mathcal{T} \vk,\sigma}.
\end{align}
In particular, although the projected interaction Hamiltonian is $C_{3z}$ symmetric throughout the entire Brillouin zone, the projected interaction tensor does not transform trivially under $C_{3z}$. 
The momentum-independent phase $e^{i\pi\sigma/3}$ acquired by the band operators away from the $\Gamma_M$ point cancels between the creation and annihilation operators associated with each valley. 
However, the change in the sewing phase at $\Gamma_M$ gives
\begin{equation}\label{app:eq:C3action}
    \tilde{V}_{\sigma\sigma'} \left( C_{3z} \vq, C_{3z} \vk, C_{3z} \vk' \right)
    =
    e^{
        i \frac{2\pi}{3}
        \left[
            \sigma \left(
                \delta_{\vk,\vec{0}}
                -
                \delta_{\vk+\vq,\vec{0}}
            \right)
            +
            \sigma' \left(
                \delta_{\vk'+\vq,\vec{0}}
                -
                \delta_{\vk',\vec{0}}
            \right)
        \right]
    }
    \tilde{V}_{\sigma\sigma'} \left( \vq, \vk, \vk' \right).
\end{equation}
Here, equality of a momentum to $\vec{0}$ is understood modulo a moir\'e reciprocal lattice vector.

After expanding the density operators, relabeling the momenta according to \cref{eq:app:channels_bare} and bringing the fermionic operators into normal order, the interaction becomes
\begin{align}
    \mathcal H_{\mathrm{int}}^{\rm{band}}
    =
    \frac{1}{2 N_\vq}
    \sum_{\substack{\mathbf k_1,\ldots,\mathbf k_4\\
    \sigma,\sigma'\\
    m,n,m',n'}}
    &
    \delta_{\mathbf k_1+\mathbf k_2,\mathbf k_3+\mathbf k_4}
    V_{\sigma\sigma'}^{nm;n'm'}(\mathbf k_1-\mathbf k_3; \vk_2, \vk_3)
    \Lambda^\sigma_{nm}
    (\mathbf k_3,\mathbf k_1-\mathbf k_3)
    \Lambda^{\sigma'}_{n'm'}
    (\mathbf k_2,\mathbf k_4-\mathbf k_2)
    \nonumber \\
    &\times
    \tilde{\gamma}^\dagger_{\mathbf k_1,\sigma,m}
    \tilde{\gamma}^\dagger_{\mathbf k_2,\sigma',m'}
    \tilde{\gamma}_{\mathbf k_3,\sigma',n'}
    \tilde{\gamma}_{\mathbf k_4,\sigma,n}.
\end{align}
Extracting the 2-particle interaction in band space
\begin{align}
    V_{\sigma,\sigma'}(\vk_1,\vk_2,\vk_3,\vk_4)
    &=
    \sum_{n,m,n',m'}
    V_{\sigma\sigma'}^{nm;n'm'}(\mathbf k_1-\mathbf k_3; \vk_2, \vk_3)
    \Lambda^\sigma_{nm}
    (\mathbf k_3,\mathbf k_1-\mathbf k_3)
    \Lambda^{\sigma'}_{n'm'}
    (\mathbf k_2,\mathbf k_4-\mathbf k_2)
\end{align}
which can then be rewritten in the channel-dependent bosonic parametrization \cref{eq:app:channels_bare} into the initial interactions $V_{\sigma_1,\sigma_2,\sigma_3,\sigma_4}^{X,\Lambda_0}(\vq_X;\vk_X,\vk'_X)$ where we extended the valley indices $(\sigma,\sigma') \rightarrow (\sigma_1,\sigma_2,\sigma_3,\sigma_4)=(\sigma,\sigma',\sigma,\sigma')$
\begin{align}
    V_{\sigma_1,\sigma_2,\sigma_3,\sigma_4}^{X,\Lambda_0}(\vq_X;\vk_X,\vk'_X)
    &=
    V_{\sigma,\sigma'}^{X,\Lambda_0}(\vq_X;\vk_X,\vk'_X)
    \delta_{\sigma_1 \sigma} \delta_{\sigma_3 \sigma}
    \delta_{\sigma_2 \sigma'} \delta_{\sigma_4 \sigma'}
\end{align}

\subsection{Initial Conditions}\label{sec:app:initial_interactions}
The initial conditions for the FRG calculations, obtained from ab-intio calculations, are crucial for accurately capturing the electronic properties of twisted WSe$_2$.
They are split into two parts, the kinetic part, described by the band structure $E_1^\sigma(\vk)$, and the interaction part $\tilde{V}^{\Lambda_0}_{\sigma,\sigma'}(\vq,\vk,\vk')$ (which also contains the dependence on the original single-particle wavefunctions via the overlap matrices). 
Motivated by experiment, we focus on a twist angle of $\theta=3.65^\circ$ and filling $\nu=-1$, and use a displacement field $D$ applied perpendicular to the layers as a tuning parameter, see \eqref{eq:displacement_field_dependency}. 
We investigate a range of displacement fields from $D=0$~meV to $D=33$~meV in steps of $3$~meV. 
We supress the index for the displacement field in the following.
In addition, we tune the interaction strength to assess weak vs. strong coupling effects and compare to the tuning via twist angle. 

\subsection{Band Structure}\label{sec:app:band_structure}
As the single-particle input $\mathcal H_{\text{kin}}^{\text{orb}}$ for the FRG calculations we use a three-orbital model on the triangular lattice, which is obtained from a Wannierization of a DFT-inspired lattice model of tWSe$_2$ \cref{app:DFT3Tri}. 
Due to the strong Ising spin-orbit coupling in WSe$_2$, the low-energy states in a given valley are fully spin-polarized with opposite spins in opposite valleys. 
Consequently, spin and valley degrees of freedom are locked into a single pseudospin/spin-valley index, which we denote as $\sigma$.
The three-orbital (3Tri) model contains hopping parameters $t^{\sigma}_{n,m}(\mathbf{R})$ from the Wannierization between sites with distance $\mathbf{R}$ and orbital $n$ and $m$ in valley $\sigma$ up to a distance of $6$ moiré lattice vectors in order to capture the continuum-model band structure accurately. 
After a Fourier transformation
\begin{equation}
    h^\sigma_{n,m}(\mathbf{k}) = 
    \sum_{\mathbf{R}} 
    e^{i \mathbf{k} \cdot \mathbf{R}} \ t^\sigma_{n,m}(\mathbf{R})\,,
    \label{eq:app:hopping_to_dispersion_fourier_transformation}
\end{equation}
the kinetic Hamiltonian in orbital space can be represented as a $6 \times 6$ matrix with block-diagonal structure
\begin{equation}\label{eq:app:kinetic_hamiltonian_block_structure}
    \mathcal{H}_{\text{kin}}^{\text{orb}}(\mathbf{k}) = 
    \begin{pmatrix}
        \{h^{+}_{i,j}(\mathbf{k})\}_{n,m} &  0_{3x3} \\
         0_{3x3} &  \{h^{-}_{n,m}(\mathbf{k})\}_{n,m}  
    \end{pmatrix}\,.
\end{equation}
Here, the two $3 \times 3$ blocks correspond to the two valleys $\sigma$ and $\sigma'$. 

We diagonalize the Hamiltonian $\mathcal{H}_{\text{kin}}^{\text{orb}}(\mathbf{k})$ to obtain the energy bands \( E_n^\sigma(\mathbf{k}) \) 
\begin{equation}\label{eq:app:kinetic_hamiltonian_diagonalization}
    \mathcal{H}_{\text{kin}}^{\text{orb}}(\mathbf{k})
    \mathbf{Y}(\mathbf{k}) 
    = 
    \mathcal{H}_{\text{kin}}^{\text{band}}(\mathbf{k}) 
    \mathbf{Y}(\mathbf{k}),
\end{equation}
where \( \mathbf{Y}(\mathbf{k}) \) are the corresponding eigenstates. 
The resulting diagonal Hamiltonian $\mathcal{H}_{\text{kin}}^{\text{band}}(\mathbf{k})$ can be expressed as
\begin{equation}\label{eq:app:kinetic_hamiltonian_bandspace}
    \mathcal{H}_{\text{kin}}^{\text{band}}(\mathbf{k}) 
    = 
    \rm{diag}(E^\sigma_\alpha(\mathbf{k}), E^{\sigma'}_\alpha(\mathbf{k}))\quad  \alpha=1,2,3.
\end{equation}
The energy bands are used in the calculation of the FRG bubbles $\Pi_{ph}$ and $\Pi_{pp}$, whose number scales quadratically with the number of bands and represents the computational bottleneck at low temperatures, see Sec.~\ref{sec:app:bubbles}.
As above, we limit further analysis on the topmost band $\alpha=1$ and drop this index correspondingly $E_1^\sigma(\vk)\rightarrow E^\sigma(\vk)$.

For the adaptive integration routine, we require $E^{\sigma}(\mathbf{k})$ evaluated on an unpredictable grid structure. 
Direct Fourier transformation for each evaluation is computationally expensive as this has to be called many times during the integration procedure, especially at low temperatures where the integrand becomes sharply peaked and a high resolution is required. 
To improve efficiency, we precalculate $E^\sigma(\mathbf{k})$ on a 2D node grid of Chebyshev points, construct an interpolating function $E^\sigma_{\mathrm{interp}}(\mathbf{k})$, and use this during the integration procedure. 
By employing a sufficient number of Chebyshev nodes \cite{boydChebyshevFourierSpectral2001,grossiFastOPInterpolationjl2023}, namely 380 for each sixth of the BZ, we maintain a relative error $\sim 10^{-6}$ which is below that reported in the underlying Wannierization.
This precalculation step needs to be performed only once. As a result, this approach provides more than an order of magnitude speedup compared to direct evaluation of $E^\sigma(\mathbf{k})$ during integration.

\begin{figure}
    \centering
    \includegraphics[width=1\linewidth]{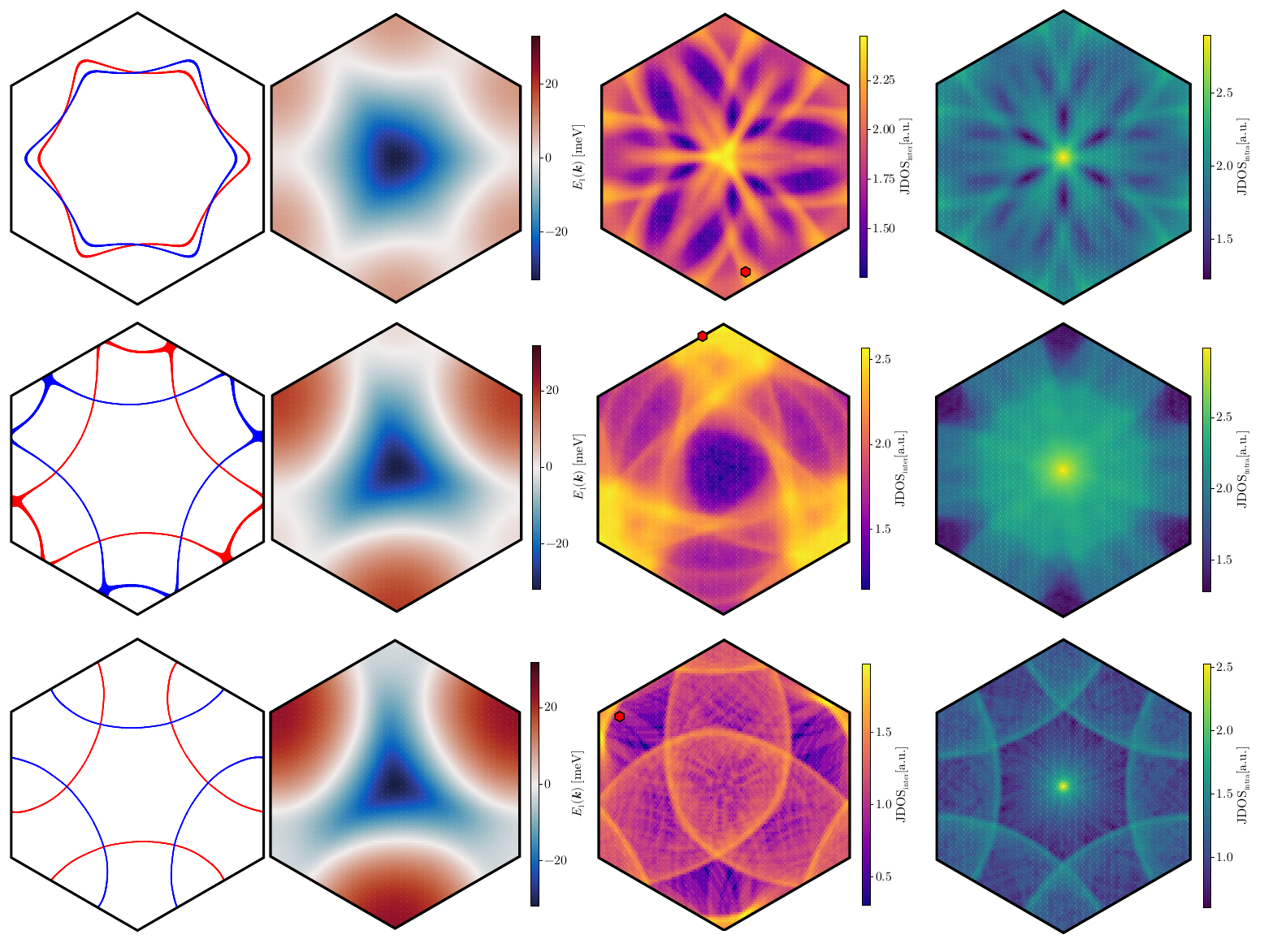}
    \caption{Single particle properties for the top bands $E^{\sigma}$ for $D=\{0,18,33\}$~meV (top to bottom) with Fermi surfaces, color plot of $E^\sigma(\vk)$, inter- and intra-band JDOS (left to right). The red marker in the inter-band JDOS indicate the wave vector of the IVC-SDW instability we find at these displacement fields.}
    \label{fig:app:single_particle_prop}
\end{figure}
The Fermi surface and energy landscape for displacement field $D=0,18$ and $33$~meV are shown in Fig.~\ref{fig:app:single_particle_prop}.
As the filling is set to one carrier per moiré unit cell ($\nu = -1$)~\cite{Guo.2026}, counting the two valleys corresponds exactly to half-filling of the moiré bands.
This filling is of particular interest to us since it has been experimentally shown to host superconductivity in the vicinity of an insulator for small twist angles~\cite{xiaSuperconductivityTwistedBilayer2025,xiaBandwidthtunedMottTransition2026,Guo.2026}. 
At the same time, these states are in close proximity to a peak in the DOS at $E^{\sigma}=0$.
As described in \cref{sec:gen_hf:band_projected}, the rotational symmetry of tWSe$_2$ is $C_3$. 
However, the Fermi surface of the topmost band is approximately invariant under $\hat C_6$ operations at $D=0$.
This symmetry becomes more strongly broken once a displacement field $D \neq 0$~meV is applied, see \cref{fig:app:single_particle_prop}.

To identify potential nesting vectors, we determine which transfer momenta $\mathbf q$ connect Fermi surface states of band $E^\sigma$ to states in the same band or in $E^{\sigma'}$ within an energy window $\delta$ via the intra-valley ($\sigma=\sigma'$) and inter-valley ($\sigma\neq\sigma'$) joint density of states
\begin{align}\label{eq:app:JDOS}
    \mathrm{JDOS}_{\mathrm{inter/intra}}(\vq)= 
    \frac{1}{V_{\rm{BZ}}}
    \int_{\vk}
    \delta(E^\sigma(\vk))
    \delta(E^{\sigma/\sigma'}(\vk+\vq))\,.
\end{align}
In practice the integration is performed by summing over $N_\vk$ weighted momentum points $1/V_{\rm{BZ}}\int_\vk \rightarrow 1/N_\vk\sum_\vk$ with a numerical acceptance window for the $\delta$ functions set to $10^{-1}$~meV. 
For displacement field $D=0,\, 18$ and $33$~meV, the resulting intra- and inter-valley joint density of states are shown in Fig.~\ref{fig:app:single_particle_prop} 

While this nesting analysis indicates important transfer momenta $\vq$ for intra- and inter-valley susceptibility calculations, it lacks the temperature dependence essential for and RPA or FRG analysis which will be provided later.
\begin{figure}
    \centering
    \includegraphics[width=0.45\textwidth]{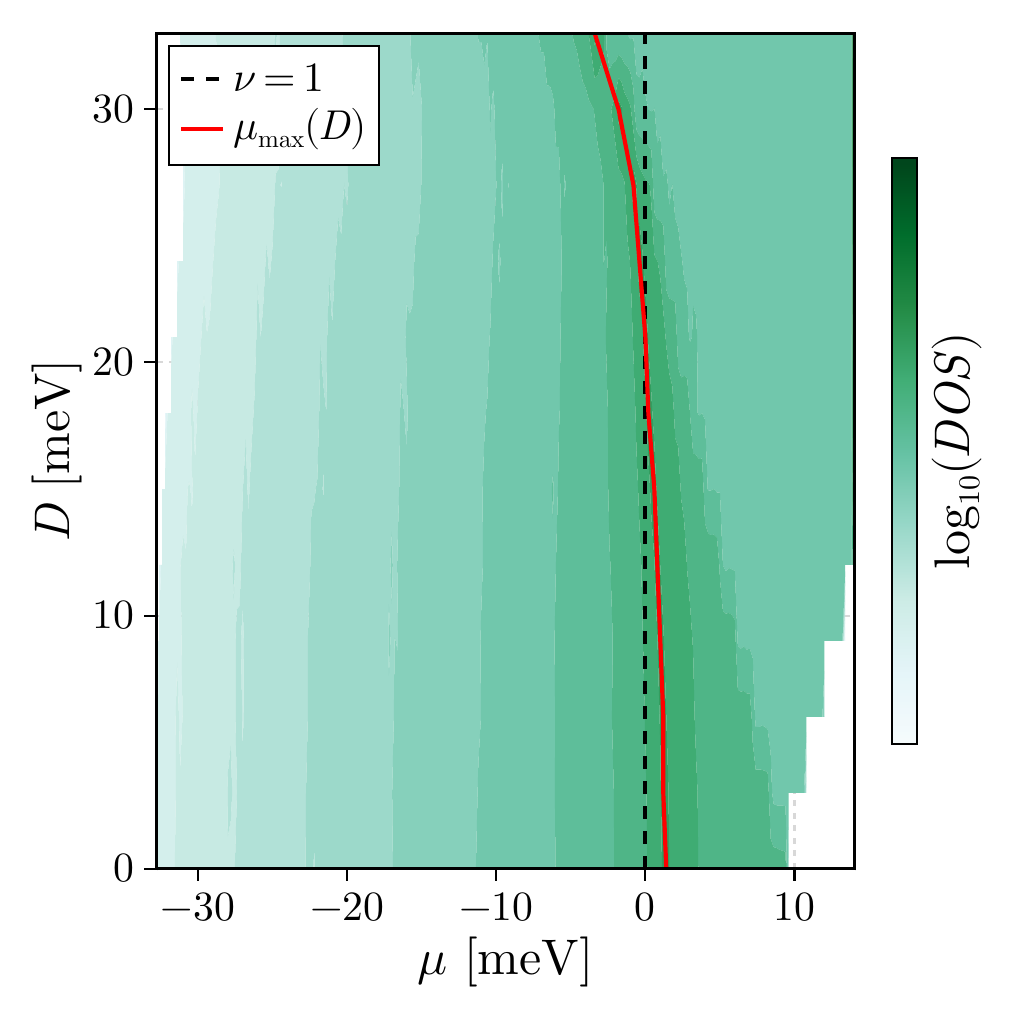}
    \caption{
    Density of states (DOS) for the top band $E^{\sigma}$ as function of displacement field $D$ and chemical potential $\mu$. 
    The red line marks the maximal DOS, the dashed line indicates filling $\nu=-1$ and is plotted in \cref{fig:model} b).}
    \label{fig:app:density_of_states}
\end{figure}
Figure~\ref{fig:app:density_of_states} displays the density of states for the top band $E^{\sigma}$ as a function of displacement field $D$ and chemical potential $\mu$.
At filling $\nu=-1$ (dotted line), the DOS exhibits a peak around $D_{\rm{VH}}=21$~meV.
This suggests strong correlation effects in this region even for moderate interaction strengths, which may lead to instabilities towards ordered phases.
Furthermore, the DOS drops rapidly for displacement fields $D>21$~meV at $\nu=-1$. 
This decline is significantly steeper than for $D<21$~meV, indicating a pronounced asymmetry around the Van Hove singularity.

\subsection{Interactions}\label{sec:app:interactions}

We consider three types of bare interactions: a screened Coulomb interaction $V_{\text{Coul}}(\mathbf{q})$, an on-site Hubbard interaction $U$, and an interaction $g$ that is momentum-independent in band space.
Our main results are obtained from $V_{\text{Coul}}(\mathbf{q})=e^2/(2\epsilon_0\epsilon|\vq|\Omega_0)\tanh(|\vq|d/2)$, where $e$ is the electron charge, $\epsilon$ the effective permittivity, $\Omega_0$ the unit cell area and $d$ the distance to the gates. 
The other two types of interaction serve as a comparison to assess the effect of the long range character and common approximations for the interaction. 

\subsubsection{Coulomb interaction} \label{sec:app:intial_interaction_abinitio}

To obtain the initial interaction input in the FRG calculation, we project the screened Coulomb interaction $V_{\text{Coul}}(\vq)$ first into Wannier orbital space, then into the highest valence bands $E^{\sigma}$ using the eigenfunctions of the 3Tri model, see \cref{app:sec:HF:Wannier_model} and finally into form factor basis.

\subsubsection{Hubbard interaction}\label{sec:app:initial_interaction_Hubbard}
For comparison, we also use a local Hubbard interaction as bare interaction in the FRG. 
It is defined in orbital space of the Wannierized 3Tri model.
We then project this interaction to highest valence band using the orbital-to-band transformation from Sec. ~\ref{sec:app:initial_interactions} , which is given by the eigenvectors $\mathbf{Y}_{n,1}(\mathbf{k})$ of the kinetic 3Tri Hamiltonian.

This transformation converts a local (momentum-independent) interaction in orbital space into a momentum-dependent interaction in band space, which is crucial for capturing the physics of the moiré system.
We start with a local Hubbard interaction $U$ in orbital space
\begin{equation}
    \mathcal{H}_{\text{int}}^{\text{orb}}
    =
    \frac{1}{2 }
    \sum_{n,m, \sigma, \sigma'}
    U_{n,m, \sigma, \sigma'}
    n_{n, \sigma}
    n_{m, \sigma} \quad ,
\end{equation}
consisting of intra-orbital ($U^{\rm{intra}}$) and inter-orbital ($U^{\rm{inter}}$) repulsion between electrons with equal/opposite valleys
\begin{align}
    \mathcal{H}_{\text{int}}^{\text{orb}} 
    &= 
    \sum_{n,m, \sigma, \sigma'}
    \left[
        U^{\rm{intra}}
        \delta_{nm}
        (1-\delta_{\sigma \sigma'}) 
        +
        U^{\rm{inter}}
        (1-\delta_{nm})
    \right]
    :n_{n \sigma} 
    n_{m \sigma'}: 
\end{align}
with orbital indices $n,m$ and spin/valley indices $\sigma, \sigma'$,
where
\begin{equation}
    n_{n \sigma}
    =
    \frac{1}{N_\vk}
    \sum_{\mathbf{k}}
    d^\dagger_{\mathbf{k},\sigma,n}
    d_{\mathbf{k},\sigma,n}
\end{equation}
is the number operator for orbital $n$ and valley/spin $\sigma$. 
In the case of inter-orbital coupling $U^{\rm{inter}}$, the valley/spin indices of all four fermions can be identical as the Pauli principle is fulfilled by the different orbitals.

We expand the number operators and reorder the fermionic operators into normal odering
\begin{align}
    \mathcal{H}_{\text{int}}^{\text{orb}} 
    &=
    \frac{1}{2 N_\vq}
    \sum_{\substack{\mathbf{k}_1, \dots, \mathbf{k}_4 \\n,m, \sigma, \sigma'}}
    \delta_{\mathbf{k}_1 + \mathbf{k}_2,\, \mathbf{k}_3 + \mathbf{k}_4} \notag\\
    &\left[
        U^{\rm{intra}}  
        (1-\delta_{\sigma \sigma'})
        d_{\mathbf{k}_1,\sigma,n}^{\dagger}
        d_{\mathbf{k}_2,\sigma',n}^{\dagger}
        d_{\mathbf{k}_3,\sigma,n}
        d_{\mathbf{k}_4,\sigma',n} 
        +
        U^{\rm{inter}}
        (1- \delta_{nm})
        d_{\mathbf{k}_1,\sigma,n}^{\dagger}
        d_{\mathbf{k}_2,\sigma',m}^{\dagger}
        d_{\mathbf{k}_3,\sigma,n}
        d_{\mathbf{k}_4,\sigma',m}
    \right],
    \label{eq:app:hubbard_hamiltonian_orbital_space}
\end{align}
where the Kronecker delta enforces momentum conservation.
The orbital-to-band transformation for the fermions into the top band is again given by the eigenvector matrices $\mathbf{Y}_{n,1}^\sigma(k)$, see \cref{eq:app:Wannier_orbital_to_top_band}.
Analogous to the projection of the Coulomb interaction, we next express the interaction in band basis, i.e., we apply the orbital-to-band transformation given by \cref{eq:app:Wannier_orbital_to_top_band} to all fermionic operators in \cref{eq:app:hubbard_hamiltonian_orbital_space}.
This substitution introduces sums over band indices $n_1, \ldots, n_4$ and factors of the eigenvector matrix $\mathbf{U}(\mathbf{k})$ 
\begin{align}
    \mathcal{H}_{\text{int}}^{\text{band}} 
    &= 
    \frac{1}{2 N_\vq}
    \sum_{\substack{\mathbf{k}_1, \dots, \mathbf{k}_4 \\\sigma, \tau,\sigma_1,\dots, \sigma_4\\n,m}}
    \delta_{\mathbf{k}_1 + \mathbf{k}_2,\, \mathbf{k}_3 + \mathbf{k}_4}
    \delta_{\sigma_1\sigma}\delta_{\sigma_3\sigma}
    \delta_{\sigma_2\sigma'}\delta_{\sigma_4\sigma'}
    \nonumber \\
    &\quad\times
    \left[
        U^{\rm{intra}} \cdot \delta_{nm}(1-\delta_{\sigma \sigma'}) 
        + 
        U^{\rm{inter}} \cdot (1-\delta_{nm})
    \right]
    \nonumber \\
    & \quad\times
    \left(\mathbf{Y}_{n, 1}^{\sigma_1}(\mathbf{k}_1) \right)^\dagger
    \left(\mathbf{Y}_{m, 1}^{\sigma_2}(\mathbf{k}_2) \right)^\dagger
    \mathbf{Y}_{n, 1}^{\sigma_3}(\mathbf{k}_3)
    \mathbf{Y}_{m, 1}^{\sigma_4}(\mathbf{k}_4)
    \nonumber \\
    &\quad\times
    \tilde{\gamma}_{\mathbf{k}_1,\sigma_1}^{\dagger}
    \tilde{\gamma}_{\mathbf{k}_2,\sigma_2}^{\dagger}
    \tilde{\gamma}_{\mathbf{k}_3,\sigma_3}
    \tilde{\gamma}_{\mathbf{k}_4,\sigma_4}
    \quad. \label{eq:app:hubbard_hamiltonian_band_space}
\end{align}
The interaction vertex in band space can be extracted from the above expression as
\begin{align}
    V_{\substack{\sigma_1,\dots,\sigma_4}}^{\rm{Hub}}(\mathbf{k}_1, \dots, \mathbf{k}_4) 
    &=
    \sum_{\substack{\sigma, \sigma',n,m}}
    \delta_{\mathbf{k}_1 + \mathbf{k}_2,\, \mathbf{k}_3 + \mathbf{k}_4}
    \delta_{\sigma_1\sigma}\delta_{\sigma_3\sigma}
    \delta_{\sigma_2\sigma'}\delta_{\sigma_4\sigma'}
    \nonumber \\
    &\quad \times
    \left(\mathbf{Y}_{n 1}^{\sigma_1}(\mathbf{k}_1)\right)^\dagger
    \left(\mathbf{Y}_{m 1}^{\sigma_2}(\mathbf{k}_2)\right)^\dagger
    \mathbf{Y}_{n 1}^{\sigma_3}(\mathbf{k}_3)
    \mathbf{Y}_{m 1}^{\sigma_4}(\mathbf{k}_4) 
    \nonumber \\
    & \quad \times
    \left[
        U^{\rm{intra}} 
        \cdot 
        \delta_{nm}(1-\delta_{\sigma \sigma'}) 
        +
        U^{\rm{inter}} 
        \cdot 
        (1-\delta_{nm})
    \right]
    \quad . \label{eq:app:hubbard_interaction_band_space}
\end{align}
The interaction vertex in band space depends again on all four momenta $\mathbf{k}_1, \ldots, \mathbf{k}_4$ through the eigenvector matrices $\mathbf{Y}^\sigma_{n,1}(\mathbf{k})$. 
Even though, the eigenvectors now come from the Wannierized model, not from the continuum model, they are approximately identical because of the nearly perfect overlap $>99\%$ between the wave functions of the first band of both models. 
Additionally, the band and spin index structure becomes nontrivial, reflecting the mixing of orbitals in the band eigenstates. 
Of course, $U^{\rm{intra}}$ and $U^{\rm{inter}}$ also depend on the orbital itself. 
We list their values in Sec.~\ref{app:DFT3Tri}. 
However, for simplicity, we set $U^{\rm{intra}}=U^{\rm{inter}}$, since this is not the main focus of our study. 
Note that in Ref.~\cite{Guo.2026}, the orbital dependence of $U^{\rm{intra}}$ was kept, while $U^{\rm{inter}}$ was set to zero. Furthermore, we truncate the Fourier expansion used in TUFRG after the first three shells because the projected interaction decays rapidly (see Fig.~\ref{fig:model} c and \cref{fig:app:initial_interactions_2D_HUB})).
The fourth shell of the initial Hubbard interaction was left empty, but still included in the flow to be filled by fluctuations.

\subsubsection{Constant interaction in band space} \label{sec:app:initial_interaction_constant}
As the simplest initial condition, we consider an intra-band interaction which is constant in band space, i.~e., completely lacks non-trivial quantum geometric corrections,
\begin{align}
    \mathcal{H}_\mathrm{int}^\mathrm{band}
    &= 
    \frac{1}{2 N_\vq}
    \sum_{\substack{\mathbf{k}_1, \dots, \mathbf{k}_4 \\\sigma_1,\dots, \sigma_4}}
    V_{\substack{\sigma_1,\dots,\sigma_4}}^{\rm{const}}(\mathbf{k}_1, \dots, \mathbf{k}_4) 
    \tilde{\gamma}_{\mathbf{k}_1,\sigma_1}^{\dagger}
    \tilde{\gamma}_{\mathbf{k}_2,\sigma_2}^{\dagger}
    \tilde{\gamma}_{\mathbf{k}_3,\sigma_3}
    \tilde{\gamma}_{\mathbf{k}_4,\sigma_4} \,,
\end{align}
with
\begin{equation}
    V_{\substack{\sigma_1,\dots,\sigma_4}}^{\rm{const}}(\mathbf{k}_1, \dots, \mathbf{k}_4) 
    =
    g\,
    \delta_{\mathbf k_1+\mathbf k_2,\mathbf k_3+\mathbf k_4}
    \delta_{\sigma_1\sigma_3}
    \delta_{\sigma_2\sigma_4}.
\end{equation}

\subsubsection{Anti-Symmetrization}\label{sec:app:vertex_antisymmetrization}
The flowing FRG vertex is antisymmetric under exchange of ingoing or outgoing fermions. 
Accordingly, the initial condition is obtained by antisymmetrizing the bare interaction,
\begin{align}
    V^{\Lambda_0}(k_1,k_2,k_3,k_4)
    =
    V(k_1,k_2,k_3,k_4)
    -
    V(k_1,k_2,k_4,k_3)\quad. \label{eq:app:antisymmetrization}
\end{align}

\subsubsection{Implementation} \label{sec:app:implementation}
\begin{figure}
    \centering
    \includegraphics[width=.5\linewidth]{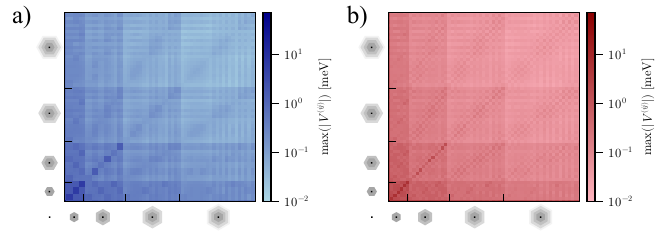}
    \caption{Range of different projected interactions similar to Fig. \ref{fig:model}: maximal magnitude of a) particle-particle (PP, blue) and b) the bare particle-hole (PH, red) and  interaction $V^{ij}$ at distance $\textbf{R}_i$ for $D=0$~meV, $10/\epsilon=1.6$ using the Coulomb interaction.}
    \label{fig:app:initial_interactions_2D_DFT}
\end{figure}

\begin{figure}
    \centering
    \includegraphics[width=.5\linewidth]{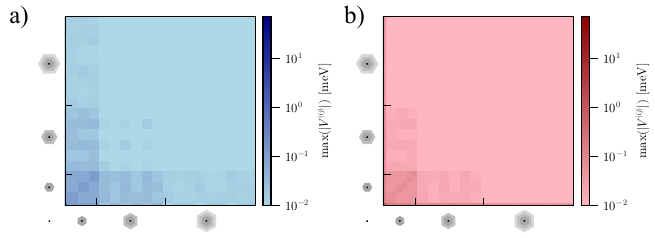}
    \caption{Range of different projected interactions similar to Fig. \ref{fig:model}: maximal magnitude of a) particle-particle (PP, blue) and b) the bare particle-hole (PH, red) and  interaction $V^{ij}$ at distance $\textbf{R}_i$ for $D=0$~meV, $\alpha=1.67$ using the Hubbard model in the 3Tri orbital basis for the projection.}
    \label{fig:app:initial_interactions_2D_HUB}
\end{figure}

Next, we decompose the interaction vertex in the scattering channels (particle-particle, crossed particle-hole, direct particle-hole) and project them onto form factor (Fourier) space. This results in 
\begin{align}
    &
    V^{X,l^{\phantom{\prime}},l^\prime}_{\sigma_1, \dots, \sigma_4}(\mathbf{q}_X) 
    = 
    \int_{\vk_X \vk_X'}
    f_{l}^{\phantom{\ast}}(\mathbf{k}_X^{\phantom{\prime}}) 
    f_{l^\prime}^\ast(\mathbf{k}_X^\prime) 
    V^{X}_{\sigma_1, \dots, \sigma_4}
    (
    F_{1,X}(\mathbf{q}_X, \mathbf{k}_X^{\phantom{\prime}}, \mathbf{k}_X^\prime
    ), \dots, 
    F_{4,X}(\mathbf{q}_X, \mathbf{k}_X^{\phantom{\prime}}, \mathbf{k}_X^\prime)
    )
    \label{eq:app:band_interaction_ff}
\end{align}
where $F_{i,X}(\mathbf{q}_X, \mathbf{k}_X^{\phantom{\prime}}, \mathbf{k}_X^\prime)$ is the function that transforms $\vk_i$ into the appropriate momenta $\vk_X,\vk'_X,\vq_X$ for each channel $X$, see \cref{eq:app:vertex_decomposition_bosonic_representation}: 
\begin{align}
    \vk_1 &= 
    F_{1,\rm{PP}}(\mathbf{q}_{\rm{PP}}, \mathbf{k}_{\rm{PP}}^{\phantom{\prime}}, \mathbf{k}_{\rm{PP}}^\prime) 
    = \mathbf{q}_{\rm{PP}} + \mathbf{k}_{\rm{PP}}
    \nonumber \\
    \vk_2 &= 
    F_{2,\rm{PP}}(\mathbf{q}_{\rm{PP}}, \mathbf{k}_{\rm{PP}}^{\phantom{\prime}}, \mathbf{k}_{\rm{PP}}^\prime) 
    = - \mathbf{k}_{\rm{PP}}
    \nonumber \\
    \vk_3 &= 
    F_{3,\rm{PP}}(\mathbf{q}_{\rm{PP}}, \mathbf{k}_{\rm{PP}}^{\phantom{\prime}}, \mathbf{k}_{\rm{PP}}^\prime) 
    = \mathbf{q}_{\rm{PP}} + \mathbf{k}_{\rm{PP}}'
    \nonumber \\
    \vk_4 &= 
    F_{4,\rm{PP}}(\mathbf{q}_{\rm{PP}}, \mathbf{k}_{\rm{PP}}^{\phantom{\prime}}, \mathbf{k}_{\rm{PP}}^\prime) 
    =- \mathbf{k}_{\rm{PP}}'
    \nonumber \\
    \nonumber \\
    \vk_1 &= 
    F_{1,\rm{PH}}(\mathbf{q}_{\rm{PH}}, \mathbf{k}_{\rm{PH}}^{\phantom{\prime}}, \mathbf{k}_{\rm{PH}}^\prime) 
    = \mathbf{q}_{\rm{PH}} + \mathbf{k}_{\rm{PH}}
    \nonumber \\
    \vk_2 &= 
    F_{2,\rm{PH}}(\mathbf{q}_{\rm{PH}}, \mathbf{k}_{\rm{PH}}^{\phantom{\prime}}, \mathbf{k}_{\rm{PH}}^\prime) 
    =  \mathbf{k}_{\rm{PH}}'
    \nonumber \\
    \vk_3 &= 
    F_{3,\rm{PH}}(\mathbf{q}_{\rm{PH}}, \mathbf{k}_{\rm{PH}}^{\phantom{\prime}}, \mathbf{k}_{\rm{PH}}^\prime) 
    =  \mathbf{k}_{\rm{PH}}
    \nonumber \\
    \vk_4 &= 
    F_{4,\rm{PH}}(\mathbf{q}_{\rm{PH}}, \mathbf{k}_{\rm{PH}}^{\phantom{\prime}}, \mathbf{k}_{\rm{PH}}^\prime) 
    = \mathbf{k}_{\rm{PH}}'+\mathbf{q}_\mathrm{PH}
\end{align}
\begin{figure}
    \centering
    \includegraphics[width=0.2\linewidth]{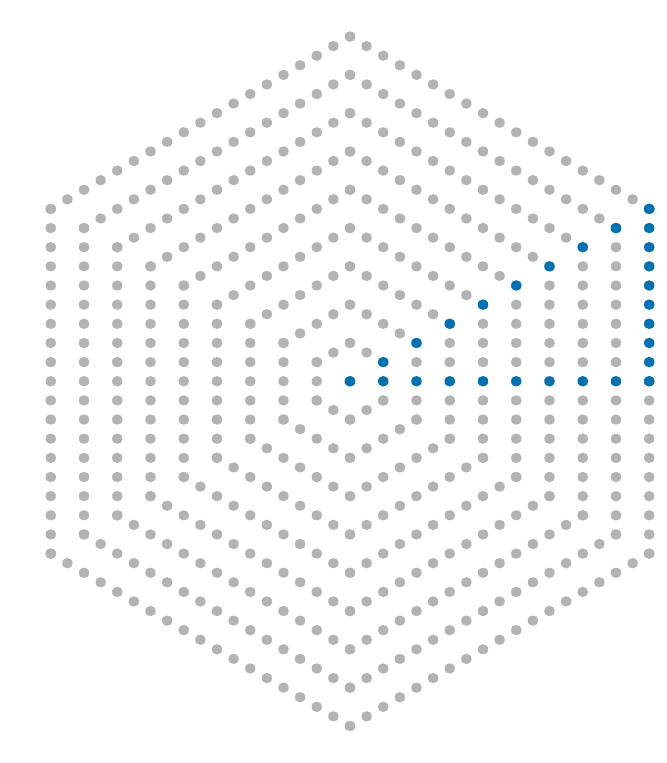}
    \caption{The momentum space grid used for the momenta $\vq$.}
    \label{fig:app:bosonic_grid}
\end{figure}
The grid in momentum space $\mathbf{q}$ is constructed such that $\Gamma \rightarrow K$ and $\Gamma \rightarrow M$ directions are covered by 10 nodes.
This results into $N_\vq=545$ grid points for the entire BZ, see \cref{fig:app:bosonic_grid}.
The form factor basis consists of the 61 plane waves $f_l(\mathbf{k})=e^{i \mathbf k \cdot \mathbf{R}_l}$ labeled according to the real space lattice vectors $\mathbf{R}_l$.
The amount of form factors is chosen such that 4 shells of real space lattice vectors, i.e., up to the eighth nearest-neighbor vectors, are included (cf. Sec.~\ref{sec:app:FRG} and Fig.~\ref{fig:app:triangular_lattice_basis}).
\cref{fig:app:initial_interactions_2D_DFT,fig:app:initial_interactions_2D_HUB} show $\rm{max}(V^{X,(i,j)})$ over all form factor indices $i,j$ for $X=\rm{PP}$ (blue) and PH (red) and  for the Coulomb and Hubbard interaction at comparable strength.
This is essentially a 2D version of \cref{fig:model} c) and visualizes the amount of long-range structure and inter-site variation present in both interactions.
We note in passing that the bare particle-particle interaction is completely repulsive.

To vary the interaction strength, we define $U=\alpha W$ and $g=\alpha W$ for the constant and Hubbard initial interactions, respectively, where $W$ is the bandwidth and $\alpha$ is a phenomenological prefactor mimicking the dependence of the screened Coulomb interaction on the effective dielectric constant $10/\epsilon$. 

For the postprocessing analysis of vertices, we project the vertex' spin/valley dependence onto a Pauli matrix basis to analyze the spin/valley structure of an instability
\begin{align}\label{eq:app:vertex_pauli_basis}
    V^{\Lambda_c, \text{PP},(l,l')}_{\sigma_1, \sigma_2, \sigma_3, \sigma_4}(\mathbf{q}) 
    &= 
    \sum_{i,j=0}^3 
    \tilde{V}^{\Lambda_c, \text{PP},(l,l')}_{i,j}(\mathbf{q})
    t^i_{\sigma_1 \sigma_2} 
    (t^j_{\sigma_3 \sigma_4})^\dagger
    \\
    V^{\Lambda_c, \text{PH},(l,l')}_{\sigma_1, \sigma_2, \sigma_3, \sigma_4}(\mathbf{q}) 
    &= 
    \sum_{i,j=0}^3 
    \tilde{V}^{\Lambda_c, \text{PH},(l,l')}_{i,j}(\mathbf{q}) \sigma^i_{\sigma_1 \sigma_3} 
    \sigma^j_{\sigma_2 \sigma_4}
\end{align}
with Pauli matrices $\sigma^i$, $\sigma^0=\mathbf 1$, and $t_i=i \sigma_i \sigma_y$ (i.e., in this context $t_i $ is unrelated to the hopping amplitudes).
In the pairing case, diagonalizing the diverging vertex $ \tilde{V}^{\Lambda_c, \text{pp},(l,l')}_{i,j}(\mathbf{q})$ and analyzing the leading eigenvector $\Delta(\mathbf{k}, \sigma_i)$ allows us to determine the leading instabilities momentum and spin/valley structure, which reveals the symmetry of the pairing function.

\subsection{Bubbles} \label{sec:app:bubbles}
\begin{figure}
    \centering
    \begin{tikzpicture}
        \begin{feynman}
            \vertex (a);
            \vertex [right=1cm of a] (b);
            \vertex [right=1cm of b] (c);
            \vertex [right=1cm of c] (d);
            \diagram* {
                (a) -- [photon] (b) -- [fermion, half left, looseness=1.5, edge label = k + q] (c),
                (c) -- [fermion, half left, looseness=1.5, edge label' = -k] (b),           (c) -- [photon] (d),
            };
        \end{feynman}
    \end{tikzpicture}
    \hspace{1cm}
    \begin{tikzpicture}
        \begin{feynman}
            \vertex (a);
            \vertex [right=1cm of a] (b);
            \vertex [right=1cm of b] (c);
            \vertex [right=1cm of c] (d);
            \diagram* {
                (a) -- [photon] (b),
                (b) -- [fermion, half left, looseness=1.5, edge label = k + q] (c),
                (b) -- [fermion, half right, looseness=1.5, edge label = k] (c),
                (c) -- [photon] (d),
            };
        \end{feynman}
    \end{tikzpicture}
    \caption{Diagrammatic representation of the particle-particle and particle-hole bubbles. Solid lines represent fermionic propagators and dashed lines represent interaction vertices.}
    \label{fig:app:bubbles_diagram}
\end{figure}
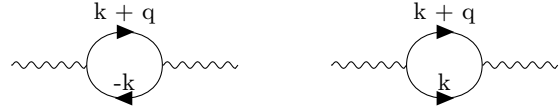

The static fermionic particle-hole or particle-particle susceptibilities, a.k.a. "bubbles",
\begin{equation}
    \Pi^{\text{PH/PP}}_{\mu_1\dots \mu_4}(\mathbf{q}) = 
    \int_\vk 
    \mathcal{G}_{0,\mu_1, \mu_3}(\vk+ \vq)\,
    \mathcal{G}_{0,\mu_2, \mu_4}(\pm \vk),
\end{equation}
with $\mathcal{G}_{0,\mu, \mu'}(\vk) = \mathcal{G}_{0,\mu} \delta_{\mu\mu'}$,
are the fundamental building blocks of the FRG flow equations and encode the temperature dependence of the flow.  
We depict them diagrammatically in Fig.~\ref{fig:app:bubbles_diagram}.

They enter the flow equations via their temperature derivatives and are the source of temperature dependence in the flow.
After performing the Matsubara frequency summation, the bubbles can be expressed in terms of the band energies $E^\sigma(\vk)$ as follows
\begin{align}
    \Pi^\text{PP}_{\mu_1\dots \mu_4}(\vq) 
    &=
    \int_\vk 
    \delta_{\mu_1\mu_3}
    \delta_{\mu_2\mu_4}
    \frac{
        1 
        - 
        n_F(E^{\mu_1}(\vk+\vq)) 
        - 
        n_F(E^{\mu_2}(-\vk))
        }{
        E^{\mu_1}(\vk+\vq) 
        + 
        E^{\mu_2}(-\vk)} \label{eq:app:pp_bubble}
    \\
    \Pi^\text{PH}_{\mu_1\dots \mu_4}(\vq) &= 
    - 
    \int_\vk 
    \delta_{\mu_1\mu_3}
    \delta_{\mu_2\mu_4}
    \frac{
        n_F(E^{\mu_1}(\vk+\vq)) 
        - 
        n_F(E^{\mu_2}(\vk))
        }{
        E^{\mu_1}(\vk+\vq) 
        - 
        E^{\mu_2}(\vk)}, \label{eq:app:ph_bubble}
\end{align}
with the Fermi-Dirac distribution function $n_F(E) = 1/(e^{(E-\mu)/T}+1)$. 
The scale derivative $d/d\Lambda$ in Eq. \ref{eq:app:bubble_derivatives} acts only on the Fermi-Dirac distribution functions, which become sharper as the temperature decreases, leading to an enhancement of the bubbles. 
The numerical evaluation of the bubbles is performed using an adaptive integration routine, which allows us to efficiently capture the sharp features of the integrand at low temperatures without spending excessive computational resources on smooth regions.
The adaptive integration is a triangular variant of the well-known cubature routine \cite{genzRemarksAlgorithm0061980}, generalized on two-dimensional simplices instead of hypercubes, which is more suitable for the hexagonal Brillouin zone of our system \cite{coolsConstructingCubatureFormulae1997,coolsEmbeddedPairCubature1988,keastNumericalIntegrationRecent2012,coolsConstructionMultidimensionalEmbedded1989a}. 
The integration in Eq.~\eqref{eq:app:ph_bubble} and \eqref{eq:app:pp_bubble} is the bottleneck of the FRG flow, as it needs to be performed for each value of $\mathbf{q}$ and for each pair of spin/valleys $\mu_1, \mu_2$ at each energy scale.
However, as self-energy contributions during the flow are neglected, the bubbles do not depend on the vertices and can be precalculated at the beginning of the flow for all values of $\vq$ and pairs of spin/valleys $\mu_1, \mu_2$.
As the energy scales are chosen adaptively to resolve the divergences, the bubbles are pre-calculated on a Chebyshev node grid \cite{boydChebyshevFourierSpectral2001,grossiFastOPInterpolationjl2023} in temperature space. 
This grid consists of 121 nodes.
The step size follows an Euler method and reads
\begin{equation}
    d\Lambda = \rm{max}(\Lambda, \rm{max}(V^\Lambda))/20 \label{eq:app:stepsize}
\end{equation}
where $V$ is the full vertex~\cite{gneistFunctionalRenormalizationSpinless2022}.
Furthermore, below a certain scale $\Lambda_{\mathrm{fr}}$, the divergent behavior of the temperature derivative of the bubbles does not change significantly anymore. 
$\Lambda_{\mathrm{fr}}$ has been chosen in dependence of the bandwidth 
\begin{equation}
   \log_{10}( \Lambda_{\mathrm{fr}})=\log_{10}(W)*1.05 -5
\end{equation}
such that the bubbles are pre-calculated for 5 orders of magnitude.
In principle, if further flow is required, the bubbles values are extrapolated, using the data points in the temperature interval $[\log_{10}(W)*1.05 -5,\log_{10}(W)*1.05 -4]$ as nodal points.
They either follow a universal singular form $\ln T$ (or a power-law in rare cases), or they freeze-out. 
We then interpolate the calculated bubble derivatives up to a temperature $T<T_{\mathrm{fr}}$ and extrapolate them to even smaller temperatures. 
During the flow, we can then use the inter-/extrapolated expressions, which provides a significant speedup compared to direct evaluation of the bubbles at each integration step. 
In practice, this is never necessary for any of our data points.

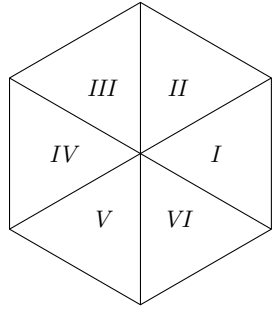
\begin{figure}
    \centering
    \begin{tikzpicture}[scale=2]
        % Define hexagon vertices
        \coordinate (center) at (0, 0);
        \coordinate (A) at (0, 1);
        \coordinate (B) at (0.866, 0.5);
        \coordinate (C) at (0.866, -0.5);
        \coordinate (D) at (0, -1);
        \coordinate (E) at (-0.866, -0.5);
        \coordinate (F) at (-0.866, 0.5);
        
        % Draw hexagon
        \draw (A) -- (B) -- (C) -- (D) -- (E) -- (F) -- cycle;
        
        % Draw triangles from center
        \draw (center) -- (A);
        \draw (center) -- (B);
        \draw (center) -- (C);
        \draw (center) -- (D);
        \draw (center) -- (E);
        \draw (center) -- (F);
        
        % Label triangles with roman numerals
        \node at (0.5, 0.) {\textit{I}};
        \node at (0.25, 0.433) {\textit{II}};
        \node at (-0.25, 0.433) {\textit{III}};
        \node at (-0.5, 0.) {\textit{IV}};
        \node at (-0.25, -0.433) {\textit{V}};
        \node at (0.25, -0.433) {\textit{VI}};
    \end{tikzpicture}
    \caption{Hexagon divided into six triangles with roman numeral labels.}
    \label{fig:app:hexagon_triangles}
\end{figure}

In order to keep the calculation of the bubbles tractable, we limit the number of momenta $\vq$ to $1/6$th of the Brillouin zone (slice \textit{I} in Fig.~\ref{fig:app:hexagon_triangles}).
The rest of the Brillouin zone is related by symmetry operations, which we take into account before inserting the bubbles into the flow equations which are defined and evaluated for the full Brillouin zone.
The symmetries used for this are the $\hat C_3$ rotation and the $\hat T$ time reversal symmetry.

Slice \textit{II} is obtained by applying two $\hat C_3$ rotations and one time reversal $\hat T$ symmetry operation to slice \textit{I}.
The slices \textit{III} and \textit{IV} (\textit{V} and \textit{VI}) are obtained by applying one (two) $\hat C_3$ rotation(s) to slice \textit{I} and \textit{II} respectively.

\begin{figure}
    \centering
    \includegraphics[width=0.3\textwidth]{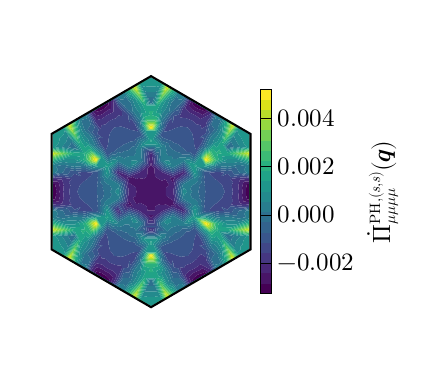}
    \includegraphics[width=0.3\textwidth]{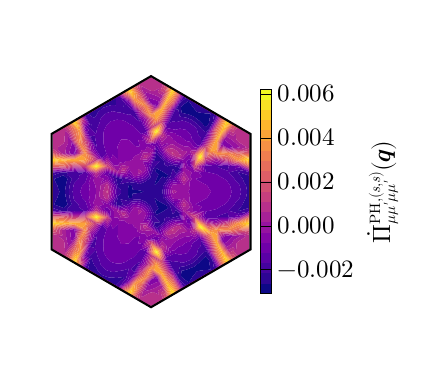}
    \caption{Temperature-dependent derivative of the bubbles for $D=0$~meV at $T=3.67$~K. Left: Intra-valley bubble for band $E^K$. 
    Right: Inter-valley bubble between bands $E^K$ and $E^{K'}$.}
    \label{fig:app:bubbles}
\end{figure}

In the truncated unity FRG, the bubbles (precisely the scale dependent derivatives thereof) are projected onto the form factor basis (see Sec. \ref{sec:app:FRG}).
The form factor dependence of the bubbles has to be respected when applying spacial symmetry operations, which leads to the following symmetry relations for the bubbles in the different slices of the Brillouin zone
\begin{align}
    \Pi^{\Lambda, \text{PP/PH},(n,m)}_{\mu_1\dots\mu_4}(\textit{II}\,) &= 
    \hat{C}_3 \hat{C}_3 \hat{T} \, (\Pi^{\Lambda, \text{pp/ph},(n,m)}_{\mu_1\dots\mu_4}(\textit{I}\,))
    \nonumber \\
    &=
    \left(\Pi^{\Lambda, \text{PP/PH},(\hat{C}_3^2(n)\, \hat{C}_3^2(m))}_{\bar{\mu}_1\dots\bar{\mu}_4}(\textit{I}\,)\right)^\ast \\
    \Pi^{\Lambda, \text{PP/PH},(n,m)}_{\mu_1\dots\mu_4}(\textit{III}\,) 
    &=
    \hat{C}_3 \, (
        \Pi^{\Lambda, \text{PP/PH},(n,m)}_{\mu_1\dots\mu_4}(\textit{I}\,)
        )
    \nonumber \\
    &=
    \Pi^{\Lambda, \text{PP/PH},(\hat{C}_3(n),\hat{C}_3(m))}_{\mu_1\dots\mu_4}(\textit{I}\,) 
    \\
    \Pi^{\Lambda, \text{PP/PH},(n,m)}_{\mu_1\dots\mu_4}(\textit{IV}\,) 
    &=
    \hat{C}_3 \hat{C}_3 \hat{T} \, (\Pi^{\Lambda, \text{PP/PH},(n,m)}_{\mu_1\dots\mu_4}(\textit{II}\,))
    \nonumber \\
    &=
    \Pi^{\Lambda, \text{PP/PH},(\hat{C}_3(n),\hat{C}_3(m))}_{\mu_1\dots\mu_4}(\textit{II}\,)\\
    &\dots \nonumber
\end{align}
where $\bar \mu$ denotes a valley index that is transformed by the time reversal symmetry operation. 
As an example, we show the intra- and inter-valley scale-derivatives of the s-wave particle-hole bubble $\dot{\Pi}^{\Lambda, \text{ph},(1,1)}_{\mathrm{inter/intra}}(\vq)$ for zero displacement field and a temperature of $T=3.67$K in Fig.~\ref{fig:app:bubbles}. 

\subsection{RPA Susceptibilities}\label{sec:app:RPA}
Before turning to the FRG results, we analyze the RPA susceptibilities for twisted WSe$_2$.
For this we use the same setup as for the FRG calculations, \textit{i.e.} the 3Tri bands, 3Tri eigenfunctions and Coulomb interactions as initial conditions, but turn off the inter-channel projections.
That way, we effectively perform an RPA calculation with the same numerical framework as the FRG.
As the energy scale $\Lambda$ is lowered, the RPA susceptibilities are enhanced by the bubbles, and diverge at a critical scale $\Lambda_c$ which can be associated with a mean-field critical temperature $T_c$ of the leading instability.
The RPA amounts to summing up the ladder diagrams in the respective channel, which is equivalent to solving the flow equations with only one channel and no inter-channel projections
\begin{align}
    \frac{d}{d\Lambda} \phi^{\Lambda,X, (l,l')}_{\sigma_1 \dots \sigma_4}&(\vq) = 
    \sum_{n,m}
    \sum_{\mu_1 \dots \mu_4}
    \phi^{\Lambda, X, (l, n)}_{\sigma_1, \sigma_2, \mu_1, \mu_2}(\vq)
    \dot{\Pi}^{\Lambda, X, (n,m)}_{\mu_1 \dots \mu_4}(\vq)
    \phi^{\Lambda, X, (m,l')}_{\mu_3, \mu_4,\sigma_3, \sigma_4 }(\vq)\, ,
\end{align}
where the energy bands enter the r.h.s. via the scale derivative of the respective particle-hole or particle-particle bubble $\dot{\Pi}_{ph/pp}$, while the wave-functions enter via the initial condition for $\phi^{\Lambda,X}$. 
Notice how the $V$'s in Eq. \ref{eq:app:TUFRG_general_flow_equations} on the r.h.s. got replaced by $\phi$'s, \textit{i.e.} there is no inter-channel feedback by $V=\sum_X \phi^X$ anymore.

As the initial interactions are purely repulsive, calculating the pairing susceptibility will never lead to an instability.

However, the particle-hole susceptibility can diverge at certain momenta, indicating a tendency towards magnetic or charge order.
\begin{figure}
    \centering
    \includegraphics[width=0.4\textwidth]{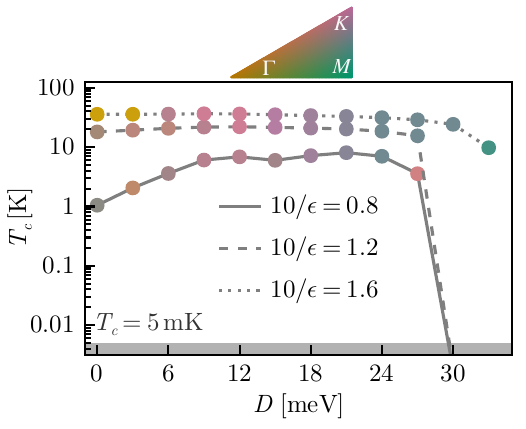}
    \caption{Mean-field critical temperatures from an RPA calculation based on the Coulomb interaction for interaction strength $10/\epsilon$ as function of displacement fields $D$. 
    }
    \label{fig:app:rpa_susceptibilities}
\end{figure}

\begin{figure}
    \includegraphics[width = 0.3\textwidth]{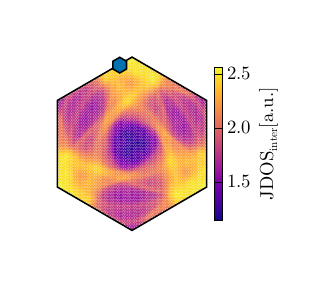}
    \includegraphics[width=0.3\textwidth]{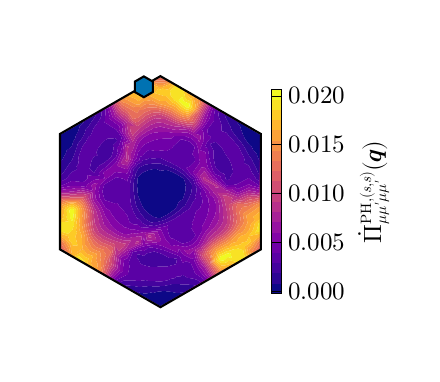}
    \caption{
    Left: JDOS for $D=21$~meV with the IVC-SDW wave vector  (blue marker) of the $10/\epsilon=0.8$ RPA calculation. 
    Right: The $D=21$~meV interpolated temperature derivative of the s-wave inter-valley particle-hole bubble at the critical temperature and the same wave vector (blue marker) of the $10/\epsilon=0.8$ RPA calculation. 
    }
    \label{fig:app:U_21_JDOS_BUB}
\end{figure}
In \cref{fig:app:rpa_susceptibilities}, we show the critical temperature $T_c$ of the leading instability in the particle-hole channel as function of displacement field $D$ for three different interaction strengths $10/\epsilon$. 
The critical temperature $T_c$ follows the density of states trend shown in \cref{fig:app:density_of_states}, with a maximum around the van Hove displacement field $D_{\mathrm{VH}}=21$~meV. 
The color of the marker indicates the transfer momenta $\vq$ of the leading instability, displayed in the hexagonal Brillouin zone inset.

For small displacement fields $D<6$~meV and $10/\epsilon = 1.6$, the transfer momenta of the leading instability are located around the $\Gamma$ point (yellow points in \cref{fig:app:density_of_states}) and they arise in the $\sigma_z$ s-wave channel. 
This indicates a transition towards a spin/valley polarized state. 
For intermediate displacement fields $6$~meV$<D<24$~meV, the leading instability transfer momenta shift to the vicinity of the $K$ points of the hexagonal Brillouin zone and to the $\sigma_{x/y}$ channel. 
This signals an inter-valley-coherent state with additional translation symmetry breaking on the moir\'e scale determined by the possibly incommensurate wave vector at or close to $K$. Such a state is sometimes referred to as inter-valley-coherent antiferromagnet \cite{devakulMagicTwistedTransition2021,FOU24,ZAN21,WAN24d,BRI26,fischerTheoryIntervalleycoherentAFM2025,Guo.2026,kleblCompetitionDensityWaves2023,tuoTheoryTopologicalSuperconductivity2025,zangHartreeFockStudyMoire2021,MUN25,PEN26}. 
For displacement fields beyond $D>21$~meV, the transfer momentum moves to the $M$ points before ending exactly there for $D=33$~meV. 
This instability also corresponds to inter-valley-coherent order, but now with wave vector $\vq$ between $K$ and $M$. 
If we lower the initial interaction strength to $10/\epsilon=1.2$ and below, the transfer momentum of the leading channel in the low displacement field regime $D<6$~meV moves also to the $K$ points, the valley structure being $\sigma_{x/y}$, indicating a transition towards an inter-valley-coherent spin density-wave phase.
The critical scale for displacement fields away from Van Hove drops faster than around Van Hove while lowering the strength of the initial interaction.
But this happens asymmetrically around $D_{\rm{VH}}$. 
The drop above $D_{\rm{VH}}$ is much more pronounced than below $D_{\rm{VH}}$, where an initial interaction regime exists where the critical scale is more than an order of magnitude smaller than around $D_{\rm{VH}}$, but not yet below the cutoff temperature $T_{\rm{cutoff}}=5$~mK.
This coincides with the DOS behaviour.

\cref{fig:app:U_21_JDOS_BUB} left shows the inter-valley JDOS of $D=21$~meV. 
For this value of $D=21$~meV, the RPA instability corresponds to an IVC-SDW order.
The red dot in the figure corresponds to the wave vector of this IVC-SDW instability from RPA and coincides with a maximum of the JDOS. 
\cref{fig:app:rpa_susceptibilities} (right) also shows the wave vector of the same RPA calculation but now with the scale derivative of the s-wave inter-valley particle-hole bubble for $D=21$~meV at the critical scale of said RPA calculation.
As can be seen, the wave vector is consistent with inter-valley Fermi surface nesting.

\subsection{FRG Results}\label{sec:app:fRG_results}
In the following, we present the results of FRG calculations for twisted WSe$_2$ using the three aforementioned initial interactions. 
We analyze the leading instabilities and critical temperatures as functions of displacement field $D$ and interaction strength. 
We have three different types of initial interactions $V^{\Lambda_0}$: a constant interaction in band space (Sec.~\ref{sec:app:initial_interaction_constant}), a local Hubbard interaction in orbital space transformed to band space (Sec.~\ref{sec:app:initial_interaction_Hubbard}), and finally the realistic Coulomb interaction also transformed to band space (Sec.~\ref{sec:app:intial_interaction_abinitio}), all projected onto the highest valence bands $E_1^{\sigma}$. 
We obtain an extended region of the phase diagram for varying interaction strength and displacement field for the Coulomb interaction, and compare for certain regions to the Hubbard and constant interaction. 
Comparing these three cases allows us to understand the role of interaction range and form factors from the projection (or so-called orbital makeup) in driving the instabilities in twisted WSe$_2$. 
To implement the varying interaction strength, the interaction is scaled with a parameter $\alpha$. 
In the case of Coulomb interaction, the interaction strength $10/\epsilon$ can be controlled by the effective permittivity which enters the screened Coulomb interaction. 
In the cases of Hubbard and constant interaction, we use the bandwidth $W$ as a scale, i.e. we set $g=\alpha W$ and $U=\alpha W$.
We set the lower temperature cutoff for the FRG flow to $T_{\text{cutoff}}=5$~mK.
The difference to the RPA calculations lies in the inter-channel projections, which enable the feedback between different channels and thus can lead to different leading instabilities and critical temperatures.
These inter-channel projections happen every flow step and are crucial for developing potential pairing instabilities, as the initial interactions are purely repulsive and thus cannot drive pairing instabilities on their own.

\subsubsection{Constant Interactions}\label{sec:app:fRG_results_constant_interaction}
To develop an initial understanding of band instabilities in twisted WSe$_2$, we begin with a repulsive constant interaction in band space of strength $ g=\alpha \cdot W$ (cf. Sec.~\ref{sec:app:initial_interaction_constant}). In this case, only the energy bands, not the eigenstates of the single-particle Hamiltonian, enter the flow equations. 
\begin{figure}
    \centering
    \includegraphics[width=0.4\textwidth]{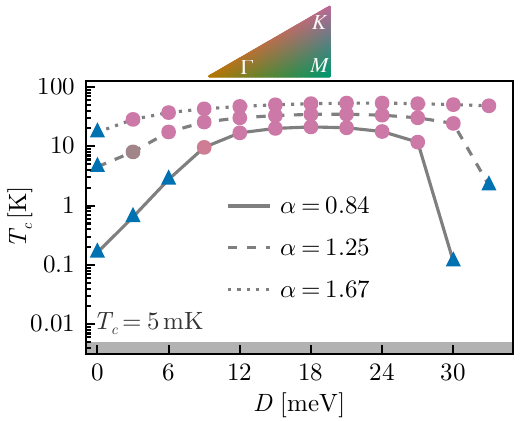}
    \caption{
    FRG critical temperature $T_c$ as function of displacement field $D$ for three different constant interactions where $g = \alpha \cdot W$ where $W$ is the bandwidth. 
    Markers indicate pairing (triangle) or particle-hole (circle) instabilities. 
    The color of the circles encodes the wave vector where the instability occurs according to the inset in Fig.~\ref{fig:fRG_phasediagram}.}
    \label{fig:app:fRG_results_constant_interaction_tc_vs_U}
\end{figure}
\begin{figure}
    \centering
    \includegraphics[trim={0 1cm 0 1cm},clip,width=0.6\textwidth]{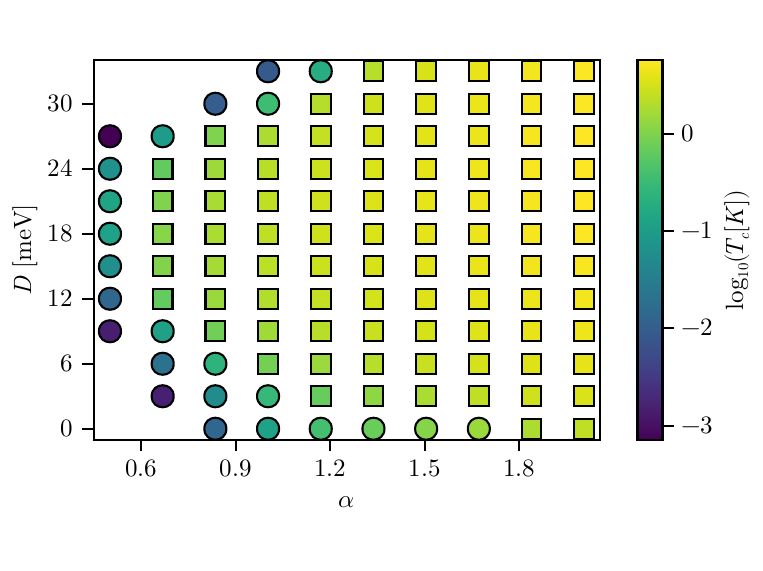}
    \caption{
    FRG phase diagram  as function of displacement field $D$ and constant interaction strength $\alpha$ where $g = \alpha \cdot W$ where $W$ is the bandwidth. 
    The marker-shape indicates the diverging channel with pp=circle and ph=square.
    The missing marker indicate where $T_{\rm{cut-off}}$ was reached.
    }
    \label{fig:app:fRG_results_constant_interaction_phasediagram}
\end{figure}
The results of FRG calculations with a constant interaction  as function of displacement field $D$ are shown in Fig.~\ref{fig:app:fRG_results_constant_interaction_tc_vs_U} and Fig.~\ref{fig:app:fRG_results_constant_interaction_phasediagram}. 
The shape of the marker indicates the leading instability channel, where circles correspond to pairing (pp) instabilities and squares to particle-hole (ph) instabilities.
The color of the marker indicates the transfer momenta of the leading instability, displayed in the hexagonal Brillouin zone inset in Fig.~\ref{fig:fRG_phasediagram}.
As expected, the critical temperature $T_c$ increases with interaction strength $\alpha$ and closely follows the density of states trend shown in Fig.~\ref{fig:app:density_of_states}.
As also expected, the regime around $D_{\mathrm{VH}}\approx 21$~meV is dominated by particle-hole instabilities due to the high DOS at the van Hove singularity. 
This means that an instability criterion for a particle-hole order, which can be satisfied by a repulsive interaction, is indeed satisfied early in the flow before any other interaction renormalisation can become effective. 
The transfer momenta of these instabilities are located at or near the K points of the hexagonal Brillouin zone.
The spin/valley structure of the order parameter indicates an inter-valley coherent phase.
Away from the van Hove singularity for $D\leq6$ depending on $\alpha$, at lower critical temperatures, particle-hole instabilities are superseded by pairing instabilities with zero transfer momentum.
If the initial interaction strength is too low, the critical scale drops below the temperature cutoff $T_{\text{cutoff}}=5$~mK.
The spin structure of the pairing instabilities indicates predominantly $t_0$ d-wave pairing with small admixtures of p-wave $t_z$ character.  
The pairing symmetry belongs to the two-dimensional irreducible representation $E$ of the model's point group $C_{3v}$.
The drop of the critical temperature for $D>D_{\mathrm{VH}}$ is much steeper than for $D<D_{\mathrm{VH}}$, which is consistent with the DOS trend in Fig.~\ref{fig:app:density_of_states}.
Another observation is the drop in $T_c$ for small displacement fields, where the pairing channel wins over the particle hole channel, compared to the RPA data.
This indicates that the inter-channel feedback in the FRG flow acts destructively for the particle-hole instabilities at small displacement fields, which allows the pairing channel to dominate at lower critical scales.
This is a clear example of the importance of inter-channel feedback for determining the leading instability and critical temperature in twisted WSe$_2$.

\subsubsection{Hubbard Interaction}\label{sec:app:fRG_results_hubbard_interaction}
As a next step, before considering Coulomb interactions, we first analyze the simpler model with onsite Hubbard interactions as initial conditions (cf. Sec.~\ref{sec:app:initial_interaction_Hubbard}). 
In this case, not only the band energy but also eigenstates of the three-orbital model are taken into account via the eigenvectors in Eq.~\eqref{eq:app:hubbard_interaction_band_space}.
\begin{figure}
    \centering
    \includegraphics[width=0.4\textwidth]{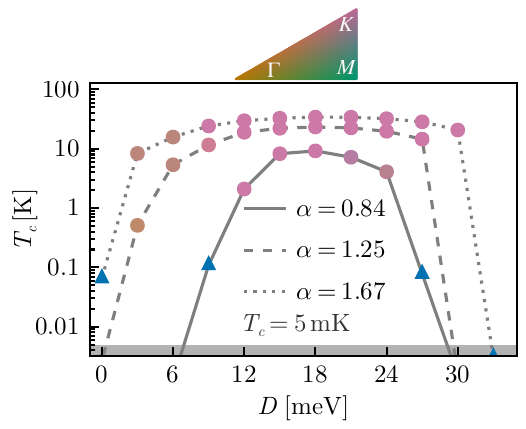}
    \caption{
    FRG critical temperature $T_c$ as function of displacement field $D$ for three different prefactors $\alpha$ for the Hubbard interaction $U = \alpha \cdot W $ with bandwidth $W$. 
    Markers indicate pairing (triangle) or particle-hole (circle) instabilities. 
    The color of the circles encodes the wave vector where the instability occurs according to the inset in Fig.~\ref{fig:fRG_phasediagram}.}
    \label{fig:app:fRG_results_hubbard_interaction_tc_vs_U}
\end{figure}
\begin{figure}
    \centering
    \includegraphics[trim={0 1cm 0 1cm},clip,width=0.6\textwidth]{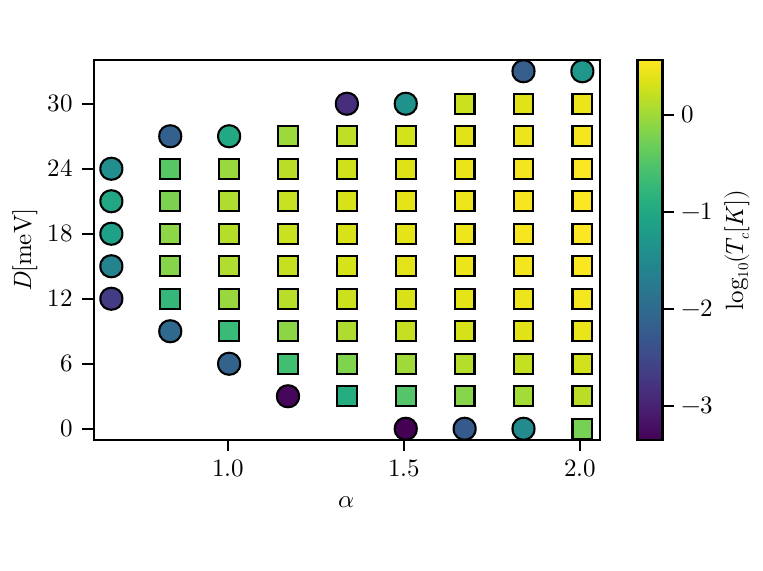}
    \caption{
    FRG phase diagram  as function of displacement field $D$ and Hubbard interaction strength $\alpha$ where $U = \alpha \cdot W$ where $W$ is the bandwidth. 
    The marker shape indicates the diverging channel with pp=circle and ph=square.
    Where markers are missing, the $T_{\rm{cut-off}}$ was reached.}
    \label{fig:app:fRG_results_hubbard_interaction_phasediagram}
\end{figure}
The initial interaction is transformed from the orbital to band basis (see Sec. \ref{sec:app:initial_interaction_Hubbard}) and again scaled by a factor $\alpha=U/W$ with bandwidth W.

Fig. \ref{fig:app:fRG_results_hubbard_interaction_tc_vs_U} and Fig.~\ref{fig:app:fRG_results_hubbard_interaction_phasediagram} show the results of the FRG calculations using this bare interaction as starting point.
Around $D_{\rm{VH}}$, there is again an instability towards an inter-valley coherent spin density wave with transfer momentum $\vq \sim K$.
Away from $D_{\rm{VH}}$, the critical scale drops, which happens again asymmetrically; for lower $D<D_{\rm{VH}}$, the $T_c$ dome prevails longer and the drop comes in softer whereas for higher $D>D_{\rm{VH}}$, the drop in $T_c$ is fast.
For $D\in \{3,6\}$~meV, the transfer momentum of the particle-hole instability moves from the vicinity of $K$ towards the center of the Brillouin zone.
Next to the IVC-SDW dome we find pairing instabilities.
The pairing instabilities (with $T_c>T_{\rm{cutoff}})$ have transfer momentum $\vq = \Gamma$ and emerge again in the $d(k)t_0 + p(k)t_z $ channel, i.e., in the irreducible representation $E$ of the point group. 
However, the extent and critical temperatures of the pairing regime are very small. This mean, in particular, they they do not extend down to zero displacement field. 

\subsubsection{Coulomb Interaction}\label{sec:app:fRG_results_abinitio_interaction}
The final phase diagram we find for the FRG calculations using the 3Tri model and Coulomb interactions as initial setup is shown in Fig.~\ref{fig:fRG_phasediagram} (top).
Starting from the displacement field that corresponds to van Hove filling $D_{\mathrm{VH}}\approx21$~meV at $\nu=-1$ for moderate interactions $10/\epsilon\gtrsim 0.96$, a regime of inter-valley particle-hole instabilities emanates towards larger interactions and occupies an extended regime 9~meV$\lesssim D \lesssim$27~meV around $D_{\mathrm{VH}}$ at $10/\epsilon=1.6$. 
The transfer momentum of those instabilities varies with displacement field $D$ and interaction strength $10/\epsilon$. For $10/\epsilon=1.6$ it is located at the $K$ point for $9$~meV $<D<21$~meV, and shifts in between $K$ and $M$ points for $21$~meV $<D<30$~meV. 
It also shifts to a wave vector between $K$ and $M$ for smaller $10/\epsilon$. 
This instability corresponds to an incommensurate inter-valley-coherent spin density wave order with translation symmetry breaking on the moir\'e and WSe$_{2}$ scale. 
The transfer momentum can be explained by the nesting properties of the Fermi surface, as can be seen in the Joint Density of States (JDOS) plot in Fig.~\ref{fig:app:rpa_susceptibilities} (middle) which experience peaks at exactly those momenta.
The critical temperature $T_c$ of the particle-hole instabilities follows the density of states trend shown in Fig.~\ref{fig:app:density_of_states}, with a maximum around the van Hove singularity. 
For even larger interaction strengths, we eventually expect that FRG also leads to a spin-valley polarized phase as RPA for even larger interactions when a Stoner criterion can be immediately satisfied leaving no room for inter-channel coupling to become important.

Around this particle-hole regime, we find  particle-particle instabilities with zero transfer momentum. 
Their critical temperature $T_c$ is significantly lower than $T_c$ of the particle-hole instabilities. 
The drop in $T_c$ is again very asymmetric around $D_{\mathrm{VH}}\approx21$~meV; it is much more pronounced for $D>D_{\mathrm{VH}}$ so that the pairing regime below $D_{\mathrm{VH}}$ is more extended than above $D_{\mathrm{VH}}$. 
However, in contrast to the cases with bare Hubbard and constant interactions, the pairing regime for $D<D_{\mathrm{VH}}$ now extends up to zero displacement field. 
Furthermore, due to the competition with the particle-hole instability, it is more extended for smaller interactions. The critical temperature decreases, the smaller the initial interaction strength $10/\epsilon$. 
Let us note that due to the small $T_c$, both the pairing regime for $10/\epsilon\approx1.6$ and $D>D_{\mathrm{VH}}$, and the one for small $10/\epsilon$ and $D\gtrsim 0$ could be below an experimental detection threshold so that they would not show up in a measurement.
The spin structure of the pairing instabilities indicates predominantly $t_0$ d-wave pairing with small admixtures of p-wave $t_z$ character (irreducible representation $E$), which is consistent with the results for the constant and Hubbard interaction case.
The $p$-wave $t_z$ admixture scales with the displacement field $D$ and becomes more prominent for larger $D$. 
Such a scaling of the degree of mixing between singlet and triplet pairing with $D$ is expected because the larger $D$ is the stronger intra-valley inversion and with it SU(2) symmetry is broken making the mixture possible.  
We note that even though the pairing symmetry belongs to the two-dimensional irreducible representation of the model's point group $C_{3v}$, we do not see exact degeneracy on the level of the pairing eigenvalues. 
The reason is that $C_3$ acts non-trivially on the projected interaction (see Eq.~\eqref{app:eq:C3action}) and we do not enforce $C_3$ symmetry during the RG flow. 
For zero displacement field $D=0$~meV and strong initial interactions, a competition between particle-hole and particle-particle instabilities interrupts the particle-particle predominance away from the van Hove singularity.
This is a difference to the constant interaction case, where the particle-particle instabilities dominate the low-displacement field regime.

\subsubsection{Comparison}\label{sec:app:comparison}
The phase diagram for the Hubbard interaction \cref{fig:app:fRG_results_hubbard_interaction_phasediagram} and for the constant interaction \cref{fig:app:fRG_results_constant_interaction_phasediagram} differ only quantitatively in the change of the critical scale with displacement field.
$T_c$ is in general higher for the constant interaction than for the Hubbard interaction, which leads to a larger IVC-SDW dome in parameter space.
But both domes are surrounded by only a thin shell of pairing states on the edges of the IVC-SDW dome before the critical scale drops below the cutoff when moving away from the IVC-SDW dome.
The IVC-SDW phase in the Coulomb interaction scenario differs qualitatively.
Even though there is a dome-like structure around $D_{\rm{VH}}$ as well, an IVC-SDW regime survives lowering the bare interaction strength $10/\epsilon$ for vanishing displacement fields.
The wave vectors of the IVC-SDW dome in all three scenarios, Coulomb interaction, Hubbard interaction and constant interaction, follow the RPA results.
They are mostly governed by single-particle properties, i.e.,  vectors that lead to a large overlap of Fermi surfaces or connect Van Hove points. 
The critical scales of the IVC-SDW domes in all cases follows the DOS closely.
The further agreement between RPA and HF indicates that self-energy does not play an important role in identifying the interaction-induced order in the considered regime.
In contrast to the other two schemes where the pairing regime is restricted to the direct vicinity of the IVC-SDW instability, the Coulomb interaction produces a pairing regime which survives moving away from the IVC-SDW dome and extends to $D=0$~meV with an extremely slow decrease in $T_c$. 
Above $D_{\rm{VH}}$, all three interactions produce similar behavior, a rapidly dropping critical scale with a thin pairing phase before dropping below $T_{\text{cut-off}}$.

\subsubsection{Convergence Check}\label{sec:app:convergence_check}
\begin{figure}
    \centering
    \includegraphics[width=0.7\linewidth]{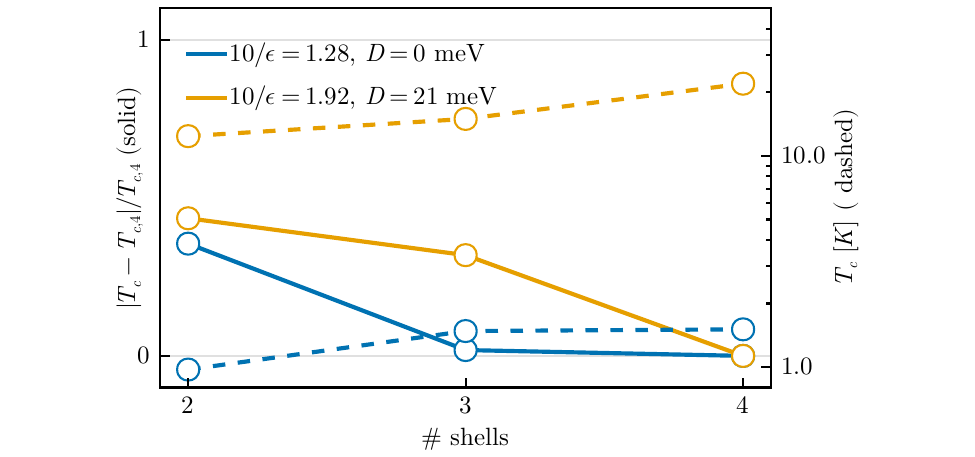}
    \caption{The dependence of $T_c$ (dashed) and the relative error in $T_c$ (solid) as function of the number of shells of Fourier form factors included in the TUFRG calculation for two example data points. Blue corresponds to a point with a pairing instability and yellow to one with an IVC-SDW instability.
    }
    \label{fig:app:convergence_examples}
\end{figure}

Overall, our results change quantitatively, but not qualitatively for varying momentum resolution.
The pairing states in the lower half of the phase diagram are robust against varying the number of shells. 
We show an example evolution of $T_c$ with the number of shells in \cref{fig:app:convergence_examples} for two points in the phase diagram, one in the IVC, one in the pairing regime. 
In both cases, the critical temperature lowers when fewer shells are taken into account, but the relative error decreases and the phase itself does not change.

The relative error in $T_c$ is the largest in the phase boundary regime where changing the number of shells has the biggest impact. In this regime, qualitative changes can occur because phase boundaries move when the number of shells is increased. We find that the IVC phase expands when more shells are taken into account. Qualitatively, however, the phase diagram does not change: the IVC phase emanates from the displacement field where the Van Hove singularity is at the Fermi level and a robust pairing regime occurs for smaller $D$ down to $D=0$.

Convergence checks in simpler models showed that the dependence on the number of points in the bosonic momentum grid $N_q$ is less severe~\cite{gneistFunctionalRenormalizationSpinless2022,gneistCompetingOrdersTriangular2022,PhysRevB.106.125141}, where $T_c$ for $N_q=545$ is already converged.

In future studies, it will be interesting to investigate the impact of other types of momentum resolutions of the two-particle vertex that do not rely on a truncated unity in systems with long-range interactions \cite{Beyer2022}. 

\subsection{Pairing Mechanism}\label{sec:app:pairing_mechanism}
To understand the pairing mechanism in twisted WSe$_2$, we analyze the effective interactions at the critical scale $\Lambda_c$ obtained from the FRG flow.
Time-reversal symmetry constrains pairing structures in twisted WSe$_2$.
We diagonalize the effective interaction in the particle-particle channel at the critical scale $\Lambda_c$ to obtain the leading pairing eigenfunction $\Delta(\vk, t_i)$, which serves as an order parameter for the superconducting instability.
We expect the pairing order parameter to be either a linear combination of $t_0$ and $t_z$  or a linear combination of $t_x$ and $t_y$, as these two sectors transform with opposite parity under time-reversal symmetry: The time-reversal operator $\mathcal{T} = i\sigma_y \mathcal{K}$ (where $\mathcal{K}$ is complex conjugation) requires the pairing order parameter $\Delta  =  \mathbf{v} \cdot (i\boldsymbol{\sigma} \sigma_y)$ to satisfy 
\begin{align}
    \Delta &= \mathcal{T}^{-1}\Delta \mathcal{T}=\sigma_y \Delta^\ast \sigma_y \, .
\end{align}
Applying this constraint to each Pauli component:
\begin{align}
\sigma_y (v_0 i\sigma_0 \sigma_y)^* \sigma_y &= \phantom{-}v_0^* i\sigma_0 \sigma_y \quad \Rightarrow \quad v_0 \text{ real} \nonumber\\
\sigma_y (v_x i\sigma_x \sigma_y)^* \sigma_y &= -v_x^* i\sigma_x \sigma_y\quad \Rightarrow \quad v_x \text{ imaginary} \nonumber\\
\sigma_y (v_y i\sigma_y \sigma_y)^* \sigma_y &= -v_y^* i\sigma_y \sigma_y\quad \Rightarrow \quad v_y \text{ imaginary} \nonumber\\
\sigma_y (v_z i\sigma_z \sigma_y)^* \sigma_y &= \phantom{-}v_z^* i\sigma_z \sigma_y \quad \Rightarrow \quad v_z \text{ real}\nonumber
\end{align}
Writing $v_x = i v'_x$ and $v_y = i v'_y$ with $v'_x, v'_y$ being real, we obtain:
\begin{equation}
\Delta = (v_0 t_0 + v_z t_z) +i (v'_x t_x + v'_y t_y)
\end{equation}
The $(t_0, t_z)$ and $(t_x, t_y)$ sectors transform with opposite parity under time-reversal and cannot mix.
\begin{figure}
    \centering
    \includegraphics[width=1\textwidth]{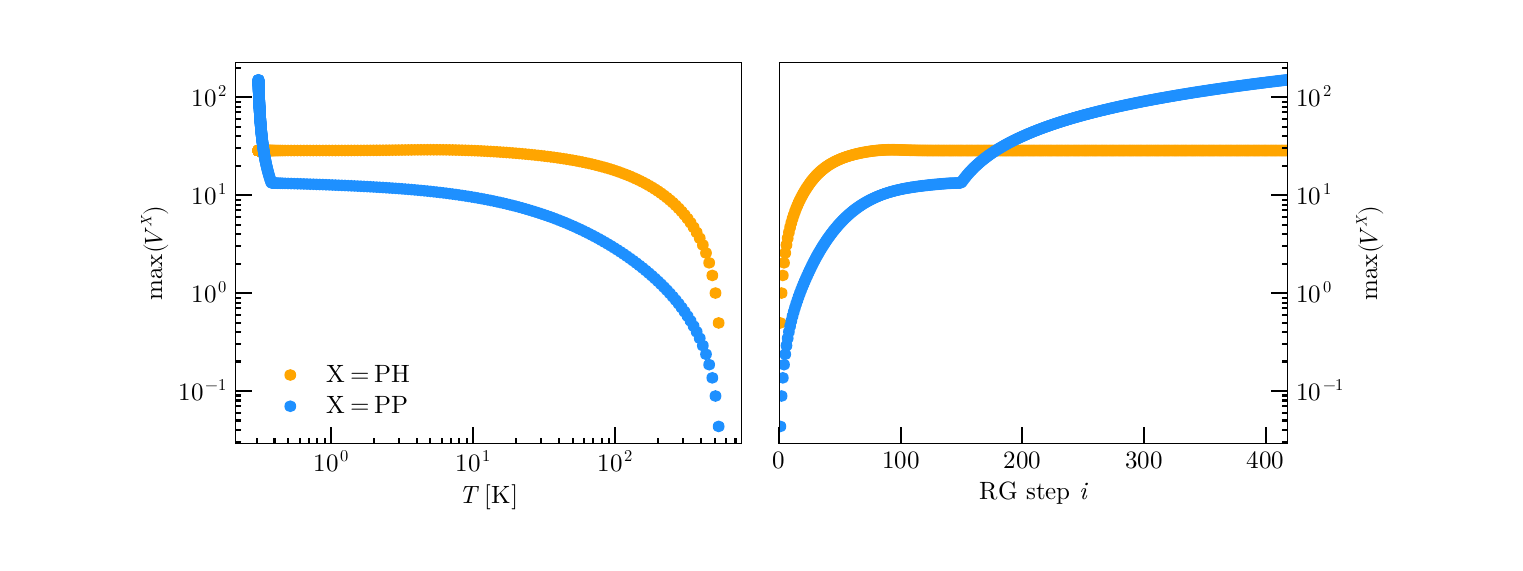}
    \caption{The flow of the absolute maximum of the pairing and the particle-hole effective interactions for $D=9$~meV and $10/\epsilon=0.8$.}
    \label{fig:app:flow}
\end{figure}
\begin{figure}
    \centering
    \includegraphics[width=.3\textwidth]{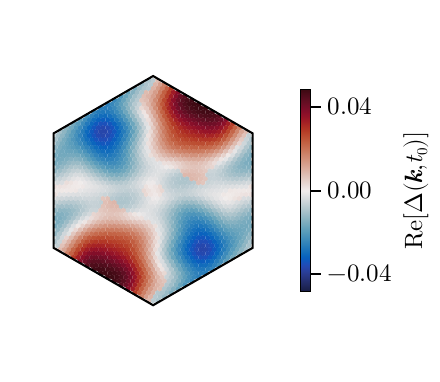}
    \includegraphics[width=.3\textwidth]{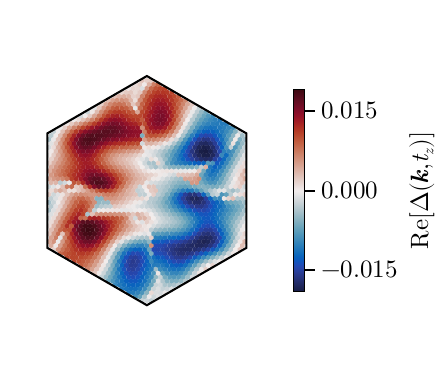}
    \includegraphics[width=.3\textwidth]{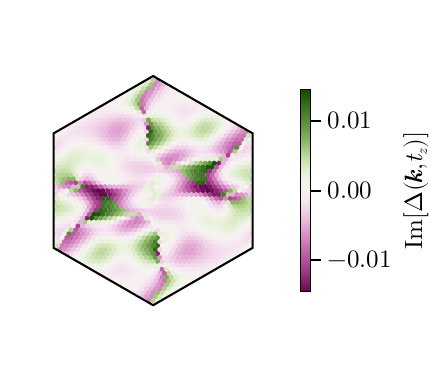}
    \caption{The leading pairing eigenfunction $\Delta(\vk, t_i)$ for the pairing instability at $D=9$~meV and $10/\epsilon=0.8$.}
    \label{fig:app:flow_and_vertex_of_pairing_instability}
\end{figure}

\begin{figure}
    \centering
    \includegraphics[trim={0 2.6cm 0 2.5cm},clip,width=0.6\linewidth]{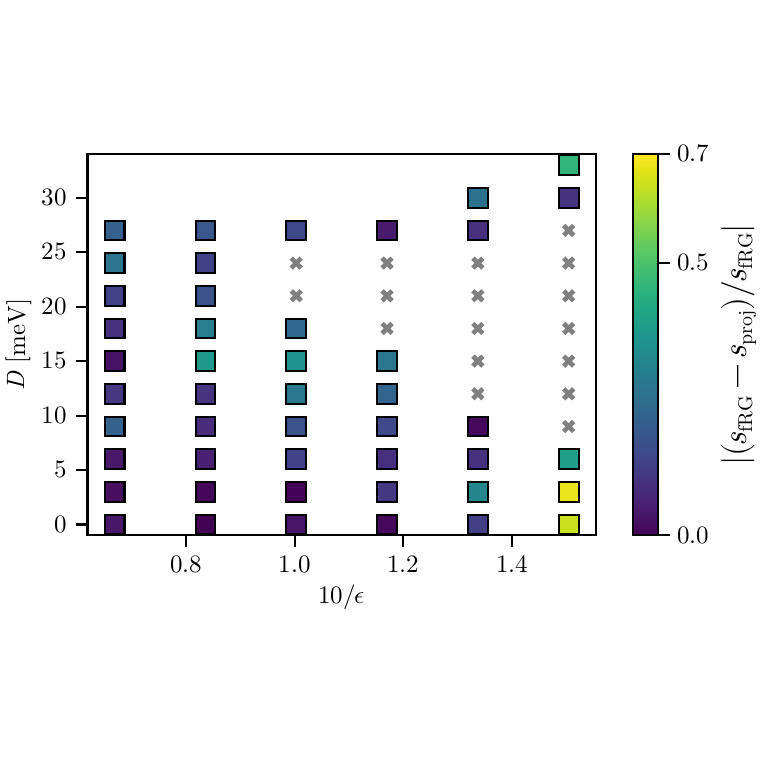}
    \caption{The relative error of the $d(k)t_0$ weight $s$ between the fRG pairing states and the pairing vertex obtained by projecting the RPA vertices into the pairing channel.
    No pairing instability emerged from fRG in the white regions because an IVC-SDW instability occurs instead or because the infrared cutoff $T_{\rm{cut-off}}$ was reached without any divergence in the flow.}
    \label{fig:app:pairing_projection_additional}
\end{figure}
In \cref{fig:app:flow} we show the flow and in \cref{fig:app:flow_and_vertex_of_pairing_instability}  and the leading pairing eigenfunction $\Delta(\vk, t_i)$ for the case of $D=9$~meV and $10/\epsilon=0.8$.
For the flow one can observe the sudden onset of the divergence in the pairing channel at $T_c$, which is a clear indication of a superconducting instability.
We gauged the eigenvector such that the $t_0$ component is purely real for every $\vk$. 
This is why there is no imaginary part in the $t_0$ component plotted.
For this example, the $t_0$ component constitutes $\sim 75\%$ of the total pairing order parameter and exhibits a $d$-wave form factor structure. 
The $t_z$ component makes up the remaining $\sim 25\%$ of the order parameter, manifesting predominantly with a $p$-wave form factor structure in its real part (its imaginary part appears as numerical noise). 
This coexistence of $d$-wave and $p$-wave pairing components is a direct consequence of the broken SU(2) symmetry in twisted WSe$_2$, which removes the protection against parity mixing in the superconducting order parameter. 

To identify the microscopic fluctuations providing the ``pairing glue'' that drives the superconducting instability, we consider a parameter region where a pairing instability occurs within FRG. 
We then disable the inter-channel feedback, run an RPA calculation in the particle-hole crossed channel, and use the resulting effective action as an initial interaction for the pairing channel. 
In a purely electronic mechanism, the effective attraction in the Cooper channel can be viewed as arising from the exchange of collective charge or spin fluctuations. 
By projecting the RPA crossed particle-hole vertex onto the zero-momentum pairing channel, we obtain exactly such a particle-hole flucutation-mediated effective pairing interaction. 
This procedure is analogous to a spin-fluctuation mechanism~\cite{maitiSuperconductivityRepulsiveInteraction2013, prelovsek_spin-fluctuation_2005}. 
The resulting pairing interaction is given by 
\begin{align}
    \hat{P}&[
    V^{\Lambda_c, \rm{RPA, PHC}, (l, l')}_{\sigma_1, \sigma_2, \mu_1, \mu_2}(\vq_{\rm{PHC}})]
     =
     \sum_L 
    \tilde{V}^{\Lambda_c, \rm{RPA, PHC}, (L,-L+l+l')}_{\sigma_1, \sigma_2, \mu_1, \mu_2}
    (-\mathbf{R}_L + \mathbf{R}_{l'})
    e^{-i\vq_{\rm{PHC}} (\mathbf{R}_L -\mathbf{R}_{l'})} 
    \label{eq:app:phc_into_pp_projection}
\end{align}
where $\tilde V(\mathbf R)$ is the Fourier transform of $V$
\begin{align}\label{eq:app:vertex_fourier_transformation}
    \tilde V^{X,(l,l')}_{\sigma_1, \dots, \sigma_4}(\mathbf{R}) 
    = 
    \sum_{\vq_X\in \mathrm{BZ}} 
    V^{(l,l')}_{\sigma_1, \dots, \sigma_4}(\vq_X) \, 
    e^{i\vq_X \cdot \mathbf R}\quad .
\end{align}
The Fourier transform appears because we use plane waves as form factors so that the change from the momentum structure of one channel to another one is precisely the Fourier transform.
The sum over $L$ covers the 4 form factor shells \cite{gneistCompetingOrdersTriangular2022}.

The RPA effective interaction encompasses the bare repulsion diagrammatically dressed by density-wave fluctuations. 
The RPA would lead to an instability of inter-valley-coherent order with wave vector at or close to $K,K'$ as described in Sec.~\ref{sec:app:RPA}.

We prefer using the RPA crossed particle-hole vertex evaluated near its own critical scale over the FRG crossed particle-hole vertex evaluated at the critical scale in the full FRG flow. 
The latter already contains complicated, mixed feedback contributions from interactions in the pairing and direct particle-hole channels, making an isolation of main contribution to the pairing glue less straightforward. 

\begin{figure}
    \centering
    \includegraphics[width=.3\textwidth]{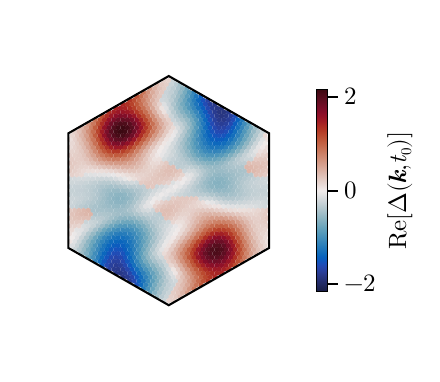}
    \includegraphics[width=.3\textwidth]{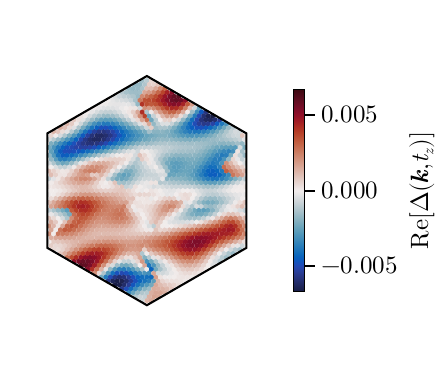}
    \includegraphics[width=.3\textwidth]{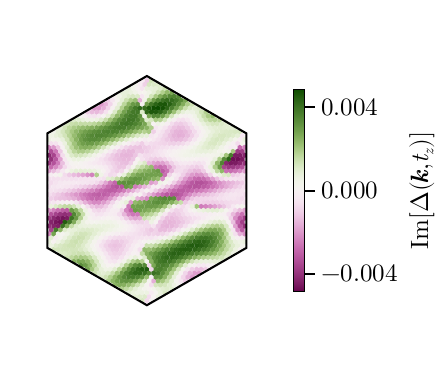}
    \caption{The leading attractive eigenvector of the projected RPA particle-hole crossed channel for $D=9$~meV and $10/\epsilon=0.8$. The weight of the $t_0$ component is $\sim 71\%$ of the total eigenvector. 
    The eigenvector has been gauged such that $\Delta(\vk, t_0)$ is real.}
    \label{fig:app:RPA_projection_plot}
\end{figure}

Through this investigation of the momentum structure of the pairing vertex from the RPA particle-hole fluctuations at zero pairing momentum, we uncover that its leading attractive eigenmode possesses a $d$-wave form factor structure with a dominant $t_0$ character and an admixture from $p$-wave in the $t_z$ component, see Fig.~\ref{fig:app:RPA_projection_plot}. 
This structure reproduces the symmetry of the eigenfunction of the leading FRG instability and the $t_0/t_z$ mixing ratio, see \cref{fig:pairing,fig:app:pairing_projection_additional}.

This analysis is a strong indication that the pairing instability in twisted WSe$_2$ is consistent with IVC spin fluctuations because these are the dominating channel of the RPA particle-hole crossed vertex. 
This mechanism holds throughout the phase diagram and successfully captures also the $t_0/t_z$ ratio behavior, see Fig. \ref{fig:pairing} and \ref{fig:app:pairing_projection_additional}.
The regime with the largest difference between the fRG pairing vertices and the ones obtained by the IVC spin fluctuation mechanism based on the RPA vertices is in the lower right corner of the phase diagram for strong interaction strengths $10/\epsilon \sim 1.6$, where the competition between the particle-particle and particle-hole instabilities in the fRG calculations is the strongest.

Furthermore, we can weight the particle-hole crossed vertex such that transfer momenta away from $\vq=K$ are punished, repeat the analysis and find similar results in the pairing channel.
This analysis can be repeated for other displacement fields $D$ and interaction strengths $10/\epsilon$, where we find the same conclusion that the pairing instability is driven by spin fluctuations with zero transfer momentum.

\begin{figure}
    \centering
    \includegraphics[trim={0 .5cm 0 .5cm},clip,width=0.36\linewidth]{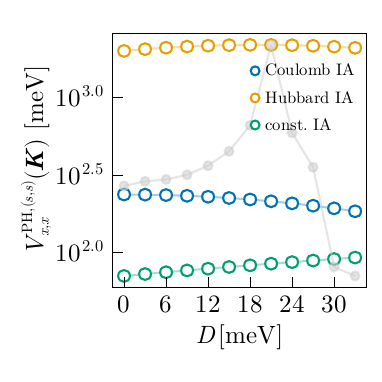}
    \caption{The bare interaction in the IVC-SDW channel for Coulomb, Hubbard, and constant interaction as function of $D$. 
    In gray the DOS as reference.}
    \label{fig:pairing_glue_bare}
\end{figure}

An analysis of the availability of this pairing glue (see \cref{fig:pairing} b), indicates why the Coulomb interaction as initial interaction provides a much more stable pairing state around small $D$.
Compared to the Hubbard interaction, the Coulomb interaction setup offers more of the aforementioned pairing around small $D$.
This observation has to be made for the renormalized IVC fluctuation, as this statement does not hold for the bare interactions, see \cref{fig:pairing_glue_bare}.
However, it is only after renormalization that the IVC fluctuations begin to drive the pairing channel, not from the bare scale on.

%TC:endignore
\end{document}